\documentclass[twocolumn,twocolappendix]{aastex631}
\usepackage{natbib}
\usepackage{chemformula}
\usepackage[T1]{fontenc}
\usepackage{times, graphicx}
\usepackage{url}
\usepackage[version=4]{mhchem}
\usepackage{multirow}
\graphicspath{}
\usepackage{CJK}

\newcommand{\reply}[1]{{#1}}
\newcommand{\replytwo}[1]{{#1}}

\usepackage[version=4]{mhchem}

\received{July 2, 2024}
\accepted{October 23, 2024}

\shorttitle{JWST/MIRI detection of a carbon-rich chemistry in a solar nebula analog}
\shortauthors{Colmenares et al.}

\begin{document}
\begin{CJK*}{UTF8}{gbsn}

\title{JWST/MIRI detection of a carbon-rich chemistry in the disk of a solar nebula analog}

\author[0000-0002-5296-6232]{Mar\'ia Jos\'e Colmenares}
\affiliation{Department of Astronomy, University of Michigan, 1085 South University Avenue, Ann Arbor, MI 48109, USA}

\author[0000-0003-4179-6394]{Edwin A. Bergin}
\affiliation{Department of Astronomy, University of Michigan, 1085 South University Avenue, Ann Arbor, MI 48109, USA}

\author[0000-0003-3682-6632]{Colette Salyk}
\affiliation{Vassar College, 124 Raymond Avenue, Poughkeepsie, NY 12604, USA}

\author[0000-0001-7552-1562]{Klaus M. Pontoppidan}
\affiliation{Jet Propulsion Laboratory, California Institute of Technology, 4800 Oak Grove Drive, Pasadena, CA 91109, USA}
\affiliation{Division of Geological and Planetary Sciences, California Institute of Technology, MC 150-21, 1200 E California Boulevard, Pasadena, CA 91125, USA}

\author[0000-0003-2631-5265]{Nicole Arulanantham}
\affiliation{Space Telescope Science Institute, 3700 San Martin Drive, Baltimore, MD 21218, USA}

\author[0000-0002-0150-0125]{Jenny Calahan}
\affiliation{Center for Astrophysics, Harvard \& Smithsonian, 60 Garden St., Cambridge, MA 02138, USA}

\author[0000-0003-4335-0900]{Andrea Banzatti}
\affiliation{Department of Physics, Texas State University, 749 North Comanche Street, San Marcos, TX 78666, USA}

\author[0000-0003-2253-2270]{Sean Andrews}
\affiliation{Center for Astrophysics, Harvard \& Smithsonian, 60 Garden St., Cambridge, MA 02138, USA}

\author[0000-0003-0787-1610]{Geoffrey A. Blake}
\affiliation{Division of Geological and Planetary Sciences, California Institute of Technology, MC 150-21, 1200 E California Boulevard, Pasadena, CA 91125, USA}

\author[0000-0002-0093-065X]{Fred Ciesla}
\affiliation{Dept. of the Geophysical Sciences, The University of Chicago, Chicago, IL 60637, USA}

\author[0000-0003-1665-5709]{Joel Green}
\affiliation{Space Telescope Science Institute, 3700 San Martin Drive, Baltimore, MD 21218, USA}

\author[0000-0002-7607-719X]{Feng Long (龙凤)}
\affiliation{Lunar and Planetary Laboratory, University of Arizona, Tucson, AZ 85721, USA}
\altaffiliation{NASA Hubble Fellowship Program Sagan Fellow} 

\author[0000-0001-9321-5198]{Michiel Lambrechts}
\affiliation{Center for Star and Planet Formation and Natural History Museum of Denmark, Globe Institute, University of Copenhagen, Oster Voldgade 5–7, 1350 Copenhagen, Denmark}

\author[0000-0002-5758-150X]{Joan Najita}
\affiliation{NSFs NOIRLab, 950 N. Cherry Avenue, Tucson, AZ 85719, USA}

\author[0000-0001-7962-1683]{Ilaria Pascucci}
\affiliation{Lunar and Planetary Laboratory, University of Arizona, Tucson, AZ 85721, USA}

\author[0000-0001-8764-1780]{Paola Pinilla}
\affiliation{Mullard Space Science Laboratory, University College London, Holmbury St Mary, Dorking, Surrey RH5 6NT, UK}

\author[0000-0002-3291-6887]{Sebastiaan Krijt}
\affiliation{ School of Physics and Astronomy, University of Exeter, Stocker Road, Exeter, EX4 4QL, UK}

\author[0000-0002-8623-9703]{Leon Trapman}
\affiliation{Department of Astronomy, University of Wisconsin-Madison, 475 N Charter St, Madison, WI 53706, USA}

\author{the JDISCS Collaboration}

\begin{abstract}

It has been proposed, and confirmed by multiple observations, that disks around low mass stars display a molecule-rich emission and carbon-rich disk chemistry as compared to their hotter, more massive \reply{solar} counterparts. In this work, we present JWST Disk Infrared Spectral Chemistry  Survey (JDISCS) \reply{MIRI-MRS} observations of the solar-mass star DoAr~33, a low-accretion rate T Tauri star showing an exceptional carbon-rich inner disk. We report detections of H$_2$O, \reply{OH}, and CO$_2$, as well as the more complex hydrocarbons, C$_2$H$_2$ and C$_4$H$_2$.
Through the use of thermochemical models, we explore different spatial distributions of carbon and oxygen across the inner disk and compare the column densities and temperatures obtained from LTE slab model retrievals. We find a best match to the observed column densities with models that have carbon enrichment, and the retrieved emitting temperature and area of \ce{C2H2} with models that have $\rm{C/O}=2-4$ inside the 500\,K carbon-rich dust sublimation line. This suggests that the origin of the carbon-rich chemistry is likely due to the sublimation of carbon rich grains near the soot line. \reply{This would be consistent with the presence of dust processing as indicated by the detection of crystalline silicates.} We propose that this long-lived hydrocarbon rich chemistry observed around a solar-mass star is a consequence of the \reply{unusually} low M-star-like accretion rate of the central star, which lengthens the radial mixing timescale of the inner disk allowing the chemistry powered by carbon grain destruction to linger.

\end{abstract}

\section{Introduction} \label{sec:intro}

The predominant approach for understanding the connections between disk composition and the atmospheres of giant planets relies on the elemental carbon-to-oxygen (C/O) ratio \citep{Oberg11_C_O, madhusudhan2012}.  At a baseline level, the temperature-dependent condensation sequence of volatile carriers of oxygen and carbon, such as CO, CO$_2$ and H$_2$O, can lead to spatially varying C/O ratios in both the gas and solid components \citep{Oberg11_C_O}. Beyond the CO snowline, the expectation is that all of the C and O carriers reside mostly in the solids, frozen as icy mantles atop dust grain cores. As dust grains move closer to the star and cross the corresponding local snowlines, the icy mantles will sublimate and alter the abundance of C and O in the gas. This implies that the material available for planet formation will change across the disk, potentially yielding a variety of planet compositions.

However, a variety of mechanisms can influence the gaseous C/O ratio.  For instance, extensive inwards pebble drift can supply abundant water ice to the inner disk.  Once this water ice sublimates it would substantially lower the gaseous inner disk C/O ratio \citep{Booth19, Banzatti20,banzatti2023}. Extensive planetesimal formation and water sequestration beyond the snow line should have the opposite effect \citep[raising inner disk gas-phase C/O;][]{Najita13}. The trapping of ices can also strongly influence the C/O ratio \citep{Ligterink2024}. These concepts focus on the most volatile and abundant components that are found as molecular ices coating the more refractory solid grain cores, which are themselves comprised of solid state organics and silicate minerals \citep{Draine_ism}. 
Closer to the star, within 1~au, there will exist additional dust sublimation fronts that return this less volatile material to the gas.  Of particular importance for the composition of planets forming within this region is the soot line.  This is the theorized sublimation front of carbonaceous organic material \citep{Kress10} that is believed to activate near $\sim$500~K \citep{Li21}.   This sublimation will flood the gas with carbon, elevating the local gaseous C/O ratio, while at the same time lowering C/O in the solids.  Interior to the soot line this would lead to the formation of rocky planets with reduced carbon content, such as the Earth \citep{Li21}.  Beyond the soot line, it would facilitate the formation of worlds  with significant carbon inventories in their mantles, which, through out-gassing, would impact the nascent atmospheric composition \citep{Bergin23}.

Within this context, spectroscopic surveys of protoplanatary disks by the Spitzer Space Telescope \citep{werner2004} isolated a dichotomy where disks surrounding very low mass stars ($<$0.2~M$_\odot$) have elevated  \ce{C2H2}/HCN and HCN/\ce{H2O} flux ratios \citep{Pascucci2009, Pascucci13} in comparison to solar mass T Tauri disks.  This is suggested to be the result of an increase in the gaseous C/O ratio in the innermost regions close to the star \citep{Pascucci13, Najita13}. \reply{Additional support of this dichotomy has also been seen in recent spectra of disks surrounding very low mass stars, obtained with the James Webb Space Telescope (JWST). The disk around 2MASS-J160532-1933159 (0.14~M$_\odot$) exhibits broad optically thick spectral features of \ce{C2H2}, \ce{^{13}C^{12}CH2}, as well as detections of benzene (\ce{C6H6}), and \ce{C4H2} \citep{tabone2023}, also found in ISO-ChaI 147 \citep{arabhavi2024} and Sz~28 \citep{kanwar2024}, along with emission from \ce{HC3N}, \ce{C3H4}, and \ce{C2H6}.} Other than \ce{C2H2}, these species were not isolated in Spitzer spectra and the chemistry surrounding these very low mass stars appears to be hydrocarbon dominated. 

The origin of the hydrocarbon chemistry is uncertain.  One possibility is the destruction of carbon-rich organics near the soot line.  However, this points towards a conundrum: if this chemistry is a result of carbon-rich grain destruction near 500~K, then these emission features should be prevalent surrounding more luminous early type stars, but they are not.
\citet{Mah23} provide an alternative explanation.  They suggest that the supply of carbon-rich gas from the outer disk may provide the fuel to generate hydrocarbons in inner disk gas.  This is more prevalent in lower mass stars which have ice lines that lie close to their stars where the viscous timescales are shorter, in addition to radial drift of particles being more efficient \citep{pinilla2013}. This scenario was proposed to explain the spectrum of Sz 114, which showed strong water emission lines despite having the same stellar mass as J160532-1933159, likely due to the difference in their pebble disks \citep{Xie2023}. Alternatively,  \citet{walsh2015} propose that a combination of weak FUV and X-ray driven chemistry in M dwarf disks is the cause of abundant \ce{C2H2} and HCN observed in the atmosphere of the disk.

With the idea that carbon-rich chemistry so far has been expected to be dominant in low mass stars, in this work we present JWST MIRI observations of the hydrocarbon-rich source, DoAr~33, a 1.1$M_\odot$ star. We detect emission of \ce{H2O}, \ce{CO2}, \reply{OH}, \ce{C2H2}, including its isotopologue \ce{^{13}C^{12}CH2}, \ce{C4H2}, and tentatively \ce{CH4} and \ce{HC3N}, for all of which we retrieve column densities, temperatures and emitting areas, as described in \S~\ref{sec:observations}. With the objective of disentangling the origin of a rich carbon and oxygen chemistry in the inner disk of solar analogs, in \S~\ref{sec:methods} we introduce thermochemical models with varying abundances of the main carbon and oxygen carriers. We calculate abundances, column densities and flux-weighted temperatures for comparison with the detected molecules in \S~\ref{sec:results}.  We then discuss (\S~\ref{sec:discussion}) how DoAr~33 may shed important light on the origin of the hydrocarbon dichotomy found within disks surrounding solar mass and very low mass stars.

\section{Observations and retrievals} 
\label{sec:observations}

\begin{figure*}[t!]
    \centering
    \includegraphics[width=18cm]{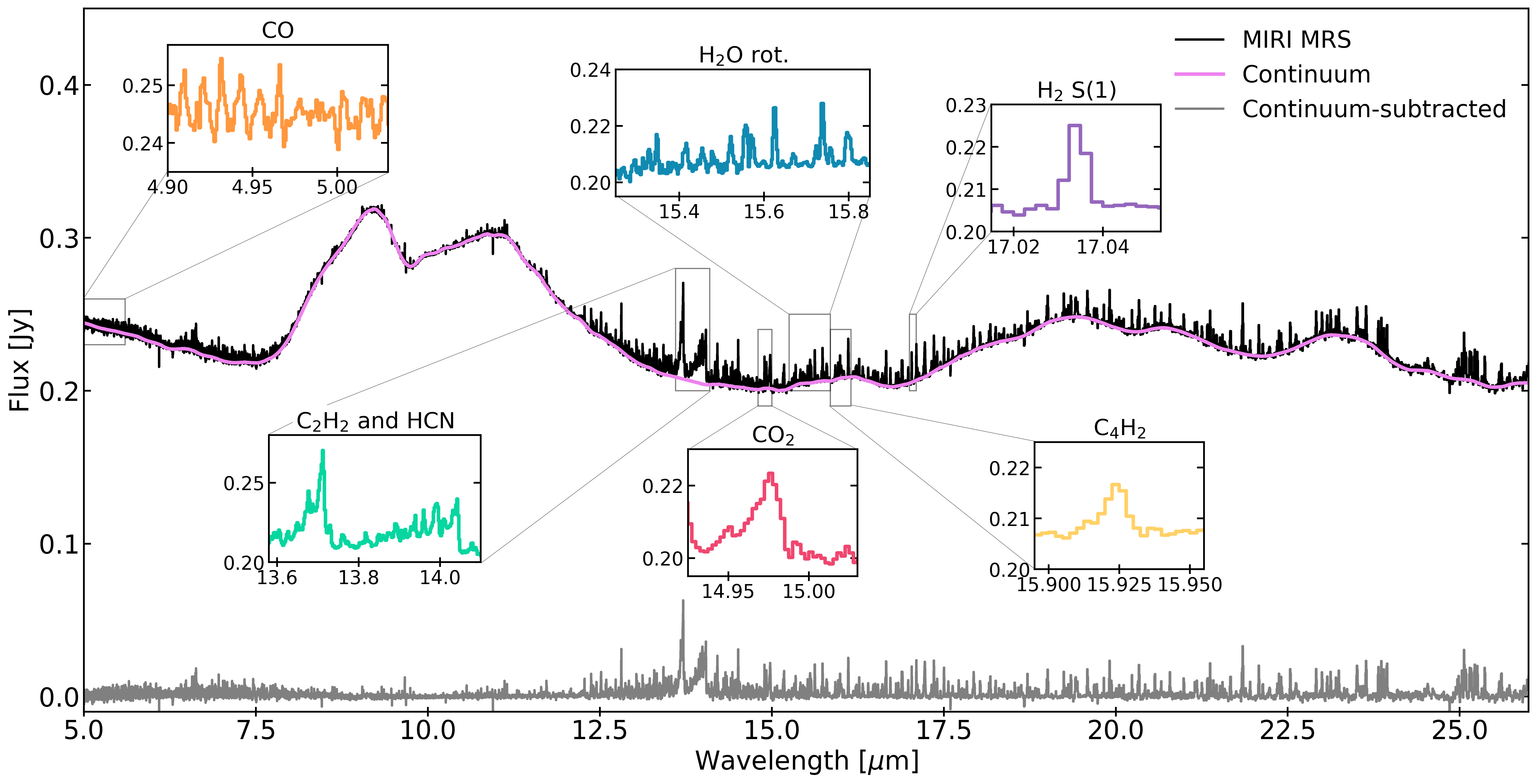}
    \caption{JWST MIRI-MRS spectrum of DoAr~33. The insets show isolated emission features of some of the molecules detected, and the calculated continuum is shown as a pink line. Most of the \reply{unmarked} emission lines correspond to water vapor. \reply{We discuss the presence of dust emission features in \autoref{app:dust}.}}
    \label{fig:spectrum}
\end{figure*}

\subsection{Observations and data reduction}
We present JWST \reply{\citep{rigby2023}} MIRI \reply{\citep{wright2023}} MRS \reply{\citep{wells2015,argyriou2023}} observations of DoAr~33, observed as a part of the JWST Cycle 1 GO1584 program (PIs C. Salyk and K. Pontoppidan), and the JWST Disk Infrared Spectral Chemistry Survey \citep[JDISCS;][]{pontoppidan2024}. This source was acquired by JWST using the MRS target acquisition procedure along with the neutral density filter. We used the four-point dither pattern in the negative direction, and observed in all three MIRI-MRS grating settings such that we had full spectral coverage (4.9-28$\mu$m). The exposure time per module was set to 1687.224\,s in order to achieve S/N$\sim$100 at the longest wavelengths. 

\begin{deluxetable}{lcc}
\label{tab:stellar_params}
\tablecaption{Stellar and disk properties of DoAr~33}
\tablehead{\colhead{Property} & \colhead{Value} & \colhead{Reference}}
\startdata
Spectral type & K4 & \cite{bouvier1992}\\
Distance & 139 pc & \cite{gaia2018}\\
$T_{\rm eff}$ & 4467 K& \cite{andrews2010}\\
Luminosity & 1.5\,$L_\odot$& \cite{andrews2010} \\ 
Mass & 1.1\,  $M_\odot$& \cite{Andrews18}\\
Inclination & 41.8$^\circ$& \cite{Huang18}\\ 
Accretion rate & $2.52\times10^{-10}\,M_\odot$/yr& \cite{cieza2010}\\
\enddata
\end{deluxetable}

The spectrum was reduced with the standard JDISCS reduction pipeline, explained in detail in \cite{pontoppidan2024}. \reply{This custom pipeline produces a higher spectral contrast than the standard JWST pipeline by using observations of asteroids as spectral response function calibrators, which produces higher SNR for Channels 3 and 4 when compared to a traditional calibration using stars.} In the case of DoAr\,33, the \reply{1.15.0} version of the JWST Calibration Pipeline \citep{bushouse_2024_1_15_0} and the \reply{jwst\_1253.pmap} Calibration Reference Data System context were used. The reduced spectrum is shown in \autoref{fig:spectrum}. The process to compute the continuum involves an iterative procedure as described in \citet{pontoppidan2024}. This method begins by applying a median filter to the spectrum to smooth it out and then comparing it with past iterations. During each iteration, wavelength channels exhibiting a lower flux density than the previous iteration are kept, while rejected ones are replaced through linear interpolation; a second-order Savitzky-Golay filter is utilized to further smooth the result. \reply{The regions 13.4-14.1$\mu$m and 14.95-15$\mu$m are excluded from the continuum calculation to avoid overestimation, because broad molecular features from the Q-branch of organic molecules produce a pseudo-continuum.}

\subsection{LTE slab model retrievals}
\label{subsec:slab_models}

\begin{figure*}
    \centering
    \includegraphics[width=18cm]{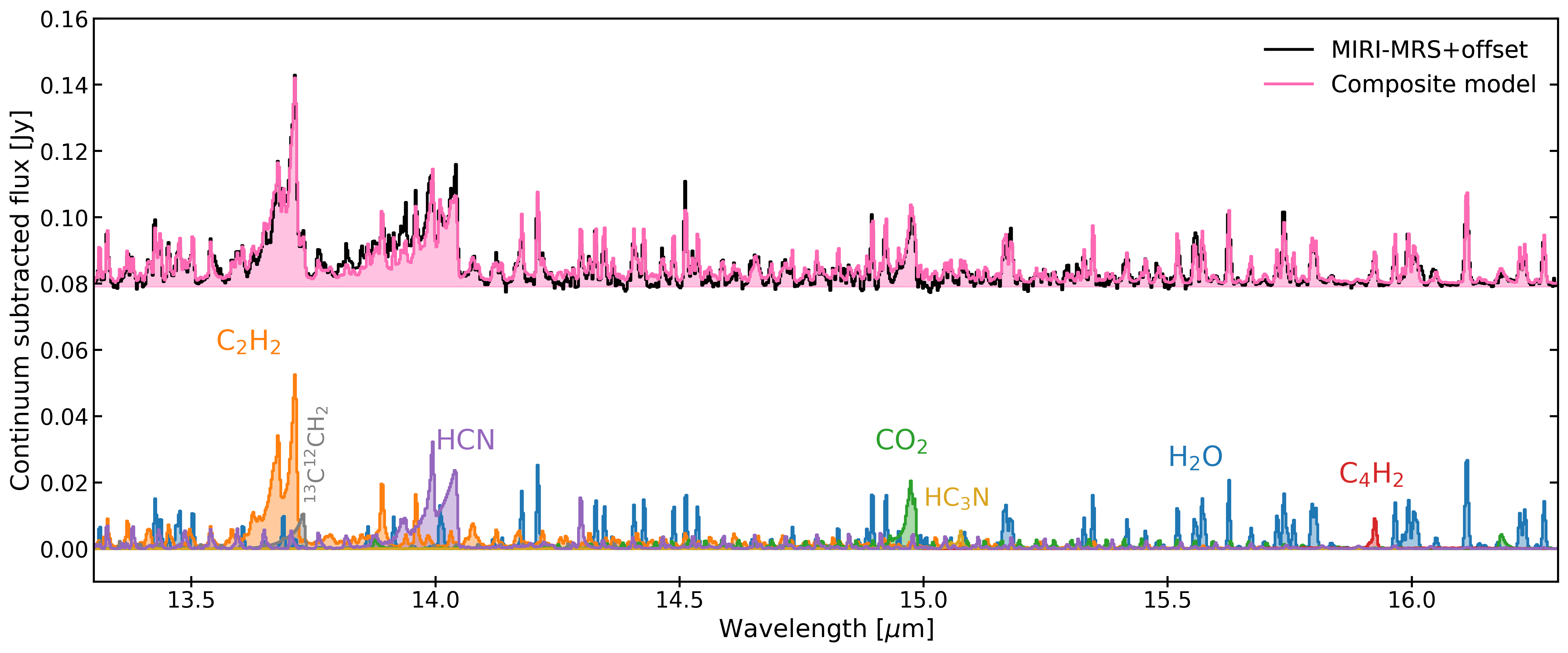}
    \caption{Continuum-subtracted spectrum and best-fit slab models for the molecules detected. The data were offset vertically for clarity. The pink shaded region shows the composite model of \ce{C2H2}, \ce{^{13}C^{12}CH2}, \ce{HCN}, \ce{CO2}, \ce{H2O}, \ce{C4H2} and \ce{HC3N}. We show a closer look of the \ce{HC3N} feature in \autoref{fig:hc3n}.}
    \label{fig:spectrum_model}
\end{figure*}

To benchmark the thermochemical models of DoAr~33, we first retrieve best-fit column densities, temperatures, and emitting areas for each detected molecule using the local thermal equilibrium (LTE) slab models in \texttt{spectools\_ir} \citep{spectools2022}. These models assume an isothermal slab of gas that can be characterized by only its temperature, line-of-sight column density, and emitting area. The emitting area can be interpreted as a circle of radius $R_{\rm slab}$, in which case an emitting radius can be obtained by assuming $A = \pi R_{\rm slab}^2$.  However, the emission may instead arise from one or more rings with total area $A$. The code uses the line parameters from the HITRAN database \citep{gordon2022} to model the emission spectra of different molecules, allowing us to directly compare to the observations. \reply{The code takes into account mutual line overlap, which has been shown to cause saturation in regions of dense line clustering for the organics \citep{tabone2023} and in ortho-para line pairs for water \citep{banzatti2024}.} Despite their simplicity, LTE models have been able to successfully reproduce infrared molecular spectra observed with Spitzer \citep[e.g.,][]{Carr11,salyk2011} and more recently JWST \citep[e.g.,][]{Gasman2023,Xie2023,munoz-romero2024,temmink2024}. 

We follow a procedure similar to previous works \citep{salyk2011,grant2023}, to obtain the parameters from the slab models. We run a grid of slab models with $N_{\rm col}=10^{14}-10^{22}\, \rm{cm}^{-2}$, in steps of $\log N_{\rm col} =0.166$ and $T_{\rm slab}=100-1500\,\rm K$, in steps of 25\,K. These models are generated within the spectral window of 12--16.5~$\mu$m and then compared with the MRS spectrum through a $\chi^2$ minimization. \reply{The only exception are the OH models, which are run until $T_{\rm slab}=3000\,\rm K$, and within a window of 12--28~$\mu$m to maximize the amount of features included in the fit.} We convolve the models at each wavelength to the corresponding MIRI-MRS resolving power, \reply{by adopting the spectral resolution measured by \citet{pontoppidan2024}, which, in the spectral range considered, is in agreement with previous measurements \citep{labiano2021,jones2023}}. Then, \reply{we select a spectral window for each molecule to calculate the reduced $\chi^2$ and find the best fit parameters. The details of the $\chi^2$ calculation are explained in \autoref{app:slab_models}, and the spectral ranges used to calculate it are shown in \autoref{fig:slab_specs} and \autoref{fig:slab_specs_oh}}. For each column density and temperature pair in the grid, we adjust emitting area to minimize $\chi^2$ for that pair.  Then, we find the lowest of those $\chi^2$ values amongst the pairs in the grid. The resulting $\chi^2$ maps for each molecule can be found in \autoref{app:slab_models}.

Molecules like C$_2$H$_2$, HCN, H$_2$O \reply{and OH} have a significant amount of overlap in their emission; therefore, we \reply{implement an additional step to further isolate their emission. First, we find the best-fit model for \ce{H2O}.} Then, prior to simulating \ce{CO2}, we subtract the best-fit \ce{H2O} model to eliminate interfering features that could influence the fitting. Similarly, we subtract the best \ce{CO2} model from the residual spectrum before fitting the rest of the molecules. This ensures that we are only taking into account the most significant feature for each species.  In the case of C$_2$H$_2$, we simultaneously fit its spectrum with $^{13}$C$^{12}$CH$_2$ and HCN, due to the overlap in wavelength of their emission features. \reply{We test all possible combinations of column densities and temperatures within the specified grid, allowing the values to vary independently.} We report the parameters of the best fit models for each molecule in \autoref{tab:slab_models}, and we show the fit of the 13--16.5\,$\mu$m region in \autoref{fig:spectrum_model}. The individual models, along with the residuals, are shown in \autoref{fig:slab_specs} \reply{and \autoref{fig:slab_specs_oh}}.  

In the case of \ce{H2O}, we only focus on the rotational transitions detected around 13--16.5\,$\mu$m, for which we found a single component with \reply{$T=600\,$K}, \reply{$N_{\ce{H2O}}=3\times 10^{18}\,\rm{cm}^{-2}$} and \reply{$R_{\rm{slab}}=0.27\,$au }is able to reproduce the features. More detailed modeling including more than one temperature components or a temperature gradient could be implemented to simultaneously fit for the rotational transitions across the MIRI bandpass \citep{banzatti2023,munoz-romero2024,schwarz2024,temmink2024b}, however, this lies beyond the scope of this work. The water rotational emission is believed to arise from larger disk radii than for the vibrational mode emission \citep{Gasman2023}, therefore, this approach allows us to isolate water emission that is co-spatial with that of the organics in the inner disk.

We fit the $Q$ branch of \ce{C2H2} and \ce{^{13}C ^{12}CH2} at $\sim$13.7$\mu$m, while limiting emitting area \reply{and temperature} of both species to be the same. \reply{We find that models with $T_{\ce{C2H2}}=500\,$K, $R_{\rm{slab}}=0.17\,$au, $N_{\ce{C2H2}} = 10^{17}\,$cm$^{-2}$, and \ce{C2H2}/\ce{^{13}C^{12}CH2}\,$=4.64$ are able to reproduce the full acetylene feature}.  Given the local interstellar medium $^{12}$C/$^{13}$C ratio of 68, this suggests that $^{12}$C$_2$H$_2$ emission is optically thick,  although not optically thick enough to blend into a pseudo-continuum as observed in other carbon-rich sources \citep{tabone2023}. Because of this, and the large degeneracies that are present when simultaneously fitting three species, we determine the actual \ce{C2H2} column from \ce{^{13}C^{12}CH2} with an assumed isotopic ratio. We discuss this further in \S~\ref{subsec:column_densities}. 

\reply{We also detect the presence of rotational lines of OH, which we fit across a larger wavelength range to obtain a better constrain of the parameters. The regions used for the $\chi^2$ fit are shown shaded in \autoref{fig:slab_specs}. The best fit temperature, $T_{\rm {OH}}=950$\,K, is higher than that of water ($T_{\rm {\ce{H2O}}}=600$\,K), with a low column density of $N_{\ce{OH}}=3\times 10^{14}\,\rm{cm}^{-2}$. Despite its relatively high flux at longer wavelengths, the column density and temperature have large uncertainties. We also checked for the presence of lines due to prompt emission between $10-12\,\mu$m caused by UV photodissociation of water \citep{Najita10}. 
For reference, we created a high-temperature (6000\, K) OH slab model, which we show in the top panel of  \autoref{fig:slab_specs_oh}. The insets show that some of the lines between $10-12\,\mu$m show the $\Lambda$-doublet asymmetry previously observed with Spitzer \citep{carr2014}, and predicted by \citet{tabone2021,tabone2024}, indicative of prompt emission. These features are very weak compared to the thermal OH observed at longer wavelengths, suggesting that this source has a low UV field.} 

In the case of di-acetylene (\ce{C4H2}), we only detect emission from the $Q$ branch at $\sim$15.9$\mu$m, which we are able to fit with a model with \reply{$N_{\ce{C4H2}} = 2\times 10^{15}\,$}cm$^{-2}$, \reply{$T=150$\,K}, and \reply{$R_{\rm{slab}}=1.36\,$au}. We also explored models where we fixed the temperature and emitting area \reply{to the one} found for C$_2$H$_2$.  This is discussed further in \S~\ref{sec:c4h2}. We also report a tentative detection of methane (\ce{CH4}) at $\sim$7.65$\mu$m and \ce{HC3N} at $\sim$15.07$\mu$m, which we further discuss in \autoref{app:ch4}. \reply{We also detect the presence of CO emission around 5\,$\mu$m, however, because these feature are from optically thin transitions at high energy levels \citep[e.g.,][]{temmink2024}, they cannot be used to determine the CO column density.}

\begin{deluxetable}{lccc}
\tablecaption{Best fit parameters from the slab models}
\tablehead{\colhead{Molecule} & \colhead{log $N_{\rm col}$ [cm$^{-2}$]} & \colhead{$T_{\rm slab}$ [K]}&\colhead{$R_{\rm slab}$ [au]}}
\startdata
\ce{H2O} &  \reply{18.5 $^{+0.4}_{-0.3}$} & \reply{600} $^{+106}_{-85}$ & \reply{0.27} \\ 
\ce{CO2} &  \reply{17.8} $^{+1.0}_{-0.5}$ & \reply{200} $^{+65}_{-57}$ &\reply{0.77}\\ 
\ce{C2H2}$^{\rm a}$ & \reply{17.0 $^{+0.3}_{-0.3}$} & \reply{500$^{+87}_{-74}$}&\reply{0.17}\\ 
\ce{^{13}C^{12}CH2}$^{\rm b}$ &  \reply{16.3} & \reply{(500)} & \reply{(0.17)}\\
\ce{HCN} &  14.0   & 600 & \reply{3.99}\\
\ce{C4H2} & \reply{15.3}  & \reply{150} & \reply{1.36} \\
\reply{\ce{OH}} &  \reply{14.5}  &\reply{950}& \reply{1.14} \\
\enddata
\tablecomments{The uncertainties are reported only for cases with a closed 1$\sigma$-contour, based on \autoref{fig:chi2}. (a) This column is uncertain, see \S~\ref{sec:results} for a discussion. (b) The emitting area \reply{and temperature} are fixed to that of \ce{C2H2} during the fitting.} 
\end{deluxetable}
\label{tab:slab_models}

\

\

\section{Methods} 
\label{sec:methods}

\subsection{Disk modeling}
\label{subsec:disk_modeling}

For modeling the disk, we used the  Dust and Lines (DALI) thermochemical code \citep{Bruderer12,bruderer2013}, with the extensions implemented by \citet{bosman2022} to reproduce the inner disk regions more accurately. These extensions include increased formation of H$_2$, water UV-shielding, and dissociation heating \citep{Glassgold15}. As DoAr~33 shares a similar spectral type classification (K4) with AS~209 (K5), we use the AS~209 stellar spectrum from \citet{Zhang21_mapsco}. 
The UV emission, generated by accretion \citep{Gullbring2000}, is an important driver of the surface chemistry in low mass T Tauri systems \citep{woitke09, Kamp2024}. Based on the lower accretion rate of DoAr~33 (2.5 $\times$ 10$^{-10}$~M$_{\odot}/$yr) compared to AS 209 (5 $\times$ 10$^{-8}$~M$_{\odot}/$yr), we decreased the FUV emission \reply{by a factor of $\sim$100} to account for the difference in accretion rates between the two sources. \reply{The low UV flux is consistent with the lack of prompt OH emission due to \ce{H2O} photodisociation, as discussed in \S~\ref{subsec:slab_models}.} We show the spectra of both sources in \autoref{fig:SED}.

\begin{figure}
    \centering
    \includegraphics[width=8.5cm]{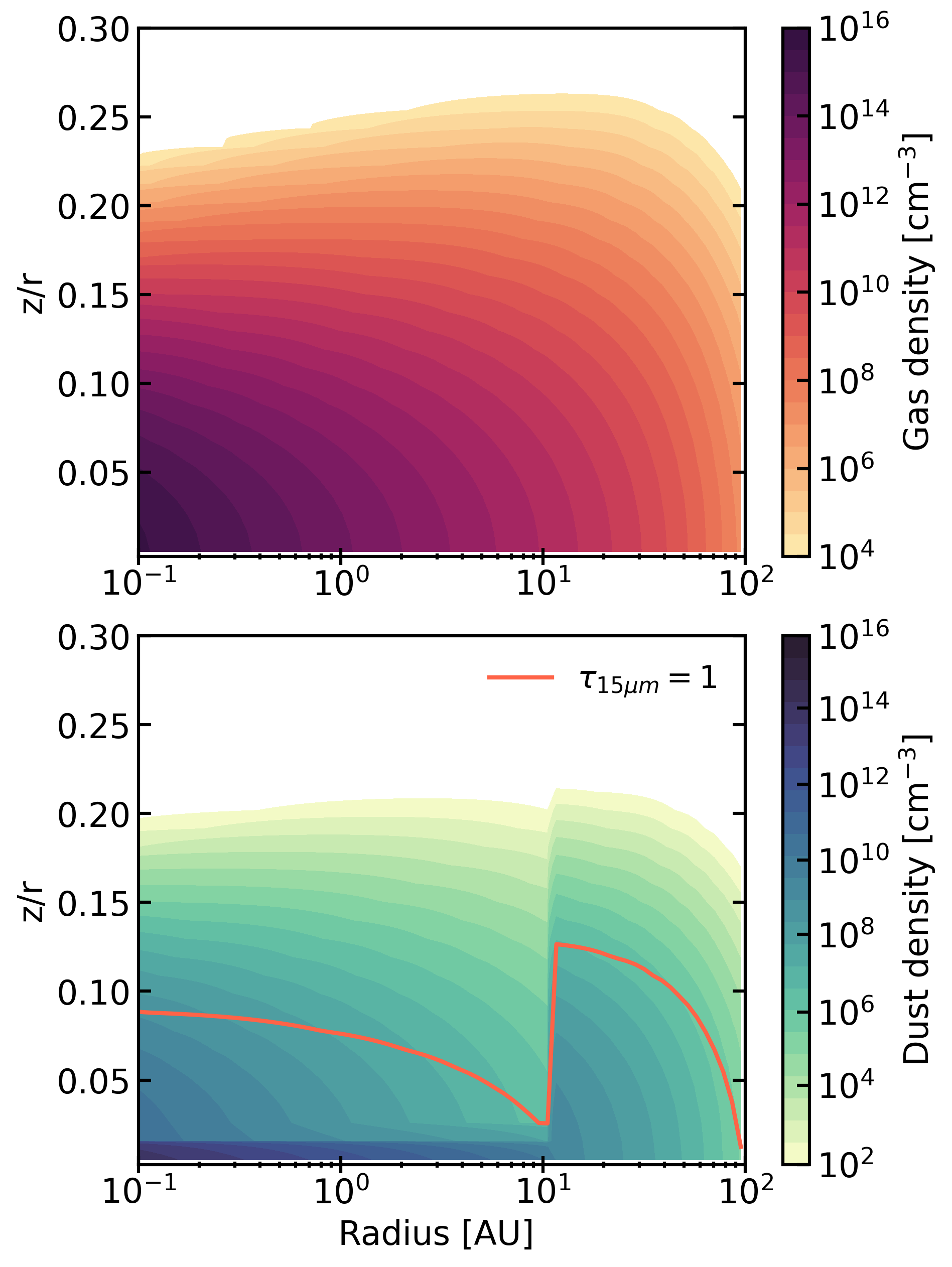}
    \caption{Gas (top) and dust (bottom) density profiles from the thermochemical models. The orange line in the bottom plot shows the continuum $\tau=1$ line at $15$\,$\mu$m. Inside 10~au, the large dust population is mostly settled in the midplane.}
    \label{fig:gas_dust_density}
\end{figure}

In terms of the physical structure of the disk, we consider a viscous accretion disk \citep{lynden_bell1974} that follows the form
\begin{equation}
    \Sigma_{\rm{gas}} = \Sigma_c\bigg(\frac{R}{R_c}\bigg)^{-\gamma}\exp\bigg[-\bigg(\frac{R}{R_c}\bigg)^{2-\gamma}\bigg]\,,
\end{equation}
\noindent where $\Sigma_{\rm{gas}}$ is the surface density of gas, $R_c$ is the critical radius, $\Sigma_c$ is the surface density at $R_c$, and $\gamma$ is the power-law slope. The scale height angle is defined by
\begin{equation}
    h = h_c\bigg(\frac{r}{R_c}\bigg)^\psi\,,
\end{equation}
\noindent where $\psi$ is the flaring angle, and the physical scale height is $H(r)\sim rh$. We include two different dust populations made up of small (0.005 -- 1$\mu$m) and large (0.005 -- 1000$\mu$m) dust grains, distributed in the disk according to
\begin{equation}
    \rho_{\rm{dust,small}} = \frac{(1-f)\Sigma_{\rm{dust}}}{\sqrt{2\pi}H(r)}\exp\Bigg[-\frac{1}{2}\bigg(\frac{z}{H(r)}\bigg)^2\Bigg]
\end{equation}
\begin{equation}
    \rho_{\rm{dust,large}} = \frac{f\Sigma_{\rm{dust}}}{\sqrt{2\pi}\chi H(r)}\exp\Bigg[-\frac{1}{2}\bigg(\frac{z}{\chi H(r)}\bigg)^2\Bigg]\,.
\end{equation}
Here, $\Sigma_{\rm{dust}}$ is the surface density of small and large dust, $f$ is the large dust fraction, and $\chi$ describes the settling of large dust. The values for all of the parameters used can be found in \autoref{tab:model_params}. The opacities of both populations follow the DSHARP opacities \citep{Birnstiel_dsharp} used by \citet{Zhang21_mapsco}. The photometry of DoAr~33 shows an emission bump around $\sim$$100\mu \rm m$ \citep{liu2022}. In order to reproduce this feature, we increased the scale height of the large dust outside 10\,au. This effectively mimics a wall of dust, consistent with the gap substructure at $\sim$9 au inferred from millimeter observations of the source \citep{Huang18}. Our spectral energy distribution (SED)-fitting procedure is further explained in \autoref{app:SED}.  The resulting 2-dimensional gas and dust density profiles are shown in \autoref{fig:gas_dust_density}

\begin{deluxetable}{lcc}
\tablecaption{Thermochemical model parameters}
\tablehead{\colhead{Parameter} & \colhead{Symbol} & \colhead{Value}}
\startdata
Critical radius & $R_c$& 20\, au\\
Gas surf. dens. at $R_c$ & $\Sigma_c$& 13\, g\,cm$^{-2}$\\
Surf. dens. power-law slope & $\gamma$ & 0.9\\
Disk opening angle & $h_c$& 0.045\\
Disk flaring angle & $\psi$& 0.08\\
Large dust fraction & $f$ & 0.999\\
\multirow{2}{*}{Large dust settling} & $\multirow{2}{*}{$\chi$}$ & 0.1 for $r\leq10\,\rm{au}$\\  
& & 0.8 for $r>10\,\rm{au}$\\
\enddata
\end{deluxetable}
\label{tab:model_params}

In order to properly model the full hydrocarbon chemistry and include the relevant species, we use an expanded chemical network as presented in \citet{duval2022}. The original network follows that implemented by \citet{walsh2015}, with the modifications added by \citet{bosman2022}.  We also vary the initial abundances in the disk in order to mimic the environment of an elevated C/O ratio. We have two different scenarios that give rise to a change in C/O: an excess of CH$_4$, or an excess of CH$_4$ accompanied with H$_2$O depletion.

\begin{deluxetable*}{lccccccc}

\tablecaption{Input abundances of the thermochemical models inside the 250 K and 500 K contours.}
\tablehead{\colhead{Name} & \colhead{Total C} & \colhead{Total O} & \colhead{C/O}&\colhead{\ce{H2O}} & \colhead{\ce{CO}}& \colhead{\ce{CO2}}& \colhead{\ce{CH4}}}
\startdata
Solar                                  & $5.00 \times 10^{-5}$ & $1.44 \times 10^{-4}$   &      0.47     & $1.20 \times 10^{-4}$                & $8.50 \times 10^{-5}$               & $4.15 \times 10^{-5}$                & $8.50 \times 10^{-6}$                \\
ch4\_excess\_co\_1\_t500               & $2.03 \times 10^{-4}$ & $4.50 \times 10^{-4}$   &         1  & $1.20 \times 10^{-4}$                & $8.50 \times 10^{-5}$               & $4.15 \times 10^{-5}$                & $1.62 \times 10^{-4}$                \\
ch4\_excess\_co\_2\_t500               & $4.91 \times 10^{-4}$ & $1.02 \times 10^{-3}$   &          2 & $1.20 \times 10^{-4}$                & $8.50 \times 10^{-5}$               & $4.15 \times 10^{-5}$                & $4.49 \times 10^{-4}$                \\
ch4\_excess\_co\_1\_t250               & $2.03 \times 10^{-4}$ & $4.50 \times 10^{-4}$   &          1 & $1.20 \times 10^{-4}$                & $8.50 \times 10^{-5}$               & $4.15 \times 10^{-5}$                & $1.62 \times 10^{-4}$                \\
ch4\_excess\_co\_2\_t250               & $4.91 \times 10^{-4}$ & $1.02 \times 10^{-3}$   &          2 & $1.20 \times 10^{-4}$                & $8.50 \times 10^{-5}$               & $4.15 \times 10^{-5}$                & $4.49 \times 10^{-4}$                \\
co\_dep\_co\_1\_t500                   & $1.20 \times 10^{-4}$ & $2.49 \times 10^{-4}$   &          1 & $1.20 \times 10^{-4}$                & $8.50 \times 10^{-6}$               & 0                & $1.20 \times 10^{-4}$                \\
co\_dep\_co\_2\_t500                   & $2.49 \times 10^{-4}$ & $5.06 \times 10^{-4}$   &          2 & $1.20 \times 10^{-4}$                & $8.50 \times 10^{-6}$               & 0                & $2.49 \times 10^{-4}$                \\
co\_dep\_co\_4\_t500                   & $5.06 \times 10^{-4}$ & $1.02 \times 10^{-3}$   &           4& $1.20 \times 10^{-4}$                & $8.50 \times 10^{-6}$               & 0                & $5.06 \times 10^{-4}$                \\
co\_dep\_co\_1\_t250                   & $1.20 \times 10^{-4}$ & $2.49 \times 10^{-4}$   &       1    & $1.20 \times 10^{-4}$                & $8.50 \times 10^{-6}$               & 0                & $1.20 \times 10^{-4}$                \\
co\_dep\_co\_2\_t250                   & $2.49 \times 10^{-4}$ & $5.06 \times 10^{-4}$   &       2    & $1.20 \times 10^{-4}$                & $8.50 \times 10^{-6}$               & 0                & $2.49 \times 10^{-4}$                \\
co\_dep\_co\_4\_t250                   & $5.06 \times 10^{-4}$ & $1.02 \times 10^{-3}$   &        4   & $1.20 \times 10^{-4}$                & $8.50 \times 10^{-6}$               & 0                & $5.06 \times 10^{-4}$                \\
co\_water\_dep\_co\_1\_t500            & $1.20 \times 10^{-5}$ & $3.25 \times 10^{-5}$   &       1    & $1.20 \times 10^{-5}$                & $8.50 \times 10^{-6}$               & 0                & $1.20 \times 10^{-5}$                \\
co\_water\_dep\_co\_2\_t500            & $3.25 \times 10^{-5}$ & $7.35 \times 10^{-5}$   &       2    & $1.20 \times 10^{-5}$                & $8.50 \times 10^{-6}$               & 0                & $3.25 \times 10^{-5}$                \\
co\_water\_dep\_co\_4\_t500            & $7.35 \times 10^{-5}$ & $1.56 \times 10^{-4}$   &       3    & $1.20 \times 10^{-5}$                & $8.50 \times 10^{-6}$               & 0                & $7.35 \times 10^{-5}$                \\
co\_water\_dep\_co\_1\_t250            & $1.20 \times 10^{-5}$ & $3.25 \times 10^{-5}$   &       1    & $1.20 \times 10^{-5}$                & $8.50 \times 10^{-6}$               & 0                & $1.20 \times 10^{-5}$                \\
co\_water\_dep\_co\_2\_t250            & $3.25 \times 10^{-5}$ & $7.35 \times 10^{-5}$   &       2    & $1.20 \times 10^{-5}$                & $8.50 \times 10^{-6}$               & 0                & $3.25 \times 10^{-5}$                \\
co\_water\_dep\_co\_4\_t250            & $7.35 \times 10^{-5}$ & $1.56 \times 10^{-4}$   &       4    & $1.20 \times 10^{-5}$                & $8.50 \times 10^{-6}$               & 0                & $7.35 \times 10^{-5}$                \\
water\_dep\_no\_co                     & $8.50 \times 10^{-5}$ & $2.55 \times 10^{-4}$   &       1.75    & $1.20 \times 10^{-5}$                & $8.50 \times 10^{-5}$               & 0                & $8.50 \times 10^{-5}$                \\
water\_dep\_co                         & $1.13 \times 10^{-4}$ & $2.55 \times 10^{-4}$   &       1.36    & $1.20 \times 10^{-5}$                & $5.67 \times 10^{-5}$               & $2.83 \times 10^{-5}$                & $8.50 \times 10^{-5}$ \\
\enddata
\tablecomments{The abundances are reported with respect to H.} 

\end{deluxetable*}

 \label{tab:dali_models}

At the same time, in order to explore the origin of excess carbon, we apply the changes in abundances  to two regions: inside a dust temperature of $T_{\rm dust}=250\,$K or, inside of $T_{\rm dust}=500\,$K.  The latter temperature has been argued to be associated with the sublimation of carbon-rich grains at the ``soot line'' by \citet{Li21}. This is a process which would locally elevate the C/O ratio as carbon grains carry $\sim$50\% of elemental carbon in the ISM \citep{Mishra15, Bergin15}. These two setups will allow us to differentiate between a scenario where the excess carbon is produced due to the sublimation of carbon grains at the soot line (500~K model), which would happen in the inner disk therefore resulting in high temperatures, and a scenario where the carbon is sublimated off icy grains in the outer disk and then transported inwards via advection resulting in a more spatially extended distribution with overall lower temperatures (250~K models).  This would be roughly consistent with the model of the origin of hydrocarbon-rich chemistry suggested by \citet{Mah23}.   Outside of the 250~K and 500~K regions, the abundances of carbon and oxygen carriers are set so that the C/O ratio is solar. As a standard for comparison, we also include a model with solar abundance everywhere in the disk. This results in a grid of 19 models, for which we indicate the names and abundances considered in \autoref{tab:dali_models}.

In this work, we refrain from comparing the observations to simulated spectra, in part due to the uncertainty that can arise from accounting for mutual line overlap, which can be significant for molecules like C$_2$H$_2$, as demonstrated in \citet{tabone2023}. But, more importantly, we aim to focus on the retrieved outputs from the data, rather than on the details of the radiative transfer. Therefore, we will compare to column densities, emitting areas, and flux-weighted temperatures (discussed below). This reduces the time required to forward-model this source using complex thermochemical models.  This solution should approximate the key effects and enable a direct exploration of why DoAr~33 is a rare K star exhibiting rich hydrocarbon emission.

\subsection{Flux-weighted temperatures}
\label{subsec:flux_temps}

\begin{figure}
    \centering
    \includegraphics[width=8.5cm]{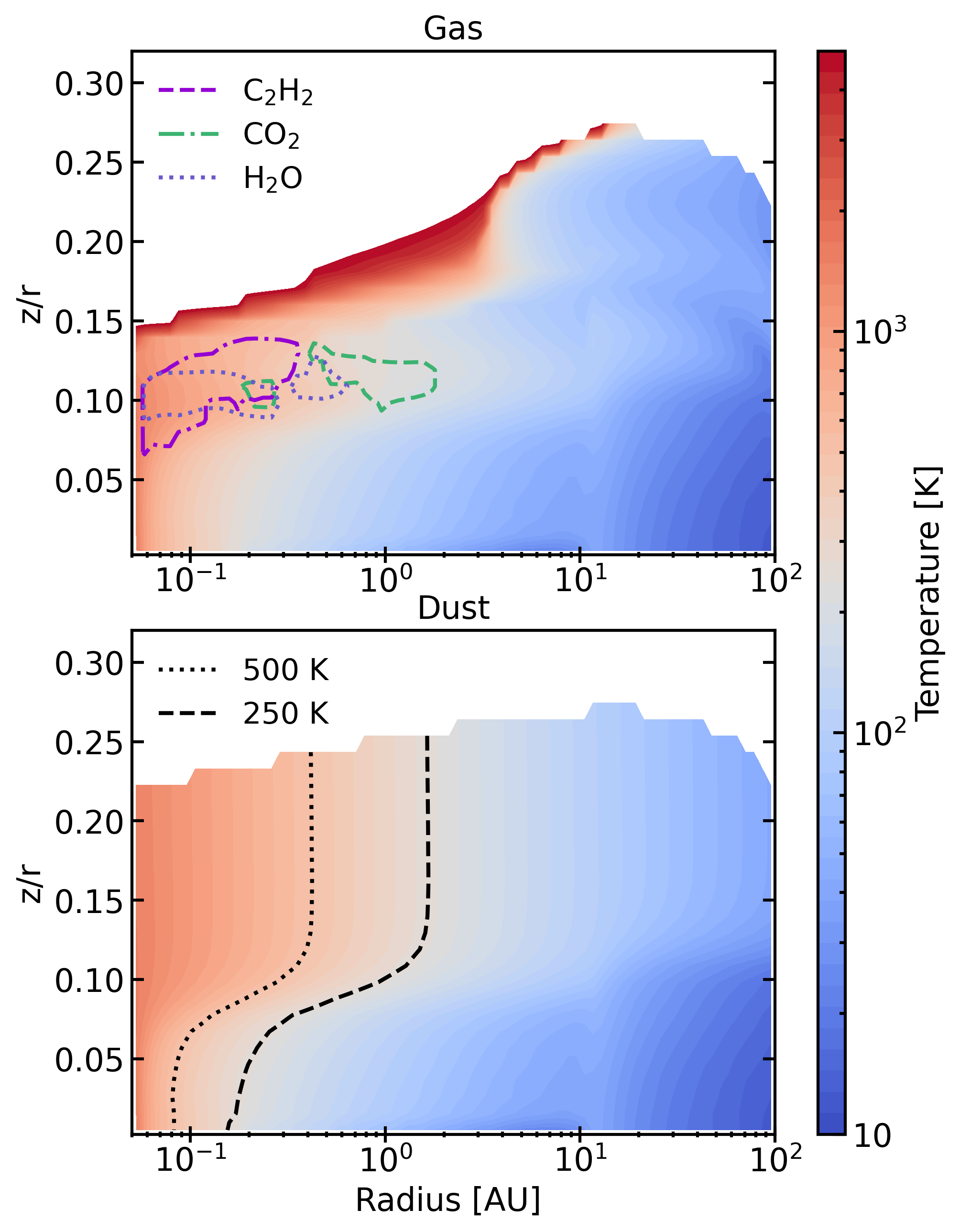}
    \caption{Gas (top) and dust (bottom) temperature profiles from the thermochemical models. The colored contours in the top figure represent the contribution functions for the \texttt{water\_dep\_co} model (see \autoref{tab:dali_models}), where 90\% of the emission is produced. The transitions for each molecule are indicated in \autoref{tab:islat_transitions}. The lines in the bottom plot represent the 250\,K and 500\,K contours, inside of which we change the abundances to increase the C/O ratio.}
    \label{fig:dali_temperatures}
\end{figure}


Column density estimates derived from observations can intrinsically reduce the uncertainty arising from not knowing how the gas is distributed vertically in the disk \reply{since they are integrated along the line-of-sight}; therefore, they can be directly compared to the column densities calculated in the thermochemical models\reply{, without having to match a 2-dimensional gas distribution}. This is not the case for the emitting temperature, which requires that we consider where in the disk this emission is coming from. From the thermochemical models, we are able to obtain a two-dimensional temperature distribution for both the gas and the dust components; we show these distributions in \autoref{fig:dali_temperatures}. However, in order to compare to the estimates from the slab models, we must calculate a representative value for each molecule in the thermochemical models. Here, we describe the procedure we applied to condense the 2D temperature structures into a single emitting temperature.

In the LTE slab models, the temperature retrieved for each species is a representative temperature at which the gas we can observe is emitting. In order to draw a similar parallel, we make use of contribution functions, which trace where in the disk an emission line is formed. Once the necessary information about energy levels, statistical weights, and Einstein A-coefficients is provided to DALI in the form of LAMDA\footnote{\url{https://home.strw.leidenuniv.nl/\~moldata/}} files \citep{schoier2005}, the code is able to perform excitation calculations for any given molecule included in the chemical network.  With this information, we can calculate the 2D contribution function of a specified transition \citep{Bruderer12}, to understand where it is originating physically. It is important to note that the region where a specific line is generated does not necessarily correspond to the region where that species is most abundant due to, e.g., the effects of optical depth. Therefore, this method allows us to mimic an emitted temperature instead of a temperature simply weighted by abundance. In the case of \ce{C2H2}, the spectroscopic information for the LAMDA file was compiled based on the ExoMol database \citep{chubb2020}.

In practice, we select a specific transition for each of the species observed in the MIRI spectrum and calculate its contribution function in DALI. To select the transition, we use the spectral-line analysis tool \texttt{iSLAT}\footnote{ \url{https://github.com/spexod/iSLAT}} \citep{jellison2024}, which takes line transition information from HITRAN \citep{gordon2022}. This tool allows us to input the parameters obtained from the fitting of slab models and generate a synthetic spectrum. Once the spectrum is generated, a spectral range can be selected for which the strongest line transition is shown. We then select the strongest transition of each molecule in the spectral range between 12-17$\mu$m for which we calculate the contribution function using DALI. The selected transitions are indicated in \autoref{tab:islat_transitions}, and plotted for the solar model in \autoref{fig:dali_temperatures}. We map these contribution functions to the gas temperature in the disk model and calculate a single contribution function-weighted temperature for each model. These temperatures are shown in \autoref{fig:flux_temps} and discussed in \S~\ref{sec:results}, where we compare them to the temperatures derived from slab models. 

\begin{deluxetable}{lccc}
\tablecaption{Strongest line transitions in the 12--16$\mu$m region for best-fit slab model parameters}
\tablehead{\colhead{Species} & \colhead{Einstein A coeff. [s$^{-1}$]} & \colhead{$E_u$ [K]}& \colhead{Wavelength [$\mu$m]}}
\startdata
\ce{H2O}      & 3.48         & 3950.8       &      14.2096     \\
\ce{CO2}      & 1.54         & 1113.3       &        14.9777    \\
\ce{C2H2}     & 6.06         & 1358.3       &       13.6992    \\
\enddata
\end{deluxetable}
\label{tab:islat_transitions}

\section{Results} 
\label{sec:results}

The thermochemical modeling process returns two-dimensional distributions of abundances in the disk for all of the species included in the chemical network. Here, we only focus on the oxygen and carbon carriers that are detected in the MIRI-MRS spectrum of DoAr~33. In order to directly compare the modeling products to the parameters retrieved from the LTE slab models, we compute column densities and flux-weighted temperatures for all of the models described in \autoref{tab:dali_models} based on the abundance maps. We calculated the radial distributions of column densities by vertically integrating the abundance from the surface of the disk down to the $\tau=1$ surface at 15$\mu$m (shown as an orange line in \autoref{fig:gas_dust_density}). For the contribution function-weighted temperature, we followed the procedure outlined in  \S~\ref{subsec:flux_temps}

In the following sections, we individually discuss the results for each molecule. We include all of the abundance maps for \ce{H2O}, \ce{CO2}, and \ce{C2H2} in \autoref{app:abundances}. To make the analysis more tractable, we did not include HCN and \ce{HC3N} and do not attempt to match \reply{the retrieved columns for} these species.

\begin{figure*}
    \centering
    \includegraphics[width=17cm]{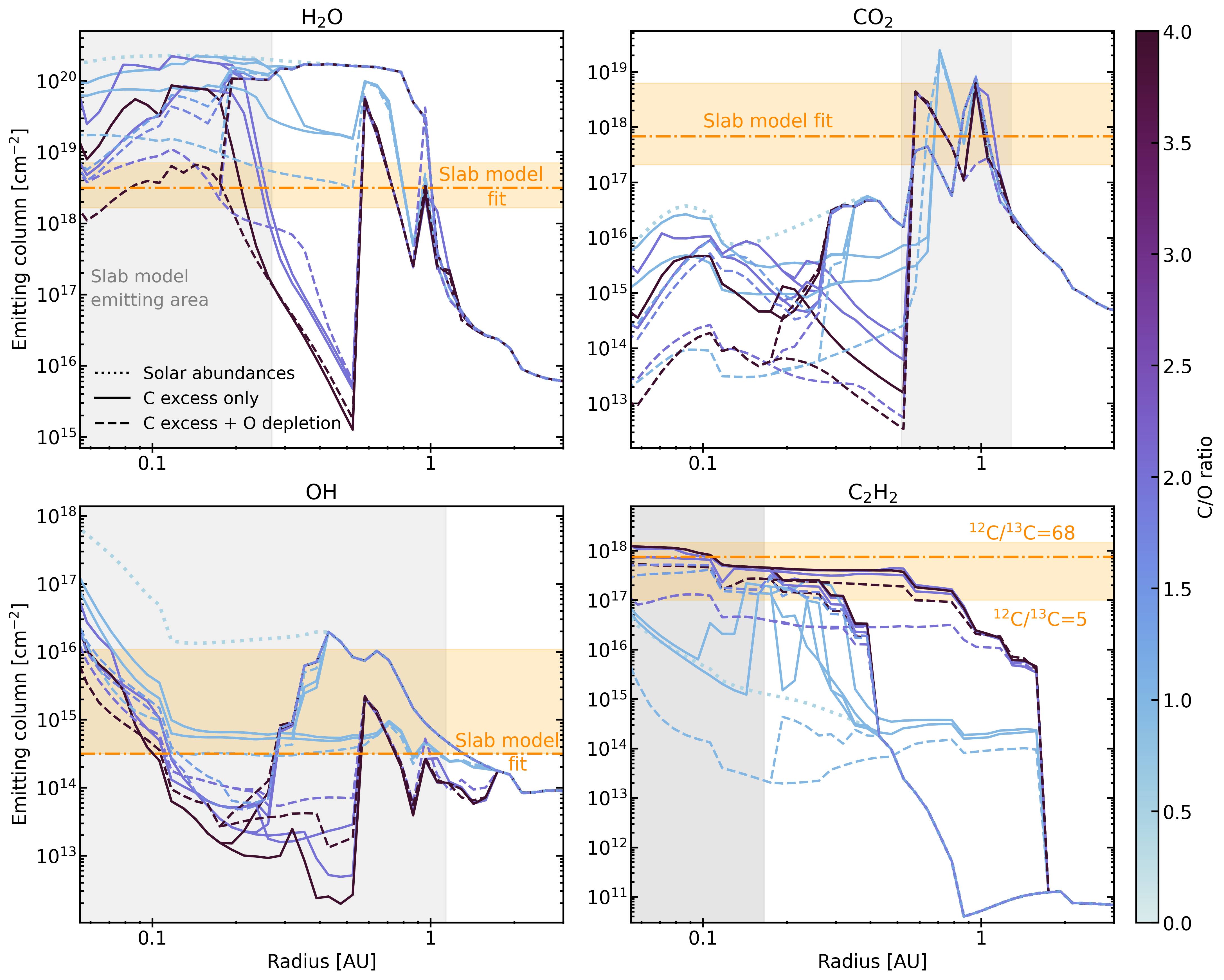}
    \caption{Column density for \ce{H2O},\ce{CO2} and \ce{C2H2} as a function of radius for the models in \autoref{tab:dali_models}. The color bar reflects value of C/O for each model, and the linestyle indicates the method for increasing the C/O ratio. The dash-dotted horizontal line shows the best fit value found from LTE slab models, along with the shaded 1$\sigma$ uncertainty. The shaded grey vertical region corresponds to the emitting area estimated from slab models. In the case of \ce{CO2}, we assumed this value to be an annulus of the same area instead of a continuous distribution, and we placed the ring where it best matched the thermochemical models.}
    \label{fig:all_columns}
\end{figure*}

\subsection{\ce{H2O}, \ce{CO2} \reply{and OH}}

In the \reply{top left} panel of \autoref{fig:all_columns}, we show the column density of water as a function of radius for all of the models, colored according to the C/O ratio assumed. The line styles indicate whether we assume the increased C/O is a product of only carbon enhancement or the combination of carbon enhancement and oxygen depletion, and we include a model with solar abundances for comparison. 

The solar abundance model predicts the overall highest column density for water, which then decreases beyond 1\,au. The high column density of the solar abundance model appears to be inconsistent with the reference values retrieved through the LTE slab models (\autoref{tab:slab_models}). In fact, all of the models that consider the default water abundance of $\sim$$10^{-4}$, shown in solid lines, over-predict the column density of water. Only the models where we include water depletion are able to match the water column. However, there are several important caveats to determining an overall water abundance through column densities. For one, we do not model the entire water spectrum, and therefore the fit results reported for water in \autoref{tab:slab_models} are only approximately valid 
\citep[see][]{banzatti2023,Gasman2023,munoz-romero2024}. We note that the retrieved water column is consistent with values obtained for other full disks like GW~Lup \citep{grant2023} and FZ~Tau \citep{pontoppidan2024}, but in this latter disk the retrieved water column of $\sim$$10^{18}$\,cm$^{-2}$ was found to be consistent with solar abundances. In our case, the models with the canonical water abundance of $\sim$$10^{-4}$ predict columns closer to $10^{20}$\,cm$^{-2}$. This discrepancy could be, in part, because some of the column we modeled might not have the right temperature for IR emission, as it has been noted by \citet{meijerink2009} and \citet{bosman2022_water}. The modeled columns depend on how far into the disk we assume we can see, which is highly dependent on the gas to dust ratio. In our models the water emission is arising from material with a gas/dust ratio of $\sim$$10^5$, \reply{which is consistent with previous findings of a well-settled dust disk \citep{Liu12}}.  It is possible that a potential solution to the water column issue \citep[e.g.,][]{meijerink2009} is a lower hydrogen column, which would also have implications for the overall carbon chemistry.  This is explored in \S~\ref{sec:total_carbon_content}. Stronger observational constrains on the total gas and dust content are needed to be able to model the water column more accurately.

We then examined the column densities of the second most significant oxygen tracer in the MIRI spectrum, \ce{CO2}. In the \reply{top right} panel of \autoref{fig:all_columns}, we show the column densities calculated for this molecule. In contrast to water, the column density of this molecule is highest around 1\,au. Both \ce{H2O} and \ce{CO2} share the same precursor, \ce{OH}.  \citet{Bosman22_co2} and \citet{duval2022} suggested that CO$_2$ formation in the warm ($>$~400~K) gas is curtailed as OH is more likely to react with H$_2$ than CO. 
The models are only able to reproduce the column density retrieved from slab models in the zone around 1\,au. For reference, the emitting area estimated from the slab models is shown as a shaded annulus of the same area. This indicates that the \ce{CO2} emission is likely coming from a ring of gas around 1\,au. One implication of this result is that it would suggest that CO$_2$ emission would trace cooler gas than regions with abundant water vapor, which has been found in some instances via analysis of JWST spectra \citep{Gasman2023, Xie2023}.

\reply{In the bottom left panel of \autoref{fig:all_columns}, we compared the retrieved OH column density. As expected, the models with a lower C/O ratio produce the highest OH columns. More specifically, the model with a solar C/O ratio predicts a column of almost $10^{18}\,\rm{cm}^{-2}$, much higher than the retrieved best-fit of $3\times10^{14}\,\rm{cm}^{-2}$. Nonetheless, the uncertainty of the column density does not allow us to rule out C/O$=1$, pointing more towards a C/O$\geq1$.} \replytwo{We note that despite the weak prompt emission at shorter wavelengths, OH emission may not be in non-LTE \citep[see discussion for \ce{H2O} by][]{tabone2021}, therefore, the best-fit parameters from the LTE models might not fully represent the local gas properties.}

\begin{figure}
    \centering
    \includegraphics[width=8cm]{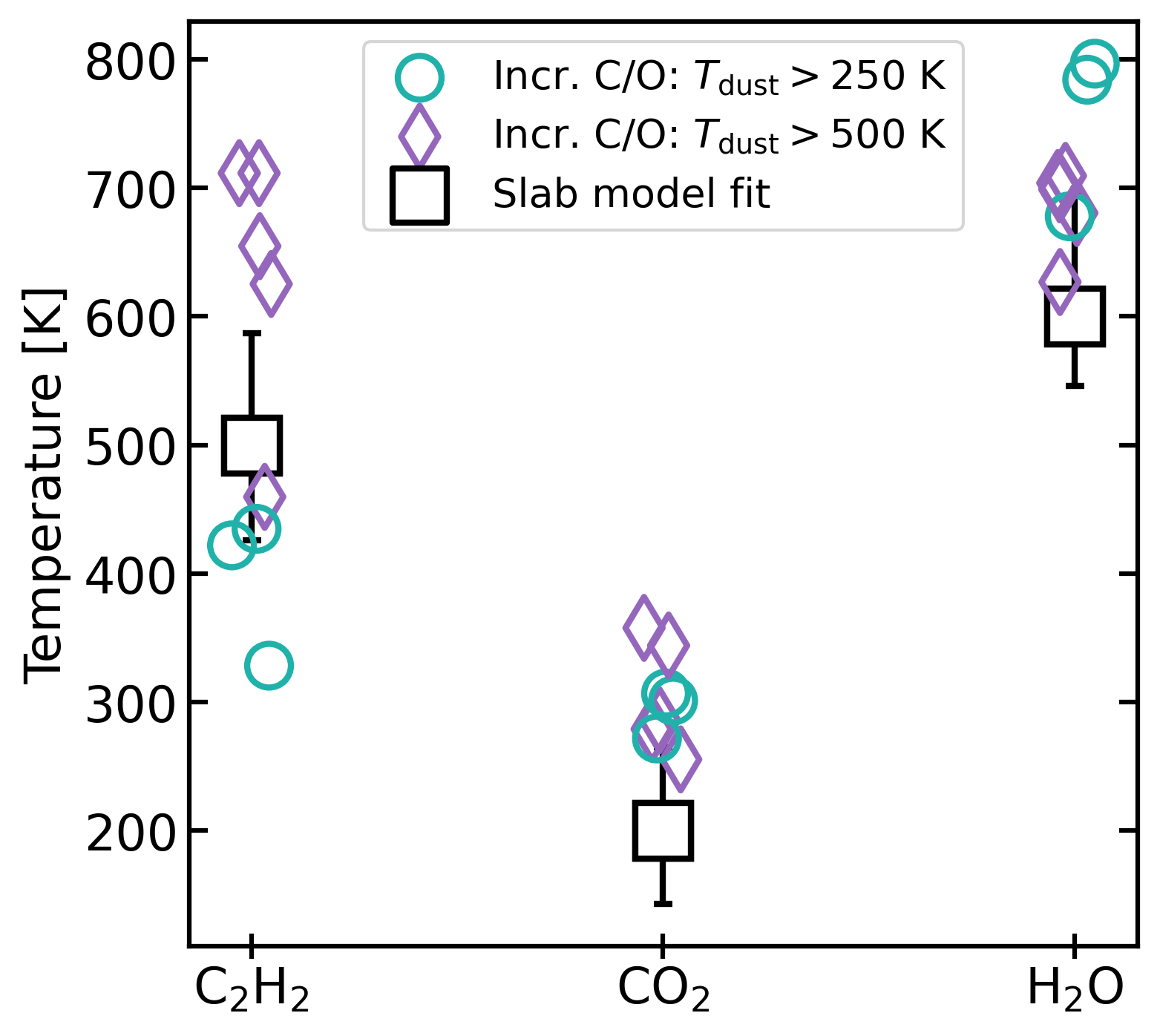}
    \caption{Contribution function-weighted temperatures of carbon and oxygen-bearing species for the water depletion models. The symbols represent where in the model the C/O ratio is changed, inside $T=250$\,K or $T=500$\,K. The markers are shifted horizontally for clarity. For comparison, we include the temperature retrieved from slab models as a black square, with errors corresponding to the $1\sigma$ uncertainty. }
    \label{fig:flux_temps}
\end{figure}

In \autoref{fig:flux_temps}, we show the model emitting temperatures for \ce{C2H2}, \ce{CO2}, and \ce{H2O}, along with the estimate from the slab model fitting. We show only the water depletion scenarios since they are consistent with the water column densities from the slab models. Overall, the thermochemical models predict water has the highest temperature and \ce{CO2} has the lowest, which is consistent with the expectation from the slab model retrievals from observations. Nonetheless, the \ce{CO2}, and \ce{H2O} temperatures do not allow us to differentiate between the two different carbon enhancement scenarios proposed (see \S~\ref{subsec:disk_modeling}). 

\subsection{\ce{C2H2} and \ce{^{13}C^{12}CH2}}
\label{subsec:column_densities}

As mentioned in \S~\ref{subsec:slab_models}, in order to derive the actual \ce{C2H2} column, we assume an isotopic ratio. This ratio is unknown and \citet{tabone2023} assumed 35 based on the models of \citet{Woods2009}. We adopt this value and report the \ce{C2H2} column density to be $35\times N_{\ce{^{13}C^{12}CH2}}$, which results in \reply{$N_{\ce{C2H2}}= 7.5\times 10^{17}\,\rm{cm}^{-2}$}. We note this value could be substantially higher as the $^{12}$C/$^{13}$C ratio of hydrocarbons and nitriles in the outer disk of TW Hya is closer to the interstellar value \citep{Hily-Blant2019, Yoshida2024, Bergin2024}. Therefore, the uncertainties of \ce{C2H2} column shown in \autoref{fig:all_columns}, stem from assuming $^{12}$C/$^{13}$C can go from \reply{$\approx$$5$}, as directly retrieved from slab models, to 68, the interstellar value. 

From the thermochemical models, it is clear to see that, similarly to water, the \ce{C2H2} emitting column peaks in the inner disk and then decreases between 0.4 and 1\, au. The difference in the location of the column density drop is due to where the C/O ratio is increased. In the models where C/O is increased within $T=500\,$K, the carbon-rich chemistry is restricted to a smaller area, whereas in the other models the \ce{C2H2} column is more spatially extended. It is clear that most models that are able to match the expectation from the slab model fit have C/O$>2$. 

Contrary to \ce{CO2} and \ce{H2O}, in the case of \ce{C2H2}, the selection of the temperature where we transition from solar to a high C/O ratio (i.e. 250\,K vs. 500\,K) plays an important role in determining the final weighted temperature; as we can see in \autoref{fig:flux_temps}, most of the models cluster around a particular value. In the models where the carbon-rich chemistry is allowed to be more spatially extended, the \ce{C2H2} has a lower temperature. This implies that a high C/O ratio extended over larger radii would produce an overall lower emitting temperature for the hydrocarbons. Similarly, by restricting the extent of the high C/O ratio (i.e. the 500\,K transition models) we limit the presence of \ce{C2H2} to the innermost disk, which in turn results in a higher emitting temperature since the material is closer to the star. Comparing to the temperature retrieved from slab models, we are not able to strictly discriminate between the two scenarios since most of the flux-weighted temperatures are within the uncertainty range. This is also true for water, where there is an overlap in the flux-weighted temperatures. 

In our approach, the key to distinguishing between a radially extended or limited carbon-rich chemistry is the emitting area. Slab models retrieve a maximum emitting area of \reply{$\pi(0.25\,\rm{au})^2$} for \ce{C2H2}, corresponding to the $3\sigma$ contour shown in \autoref{fig:chi2}. The 500 K contour for the gas temperature extends to 0.4\,au in the IR-emitting layer (\autoref{fig:dali_temperatures}), whereas it goes past 1\,au for the 250 K contour. This \reply{small emitting area}, in combination with the high the emitting temperature of \ce{C2H2}, indicates that the carbon chemistry is only released in the innermost disk, where the sublimation of carbonaceous grains is possible. If the carbon-rich material present in this system traced by C$_2$H$_2$ is caused by advection of material from further out in the disk \citep{Mah23}, then it should produce more extended emission than inferred here. 

\subsection{\reply{Other hydrocarbons}}
\subsubsection{\ce{C4H2}}
\label{sec:c4h2}

\begin{figure}
    \centering
    \includegraphics[width=8.7cm]{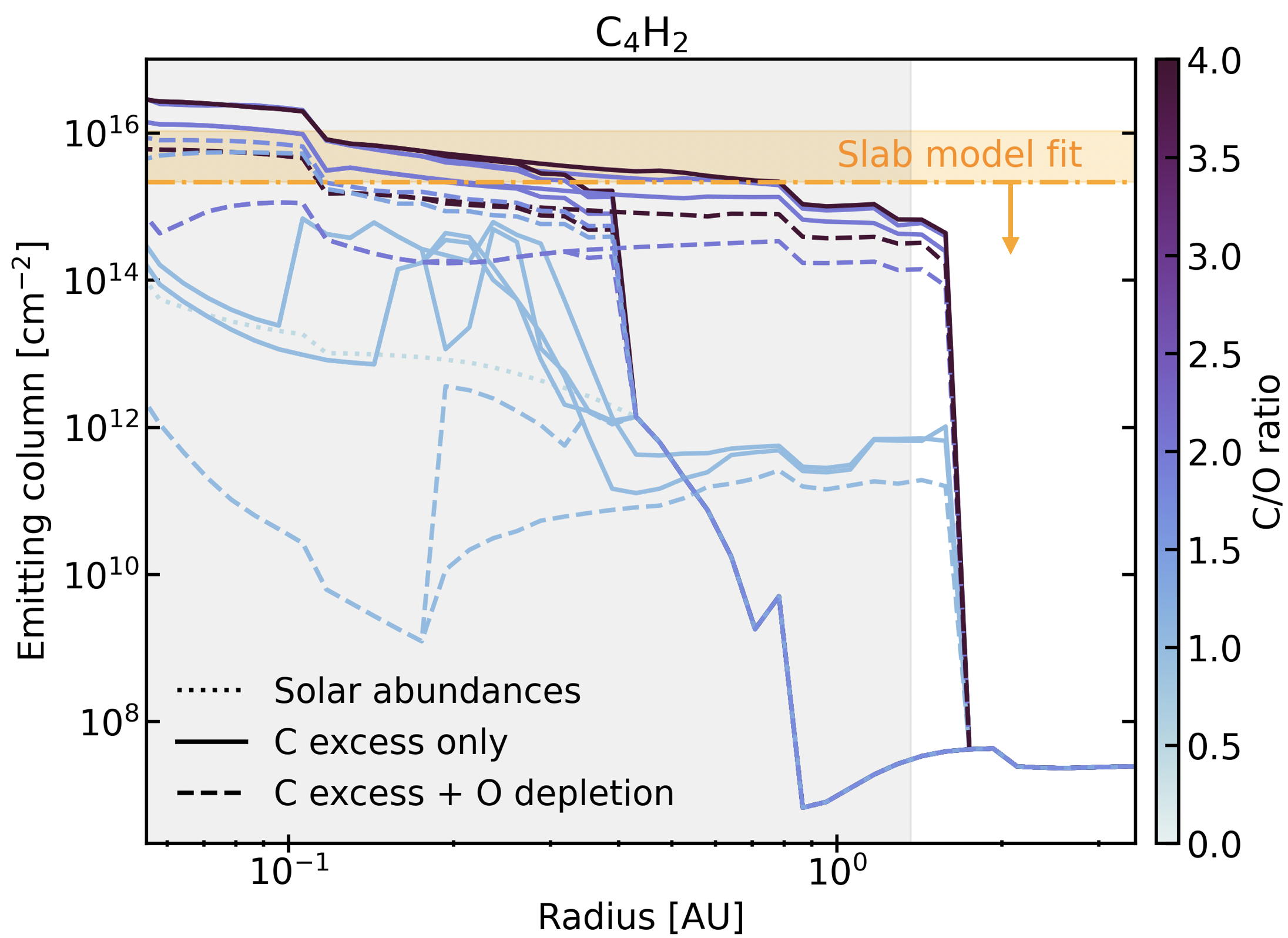}
    \caption{Same as \autoref{fig:all_columns}, but for \ce{C4H2}.}
    \label{fig:c4h2_column}
\end{figure}

We compare the retrieved \ce{C4H2} column to that predicted by our thermochemical models in \autoref{fig:c4h2_column}. Based on the $\chi^2$ maps (\autoref{fig:chi2}) from the slab model fitting, this column likely represents an upper limit. In turn, this makes it consistent with the models where we assumed C/O $=2$. The \ce{C4H2} emission is extended over a larger area than \ce{C2H2}, and is therefore compatible with a lower emitting temperature (\reply{150} K). \reply{To provide a more definite constraint, we} explored fitting \ce{C4H2} with the same emitting temperature as \ce{C2H2} in \autoref{app:c4h2_ex}, and find that the higher temperature is incompatible with the $Q$-branch feature. This indicates the presence of a warm ($<$400 K) hydrocarbon-rich reservoir, in addition to the hot ($>$400 K) \ce{C2H2}. We discuss the potential origin of these two reservoirs in \S~\ref{sec:c-origin}.

\subsubsection{\reply{Non-detections}}
\label{app:non-detections}

\reply{The chemical network used in this work is able to predict chemical abundances for over 400 species. Therefore, we checked the predictions for the abundance of hydrocarbons that have been detected in hydrocarbon-rich low-mass sources \citep{tabone2023,arabhavi2024,kanwar2024}, in order to see if we can obtain any additional constraints on the C/O ratio. In the top row of \autoref{fig:non-detections}, we show the predicted column densities for \ce{C2H4}, \ce{C2H6} and \ce{C6H6}. In the case of \ce{C2H4} and \ce{C2H6}, the predicted columns are lower than $\sim$$ 10^{14}\,$cm$^{-2}$ for all of the models. Based on this, we assumed an average column for \ce{C2H4} and \ce{C2H6} and produced LTE slab models with two different sets of temperatures and emitting areas, mimicking the ones found for \ce{C2H2} and \ce{C4H2}. The lower panel of \autoref{fig:non-detections} shows the produced spectra, as well as the parameters assumed for each one. It is clear that the emission from these molecules would be too faint to be observable in the DoAr\,33 spectrum. This indicates that our model prediction is consistent with the non-detection of \ce{C2H4} and \ce{C2H6}. The columns for these two molecules do not have any discernible trend in terms of the C/O ratio, making them poor C/O tracers for this source. Despite this predictions being consistent with our observations, we note that the predicted columns for these species are lower than the ones being detected in low-mass sources \citep{arabhavi2024,kanwar2024}. This could be because the formation of hydrocarbons larger than \ce{C2H2} might be sensitive to the explicit X-ray luminosity, and, the presence or absence of cosmic rays (Raul et al. submm), or, due to the completeness of the hydrocarbon reactions in the chemical network \citep{kanwar2024a}. 

\begin{figure*}
    \centering
    \includegraphics[width=18cm]{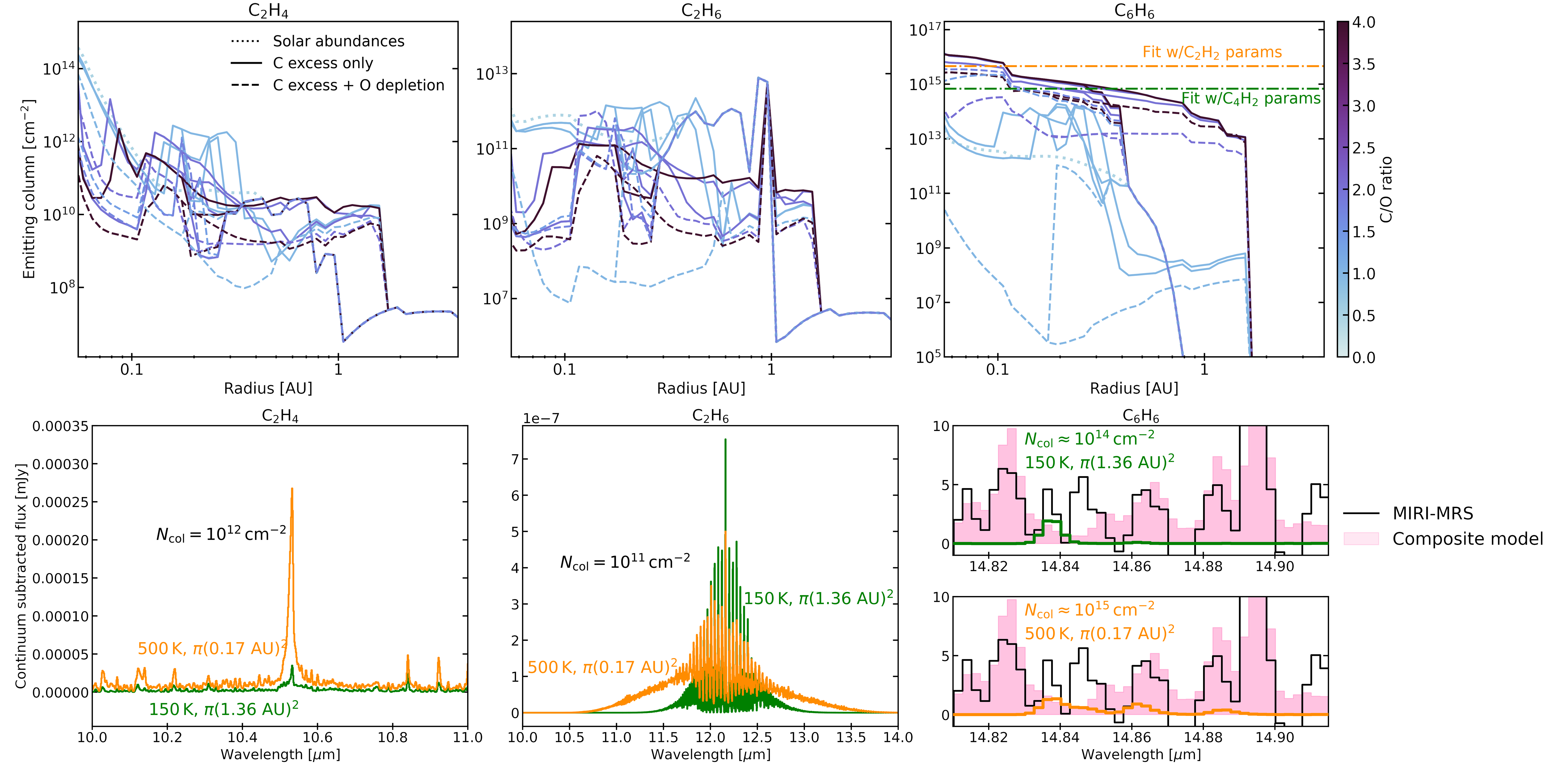}
    \caption{\reply{The top row shows the column density predictions from the thermochemical models for \ce{C2H4}, \ce{C2H6} and \ce{C6H6}. The bottom row shows slab models that simulate the conditions predicted by the thermochemical calculations. The parameters used for each slab model are listed in the figure. Two sets of slab models were tested, one co-spatial with \ce{C2H2} (orange) and one co-spatial with \ce{C4H2} (green). These models (green and orange) are shown on top of the data (black) in the bottom left.
    The pink shaded region in the bottom right panel corresponds to the same composite model as in \autoref{fig:spectrum_model}.}}
    \label{fig:non-detections}
\end{figure*}

In contrast, \ce{C6H6} follows a similar trend to \ce{C2H2}, where high column densities are predicted for the high C/O ratio models, as shown in the top right panel of \autoref{fig:non-detections}. Because of this, we attempted a retrieval in the 14.8--14.9\,$\mu$m range, where the main feature is located, using the spectroscopic files and partition functions provided by  \cite{arabhavi2024}. We first subtracted the contribution from all of the other detected molecules, adopting the parameters from the best-fit slab models \autoref{tab:slab_models}. Then, we  estimated the column density of \ce{C6H6}, assuming a given emitting area and temperature. Similar to the other hydrocarbons, we tested two cases, one where the \ce{C6H6} is co-spatial with \ce{C2H2}, i.e. a small emitting area and high temperature, and one where the \ce{C6H6} is co-spatial with \ce{C4H2}, having a lower temperature and bigger emitting area. The resulting column density retrieved for each case is overplotted with the model predictions as horizontal lines in the top right panel of \autoref{fig:non-detections}. The bottom right panel shows the spectra of both sets of best-fit parameters, over-plotted with the observed spectrum and the composite model of the detected molecules. Both sets of solutions point towards C/O$=2-4$, but we are not able to confirm the presence of \ce{C6H6} in this spectrum.}

\section{Discussion}
\label{sec:discussion}

We find that C/O $> 1$ is needed to explain the hydrocarbon-rich environment around DoAr 33. The two methods to increase the C/O ratio are to deplete oxygen and/or enrich carbon. Here, we first discuss how C/O can be increased through water sequestration. Then, we argue that C/O$>1$ can not be achieved through oxygen depletion alone and propose that additional carbon must be supplied to the gas phase. We explore how much excess carbon is inferred to be present in the inner disk of DoAr 33, and consider the two main mechanisms that have been proposed to explain the origin of this carbon excess. We also discuss the implications of a \ce{CH4} detection in the context of carbon grain sublimation.

\subsection{Depletion of water}

It has been posited that planetesimals and protoplanets forming beyond the water iceline could trap a significant fraction of the oxygen content as long as the efficiency of accretion onto larger bodies is higher than the drift rate of icy pebbles \citep{Najita13}. Modeling of dust and gas dynamics also showed that outer disk gaps, if deep enough, can reduce or stop the inward flux of icy pebbles and reduce the oxygen enrichment of inner disks by reducing water ice sublimation inside the snowline \citep{kalyaan2021,Kalyaan23}. These scenarios would raise the C/O as the inner disk accretes the remaining oxygen-poor gas.  Stellar accretion proceeds over several million years encompassing the initial phases of rocky planet assembly and the main phase of giant planet formation \citep{Hartmann16}.  Thus, if a significant population of planetesimals form, this is a possible scenario for raising C/O. However, in this instance, the volatile reservoir in the inner disk would be mostly dominated by CO and \ce{CO2} gas, meaning that, at most, the inner disk could reach a C/O ratio of $\sim$$2/3-1$. In the case of DoAr 33, we have shown that C/O$>1$ is needed to reproduce the \ce{C2H2} column. Therefore, even though planetesimal formation in the outer disk could be invoked to increase C/O, additional release of carbon is needed to go beyond C/O$=1$. 

\subsection{Total carbon content}
\label{sec:total_carbon_content}

To understand the source of the carbon enrichment, we first examine the amount of excess carbon that is consistent with the observations of DoAr 33. 
In our thermochemical models, we assume varying degrees of extra carbon being released into the system initially in the form of \ce{CH4} inside the soot line (\autoref{tab:dali_models}). At the midplane in our model the soot line is located at $\sim$$0.08$\,au (\autoref{fig:dali_temperatures}), which results in a maximum of $\sim$$0.5\times10^{-3}\,M_\oplus$ of carbon released as \ce{CH4} gas for model \texttt{water\_dep\_co} (\ce{CH4}=$8.5\times10^{-5}$), calculated by integrating the total \ce{CH4} mass in the disk inside 500\,K. Conversely, in the dust distribution modeling process (\autoref{app:SED}), we considered that the amount of refractory organics carried by the large dust population accounted for 40\% of the total dust mass \citep{Birnstiel_dsharp}. This results in $\sim$$1.5\times10^{-3}M_\oplus$ of carbon being present as solids inside 0.08 au. Therefore, our thermochemical models require that, at most, only a third of the carbon present in refractories is released into the gas.

The presence or absence of carbon grains can strongly influence the carbon content of forming planetary systems \citep{Bergin23}. In the context of the Solar System, the sublimation of carbon grains has been posited to explain the carbon depletion found in meteorites and the Earth \citep{Bergin15, Li21}. It is possible that the carbon grains that are not sublimated to go into forming planetesimals. However, we stress our models are not self-consistent.  The SED modeling, which includes pre-determination of the dust composition as part of the dust optical constants, is a separate step from the thermochemical calculations each with independent assumptions of the elemental carbon content in gas and solids.

Thus, it is entirely possible, given the similarity between the estimate of the excess carbon in the gas and the total carbon available in dust (within a factor of 3), that all of the solid-state carbon inside 0.08~au has been released to the gas. 
We note that this solution assumes the presence of a dust trap, \reply{as potentially indicated by the ALMA continuum and the SED,} and water depletion to lower the water column. Alternately, as discussed in \S~4.1, the overall gas/dust ratio may be lower than assumed by at least an order of magnitude.  This would reduce the water column and allow for solutions with ``normal or solar'' water content (i.e. the \reply{dotted and the} solid lines in Fig.~\ref{fig:all_columns}).  In this instance the C$_2$H$_2$ column would also be reduced to near the lower end of the allowed range in the column density.  Here higher C/O ratios, and carbon content, are required to bring the column closer to the middle of our retrieved estimate ($\sim$few $\times$ 10$^{17}$~cm$^{-2}$).
Future work that combines the effects of carbon destruction on the emission of both solids and gaseous species is needed to ensure consistency and derive more accurate estimates.   Finally, while carbon grains represent one potential source term for excess carbon, other scenarios have been proposed, which are discussed below.

\subsection{Origin of carbon enrichment}
\label{sec:c-origin}

Currently, there are two potential models for the origin of the excess carbon. One posits that the carbon supply can originate in the outer disk via gas accretion of material with C/O$>1$ \citep{Mah23} and the other suggests that carbon grain destruction near the soot line can increase the carbon content of the inner disk \citep{Li21}.  

We will discuss each in turn but note two things: first, in our analysis, the difference in emitting temperature and area found for \ce{C2H2} and \ce{C4H2} suggests that the carbon-rich chemistry in this source is not necessarily restricted to the innermost disk. On the same note, ALMA has observed many disks that appear to have C/O$>$1 beyond 10~au \citep[e.g.,][]{Bosman21_mapsco}, and temperatures are hot enough to sublimate carbon grains only in the inner 1~au for T Tauri disks \citep{Li21}; thus, these solutions are likely not exclusive.  However, they do have different implications for the composition of planets.

\subsubsection{Carbon rich gas advection}

The model presented by \citet{Mah23} suggests that the accretion of \ce{CH4}-rich gas into the inner disk can explain the excess carbon observed in low mass sources. This scenario works if pebbles in the outer disk are large enough to be close to the drift limit, with fragmentation velocities ($v_{\rm frag}$) higher than 5\,m/s. In this case, the system gets an early influx of water-rich pebbles crossing the water iceline, enhancing the gas. The influx of water-rich pebbles eventually runs out, and the oxygen-enriched gas gets accreted by the star within $\sim$2\,Myr. Only then, the system is left with carbon-rich gas that can be advected into the inner disk.  

In this concept, CH$_4$ is released into the gas with an abundance that exceeds both CO and CO$_2$ providing C/O =
(\ce{CH4} + CO + \ce{CO2})/(CO + 2\ce{CO2}+\reply{\ce{H2O}}) $>$ 1.  This leads to an important point. \reply{In protostellar envelopes the CO and CO$_2$ ice abundances have been determined via Spitzer spectroscopic surveys and are $\sim$60\% relative to water \citep{oberg11_c2d}.  Assuming a water ice abundance of order $\sim$2 $\times 10^{-4}$ (relative to H$_2$) \citep{Pontoppidan2004} then the typical abundance of CO and CO$_2$ ice are of order $\sim$$10^{-4}$ relative to H$_2$. This does not account for the gas phase CO and CO$_2$ content, and thus the gas \& ices carry  $>$25\% of the cosmically available carbon which is 4.28 $\times 10^{-4}$, relative to H$_2$ \citep{Nieva12}.  In sum, if CO and CO$_2$ are present at levels expected in protostellar gas (inside their respective ice lines) then to achieve C/O $>$ 1 one must destroy nearly {\em all} carbon-rich grains, which hold an equivalent amount of elemental carbon as the gas/ices.}  This seems implausible beyond 10~au as the destruction mechanisms for carbon grains in the outer disk are highly ineffective \citep{Anderson17, Klarmann18}.  However, the ALMA observations of C/O $>$ 1 also coincide with many systems where the CO gas phase abundance has been reduced \citep{Miotello_ppvii}. Simulations account for the presence of strong C$_2$H emission in these disks via the release of CH$_4$ from grains at the level of a few percent of the solid state carbon inventory \citep{Bosman21}.  These results empirically infer that   (\ce{CH4} + CO + \ce{CO2})/(CO + 2\ce{CO2}) $>$ 1.  This gas can advect inward and lead to an inner disk carbon-enrichment as suggested by \citet{Mah23}.  We note that this is not uniformly the case as, if the water and other volatiles return to the gas then the C/O ratio and molecular abundance can be reset to values consistent with ISM expectations.  In one disk surrounding a K7 star (Sz~98) there is a known elevated C/O ratio in its outer regions \citep{Miotello19} and \citet{Gasman2023} show that JWST observations are consistent with a reset in the C/O ratio  in the inner few au.   At some level this advection could potentially relate to the warm organic carbons (e.g. C$_4$H$_2$) but additional carbon supply is needed to match the elevated C$_2$H$_2$ column inside \reply{0.25}~au in DoAr 33.

One complication is that, if pebbles are small enough to be coupled to the gas ($v_{\rm frag}=1$\,m/s), the influx of icy pebbles is delayed, and the inner disk will be mostly O-rich gas for several Myr. Regardless, in this model the carbon-rich enrichment of the gas in the inner disk has a clear implication: if carbon is allowed to remain in refractory form, then any forming planets will have significant volatile carbon inventories in their mantle which can influence the atmospheric composition \citep{Bergin23}.

\subsection{Soot line and mixing timescales}

We posit that to account for the high amount C$_2$H$_2$ in the inner disk, some refractory carbon has to be released to the gas phase in order to explain the hydrocarbon-rich chemistry detected in DoAr 33 and in other sources.  Nonetheless, the majority of the mass of solids is found in disk midplanes \citep{Villenave20} and not its warm/hot atmosphere.
Thus, carbon grains need to sublimate with gaseous products diffusively transported vertically to the surface layers of the disk in order to be observed in the gas reservoir. In this approach, it is fundamental to understand whether vertical mixing processes in the inner disk are efficient enough to probe the composition of the material in the midplane. 

On the other hand, the sublimation and vertical mixing timescale must compete with the radial advection timescale, which is tightly related to transport of material from the outer to the inner disk \citep{Heinzeller11}. If radial transport is very efficient, then the C-rich gas will be accreted by the star before it can be mixed into the surface layers, and we would not be able to observe it.  With these two points in mind, we propose that the observed dichotomy between carbon-rich disks around low mass stars and oxygen-rich disks around solar-mass stars is a consequence of the interplay between radial and vertical mixing timescales. This is an extension of the concept originally proposed by \citet{Mah23}, who point out that the vertical mixing timescales are shorter in very low mass systems as the chemical transitions occur very close to the star. 

One of the primary mechanisms for vertical transport of material is turbulent diffusion, which can be driven by hydrodynamic or magnetohydrodynamic instabilities, such as the magnetorotational instability (MRI) \citep{sano2000}, or vertical shear instability (VSI) \citep{nelson2013}. These mechanisms induce turbulent eddies that can facilitate the vertical transport of both gas and dust. The extent of this mixing depends on factors such as the turbulence levels of the disk, the size and composition of dust particles, and the thermal structure of the disk \citep{youdin2007}. 

If we define a vertical mixing timescale as
\begin{equation}
    t_{\rm v} = \frac{H}{\alpha c_s}=\frac{1}{\alpha \Omega}\,,
\end{equation}
\noindent where $c_s$ is the sound speed, $H$ is the scale height of the gas disk and $\alpha$ is the viscosity parameter, and we have replaced $H=c_s/\Omega$, where $\Omega$ is the angular velocity, it is clear to see that, for a fixed viscosity, the timescale scales as $t_{\rm v} \propto R^{3/2}M_\star^{-1/2}$, where $R$ is distance from the star. A solar-mass star would have a vertical mixing timescale that is about three times faster than a lower mass star (0.1$M_\odot$) at the same radius. However, if we consider the location of the soot line in the disks around these stars, due to the lower irradiation in the 0.1\,$M_\odot$ star, the soot line will be much closer to the star. Models of the thermal structure of disks predict about a factor of 10 difference in the radial location of the soot line in the midplane of disks around lower-mass stars compared to solar-mass stars \citep{walsh2015}, which would result in a much longer mixing timescale for the solar-mass star.

This implies that the vertical mixing of sublimated carbonaceous materials would occur faster in the 0.1\,$M_\odot$ star. In contrast, the radial transport of gas and small dust is related to the mass accretion rate \citep{Aikawa1999, Heinzeller11}, so a lower mass accretion rate will result in slow radial transport of material as the radial velocity of the flow would be $\sim$ $-\dot{M}$/(2$\pi$R$\Sigma)$ \citep{Aikawa1999}. Observations isolate a correlation between accretion rate and stellar mass $\log \dot{M}_{\rm acc}\propto M_\star^{\beta}$ with $\beta\sim 1.6-2$ \citep{Manara16,Hartmann16,alcala2017}. Therefore, we expect very low mass stars, with lower accretion rates, in comparison to solar mass stars, to have a longer radial viscous evolution timescale.   In this context, DoAr~33 is just a solar mass star masquerading as a very low mass star. \reply{This could indicate that it is at a late stage of its evolutionary process. However, the age of this system was inferred to be around 4\,Myrs old \citep{andrews2010}, and more recent astrometric studies place it within the young (0.3--6\,Myr) $\rho$ Ophiuchi population \citep{grasser2021}, consistent with other T Tauri systems.}

Taking these two points into account, we can infer that, in stars with high accretion rates, the material that is released at the carbon sublimation front will likely be accreted into the star \textit{before} it can be transported to the surface layers of the disk. In turn, the carbon-rich chemistry will be more readily observable in disks around lower mass stars or in systems where the vertical mixing timescale is shorter than the radial mixing/advection/drift timescale.   We label this solution as ``burn and linger''.  That is, the carbon that is released in the midplane via sublimation (not burning) is then transported upwards so that the chemistry is able to linger over an advection timescale. The midplane carbon interior to the soot line can then be replenished to allow for the chemistry to persist. In the disk surrounding more massive stars the burning is occurring but the chemistry is not allowed to linger due to the shorter radial transport timescales.

\begin{figure}
    \centering
    \includegraphics[width=9cm]{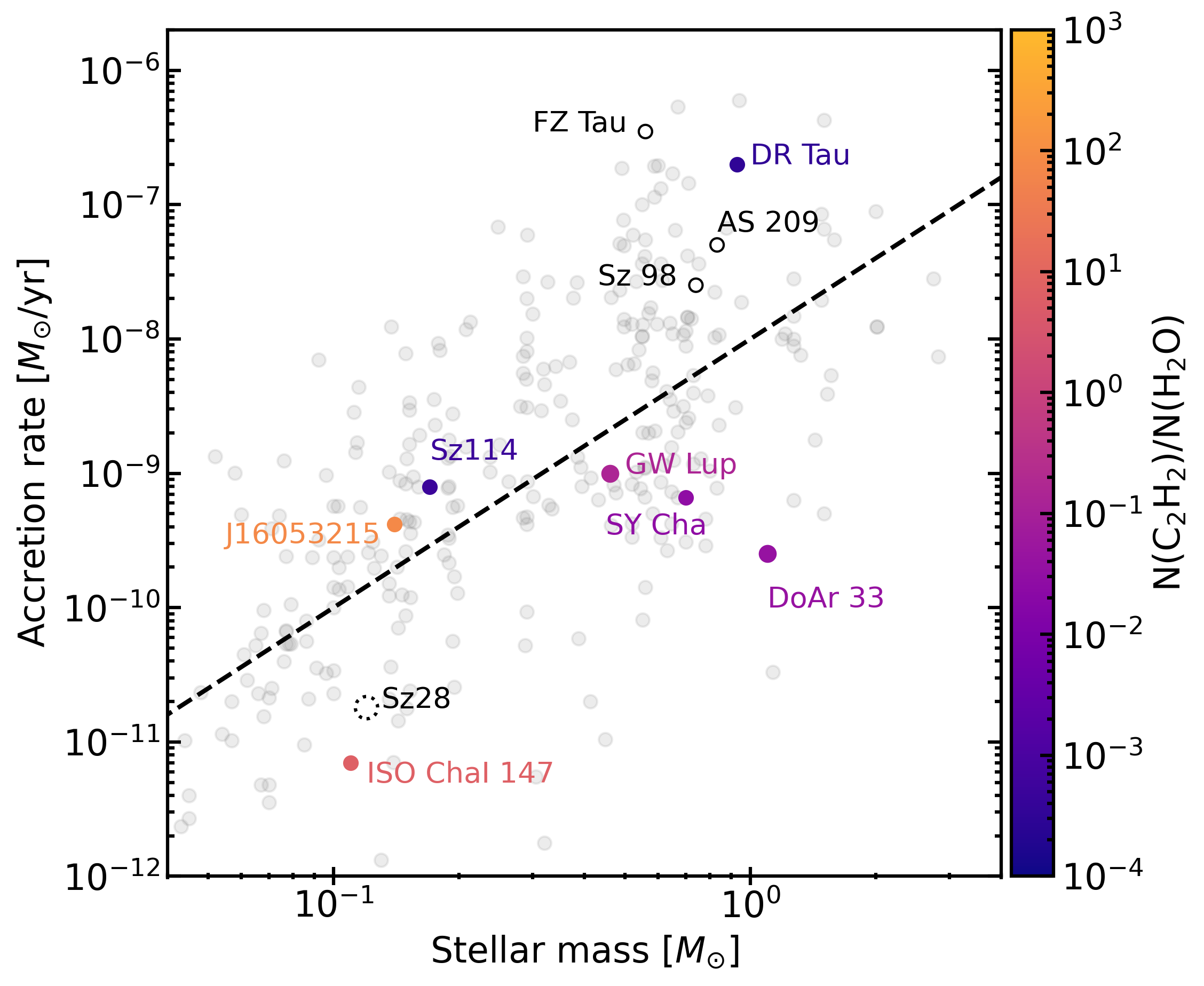}
    \caption{\ce{C2H2} to water fraction for disks surrounding a range of stellar masses and accretion rates. \reply{The grey points are taken from \cite{manara2023}}. The colored points have JWST MIRI observations and column density estimates. The color indicates the ratio of \ce{C2H2}/\ce{H2O} column. The empty \reply{solid-line} points report no \ce{C2H2} detection, and the empty \reply{dotted-line} points report no \ce{H2O} detection. The dashed line corresponds to $\dot M\propto M_\star^2$.}
    \label{fig:accretion_rates}
\end{figure}

We show in \autoref{fig:accretion_rates} the relation between mass accretion rate and stellar mass for a large sample of young stellar objects (YSOs). We highlight the sources that have JWST MIRI observations and detections of multiple molecules, with the color indicating the ratio of column densities of \ce{C2H2} and \ce{H2O}, independently estimated for each system \citep{grant2023,Gasman2023,pontoppidan2024,tabone2023,Xie2023,munoz-romero2024,schwarz2024,temmink2024, arabhavi2024, kanwar2024}. The empty circles indicate the systems where \ce{C2H2} is not detected. In some of these systems the \ce{H2O} emission is modeled using one or more components; here, we choose the components whose temperature is closest to the water emitting temperature for DoAr~33 ($\sim$600\,K), or we take the value for the component with the largest \ce{H2O} column. For AS~209, the \ce{C2H2} column is given as an upper limit on a non-detection \citep{munoz-romero2024}, meaning that the actual column density ratio could be even lower. Similarly, for J160532, the \ce{H2O} column should be taken as an upper limit, indicating a potentially higher ratio. It is clear that in general the sources with relatively low accretion rates have higher $N(\ce{C2H2})/N(\ce{H2O})$ ratios, which is consistent with our interpretation of long radial mixing timescales allowing the carbon-rich chemistry to survive. This of course depends on the specific details of the system, as can be seen by Sz 114, which has a water-rich spectrum despite having a low accretion rate \citep{Xie2023}. 

There are possible caveats to this scenario. First, one must consider the limitations of the $\alpha$-prescription for turbulence, since actual turbulence within the disk might not follow this simple parametrization, and this analysis assumes a steady-state disk. The value of $\alpha$ itself is difficult to characterize and may vary throughout the disk depending on the angular momentum transport mechanism, e.g. MRI vs.\ winds \citep{turner2014, pascucci2023}.  Further, the mixing of material from the midplane relies on the gas diffusivity which may not be the same as the viscosity \citep{Stevenson1990}, as is commonly adopted.  

Another issue is that in \autoref{tab:slab_models} the slab LTE analysis finds that \ce{C2H2} arises from a smaller emitting area and warmer temperature than \ce{C4H2}.  Thus we observe hot (T $>$ 500~K) organics that could be associated with the product of carbon grain sublimation and warm (T $\sim$ 200~K) organics that require another solution.  This could be related to the advection of outer disk gas with elevated C/O or, perhaps, could be associated with an accretion burst that burns carbon grains out to a larger radius.  When the disk material cools, the chemistry will gradually reset but the elevated C/O ratio, and its associated chemical equilibrium, will linger as discussed above. \reply{Since we are dealing with a young star, this scenario would be consistent with the accretion variability commonly observed in T Tauri stars \citep{Hartmann16}. The low accretion rate of DoAr\,33 could indicate that it is currently in a quiescent state following a high accretion event.} Another solution could be the outwards diffusion of the C-rich vapor produced at the soot line. Since this is a low accretion rate system, the inward flow of material is going to be slow, providing less of a current for this vapor to diffuse against, and because carbon-grain sublimation is posited to be irreversible \reply{\citep{Li21}}, that vapor that diffuses outward does not freeze-out again as it would at the \ce{H2O} or CO snow line. Thus organics might be found at lower temperatures. A clear implication of the soot line model is that it would foster the formation of silicate-rich worlds like the Earth inside the soot line.

\

\

\begin{figure*}
    \centering
    \includegraphics[width=17cm]{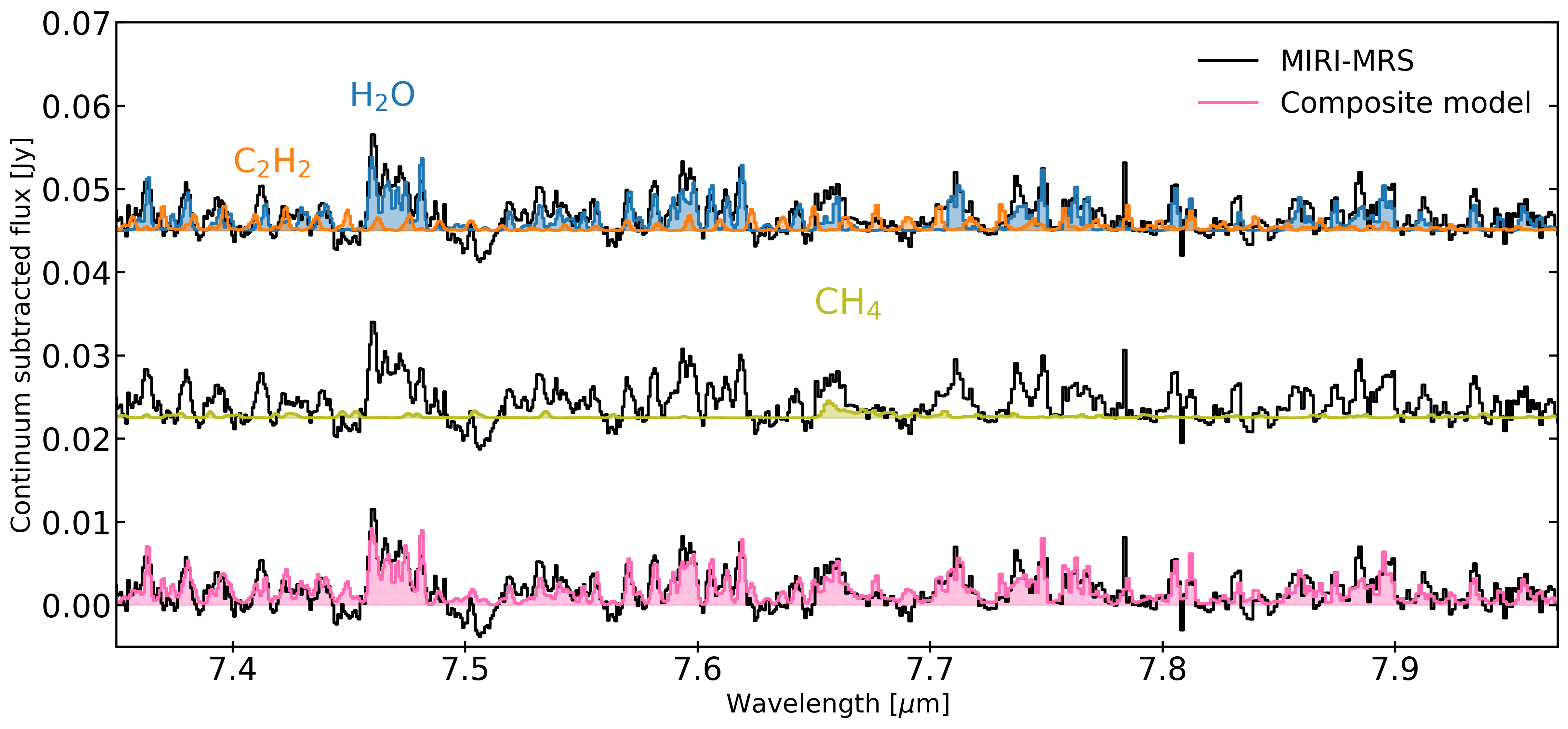}
    \caption{Tentative detection of \ce{CH4} in the MIRI-MRS spectrum of DoAr~33. The top spectra show the best-fit slab model of \ce{C2H2} in orange, along with a \ce{H2O} model in blue, with parameters indicated in the text. The MIRI spectra in the top and middle were offset for clarity. The green line shows the best fit \ce{CH4} slab model, and the bottom pink spectrum shows the composite model of the three species.}
    \label{fig:ch4_spec}
\end{figure*}

\begin{figure}
    \centering
    \includegraphics[width=8.5cm]{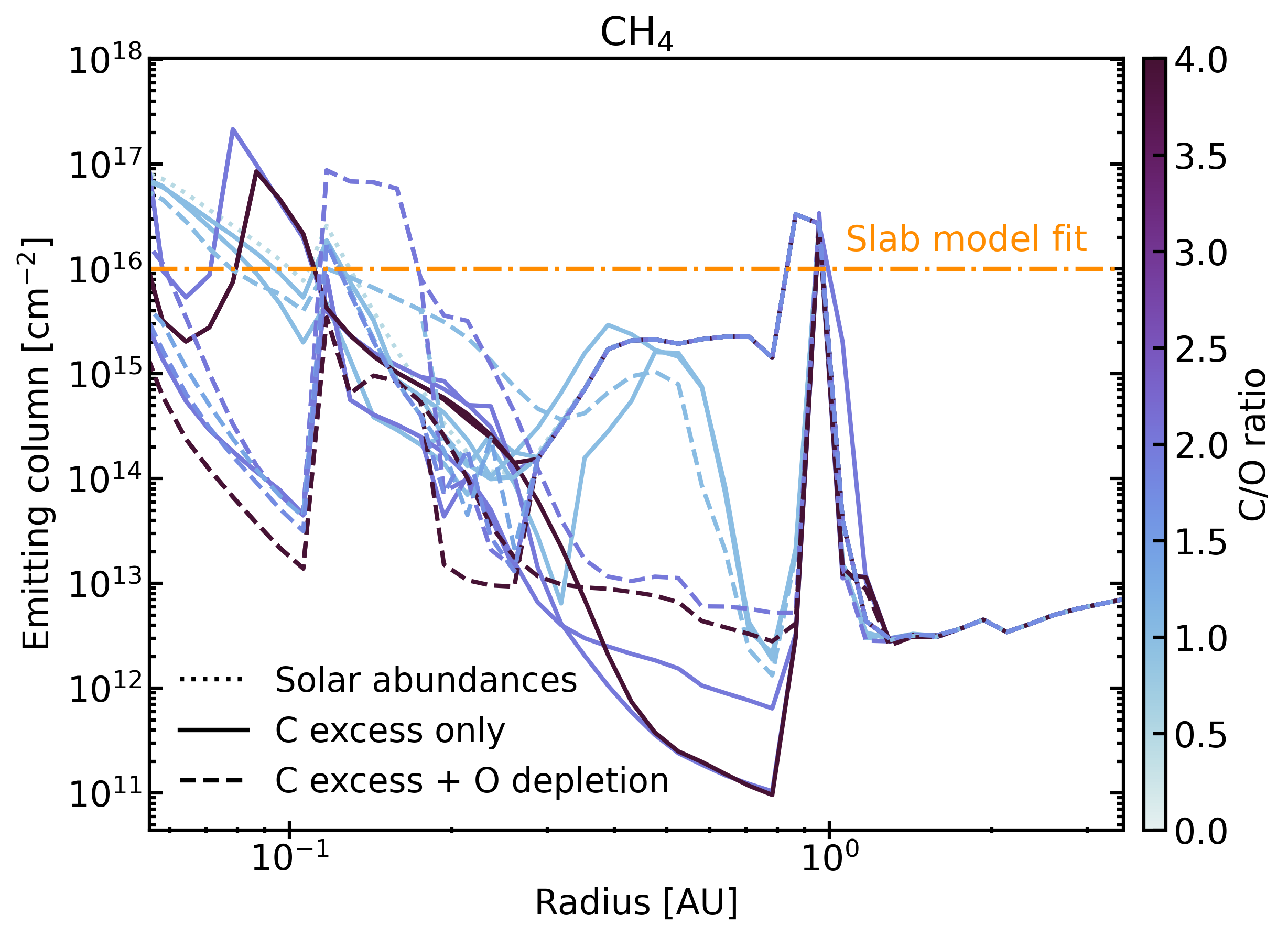}
    \caption{Column density of \ce{CH4} as a function of radius for the thermochemical models. The orange dashed line corresponds to the upper limit of \ce{CH4} column derived from slab models, assuming the same emitting temperature and area as derived for \ce{C2H2}.}
    \label{fig:methane_models}
\end{figure}

\subsection{\ce{CH4} and \ce{HC3N} detections}
\label{app:ch4}

We present a tentative detection of \ce{CH4} in the 7--8$\mu$m region of the spectrum of DoAr~33. This feature overlaps with several other \ce{H2O} and \ce{C2H2} emission lines, therefore, in \autoref{fig:ch4_spec} we show an \ce{H2O} model fitted for this specific region of the spectrum, with \reply{$T= 750\,$K}, \reply{$N=1.5\times10^{18}$\,cm$^{-2}$}, and \reply{$R= 0.08\,$au}, and the same \ce{C2H2} model retrieved for longer wavelengths (\S~\ref{sec:observations}). Assuming that \ce{CH4} is emitting at the same temperature and from the same area as \ce{C2H2}, we are able to match the feature with a slab model with a column of \reply{$N=2.2\times 10^{16}$\,cm$^{-2}$}, after removing the contributions from water and acetylene. \reply{The \ce{CH4} feature is detected with a confidence level of 2.7$\sigma$, using the continuum noise level from the 8.55--8.6$\mu$m region ($\sigma=1.49$\,mJy).} We note that we consider this an upper limit to the total methane content due to the detection being tentative.  

We compare this value from that obtained in our thermochemical modeling process in \autoref{fig:methane_models}, where we assume the carbon released from sublimation at the soot line is carried by \ce{CH4}. In general, the column densities are lower than $10^{17}\,\rm{cm}^{-2}$ everywhere in the disk, consistent with the expectation from the slab models. It is clear that despite introducing the carbon as methane, most of this excess carbon goes into generating more complex species such as \ce{C2H2} and \ce{C4H2}. Therefore, we expect the methane abundance to be low in disks with a rich hydrocarbon chemistry. 

\begin{figure*}
    \centering
    \includegraphics[width=17cm]{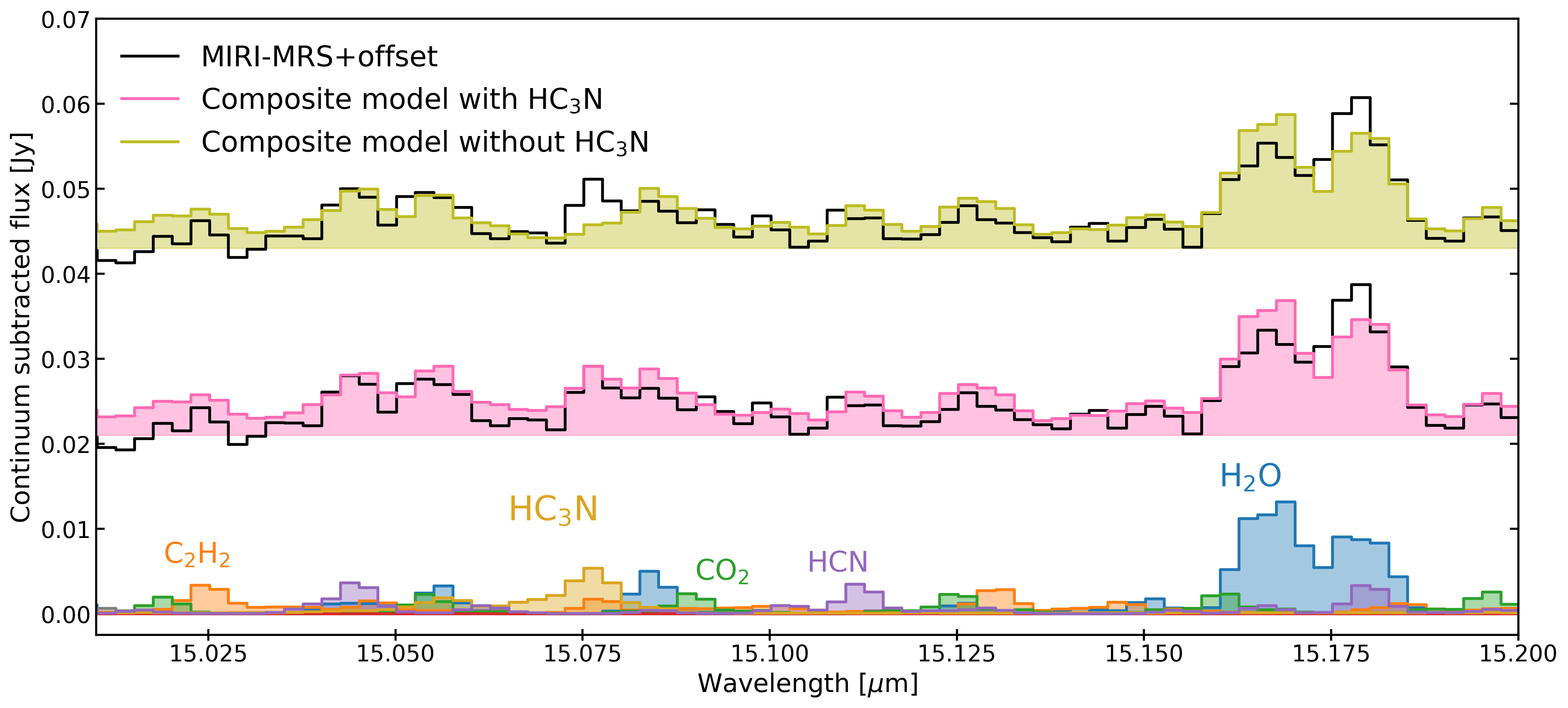}
    \caption{Tentative detection of \ce{HC3N} in the spectrum of DoAr 33. The bottom spectrum shows the individual models for each molecule, the pink spectrum in the middle shows the composite model when \ce{HC3N} is included, and the top green spectrum shows the composite model when \ce{HC3N} is not included.}
    \label{fig:hc3n}
\end{figure*}

We report a tentative detection of the $Q$ branch of \ce{HC3N} at 15.07\,$\mu$m, shown in \autoref{fig:hc3n}.  We fit this feature after subtracting the best-fit models from the spectrum, and find that we can reproduce the \ce{HC3N} feature using a slab model with $N=10^{14}\rm{cm}^{-2}$, \reply{$T=250\,\rm K$ }and \reply{$R_{\rm slab} = 1.80\,\rm{au}$}. \reply{This feature is detected with a confidence level of 7$\sigma$, using the noise level indicated in \autoref{app:slab_models}.} For comparison, we show how the composite model looks with and without the \ce{HC3N} model in \autoref{fig:hc3n}.

\section{Summary} 
\label{sec:summary}

We detect the presence of \ce{H2O}, \ce{CO2}, \reply{OH}, \ce{C2H2}, \ce{HCN}, \ce{C4H2} and tentatively \ce{CH4} and \ce{HC3N} in the JWST MIRI-MRS spectrum of the solar-mass star DoAr~33. The high \ce{C2H2} fluxes observed, and the presence of \ce{C4H2}, indicate 
that the inner disk of this source is hydrocarbon rich, contrary to the expectation of a carbon-poor environment around solar-mass stars, making this the first observed disk of its kind. To disentangle the origin of this carbon-rich chemistry, we analyze the spectrum with a combination of LTE slab models and thermochemical models.
\begin{enumerate}
    \item The thermochemical models that best fit the \ce{C2H2} emitting column have C/O$=2-4$, and emitting areas of \reply{$\lesssim$$\pi(0.25\rm\,au)^2$}. This small emitting area, as well as the high emitting temperature derived for \ce{C2H2} (500 K) suggests the carbon is being released at around 500 \,K, likely as the results of sublimation of carbonaceous grains at the soot line.
    \item The \ce{C4H2} emission appears to be originating from warm ($\sim$150\,K), more extended gas, compared to \ce{C2H2}. We argue that this carbon-rich material, in contrast to \ce{C2H2}, could be a product of advection of C-rich material from the outer disk, a result of an earlier accretion burst, or, outwards diffusion of the C-rich vapor produced at the soot line via irreversible sublimation of C-rich refractory grains. 
    \item We propose the estimated C/O ratio is a consequence of the low accretion rate of DoAr~33, which favors vertical mixing of material over radial transport, allowing us to probe the composition of the midplane where carbonaceous grains are being sublimated.  That is, the carbon is released by sublimation and is allowed to linger due to the longer radial advection timescales. We extend this idea to explain the observed carbon dichotomy between disks around low and solar-mass stars, predicting that stars with low accretion rates will promote (hydro-)carbon-rich environments. 
\end{enumerate}

The DoAr 33 source spectrum shows that the landscape of hydrocarbon-rich chemistry is not solely associated with very low mass stars.  Rather, the chemical effects observed are likely present, but \reply{hiding closer to the midplane}, in most full-disk systems.  One important aspect that will be grounding for understanding the origin of elevated C/O ratios is the gaseous CO abundance in the inner disk.  In the absence of abundant water, CO (and CO$_2$) become the dominant oxygen carriers.  Understanding their abundance is central towards delineating the total amount of carbon needed to account for the rich-hydrocarbon chemistry which, in turn, can inform on formation models and the ultimate planetary composition. As a complement, the determination of absolute abundances, which are ultimately tied to the volatile element metallicity in the gas, would be key in helping us make connections to exoplanet atmospheres and stellar photospheres. 

\begin{acknowledgements}
We thank the referee for their insightful comments, which helped improve the quality of this paper. We thank John Black for providing the LAMDA file of \ce{C2H2} spectroscopy used in the thermochemical models. M.J.C.\, and  E.A.B. acknowledge funding from the NASA's Exoplanet Research Program, grant 80NSSC20K0259. E.A.B. acknowledges support from NSF grant No. 1907653 and NASA's Emerging Worlds Program, grant 80NSSC20K0333. This work is based on observations made with the NASA/ ESA/CSA James Webb Space Telescope. The specific observations analyzed can be accessed via \dataset[https://doi.org/10.17909/1ca9-gp84]{https://doi.org/10.17909/1ca9-gp84}. The data were obtained from the Mikulski Archive for Space Telescopes at the Space Telescope Science Institute, which is operated by the Association of Universities for Research in Astronomy, Inc., under NASA contract NAS 5-03127 for JWST. A portion of this research was carried out at the Jet Propulsion Laboratory, California Institute of Technology, under a contract with the National Aeronautics and Space Administration (80NM0018D0004).
\end{acknowledgements}

\facilities{JWST.}

\software{\texttt{astropy} \citep{astropy:2013,astropy:2018,astropy:2022}, \texttt{spectools-ir} \citep{spectools2022}, \texttt{iSLAT} \citep{jellison2024,iSLAT_code}, \texttt{RADMC-3D} \citep{radmc3d}.}

\appendix

\section{Dust features}
\label{app:dust}
\begin{figure*}
    \centering
    \includegraphics[width=16.5cm]{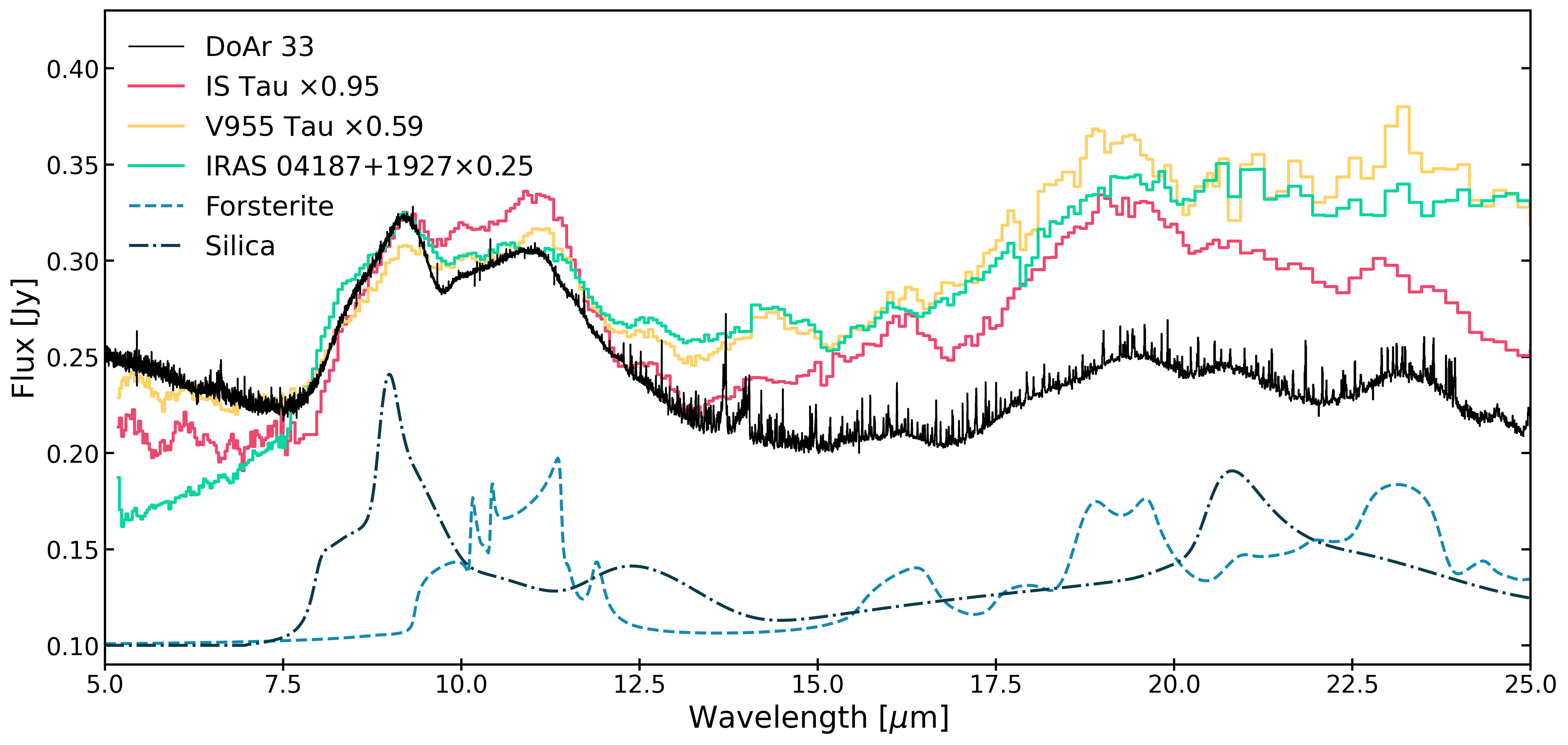}
    \caption{Dust features present in the MIRI-MRS spectra of DoAr~33. The Spitzer observations are taken from \citet{furlan2011} and have been scaled to match the peak of the 10\,$\mu$m feature. The dust opacities from \texttt{optool} are shown for reference, and have also been scaled. }
    \label{fig:dust_features}
\end{figure*}

\reply{We detect the presence of several dust features in the spectra of DoAr~33. However, we do not aim to model the exact morphology of the underlying dust continuum, rather, we provide here a comparison to systems previously observed with Spitzer and to dust opacity models, with the goal of contextualizing the hydrocarbon-rich environment around this source. In \autoref{fig:dust_features}, we show the spectrum of DoAr~33, overplotted with the Spitzer-IRS spectra of IS Tau, V955 Tau and IRAS 04187+1927, taken from \cite{furlan2011}, scaled to match the MIRI-MRS spectrum. We chose these three sources because they all have a double-peaked 10~$\mu$m feature, that has been suggested is observed due to the presence of olivines, pyroxenes and silica \citep{watson2009}. We generated default opacities for forsterite (\ce{Mg2SiO4}) and silica (\ce{SiO2}) using \texttt{optool} \citep{dominik2021}, assuming the standard power-law size distribution $N(a)\propto a^{-3.5}$. Several of the forsterite and silica features beyond 15~$\mu$m seem to be present in the spectrum of DoAr~33, along with the double-peak at 10~$\mu$m. As interstellar grains are predominantly amorophous in form \citep{kemper2004} these silicates are argued to indicate the presence of thermal annealing  at temperatures $>$500~K in the disk midplane \citep{jang2024}.  The presence of these crystalline grains is potential evidence for dust processing at high temperatures, consistent with the scenario of refractory grain sublimation. }

\section{Additional \ce{C4H2} slab models}
\label{app:c4h2_ex}

\begin{figure}
    \centering
    \includegraphics[width=8.5cm]{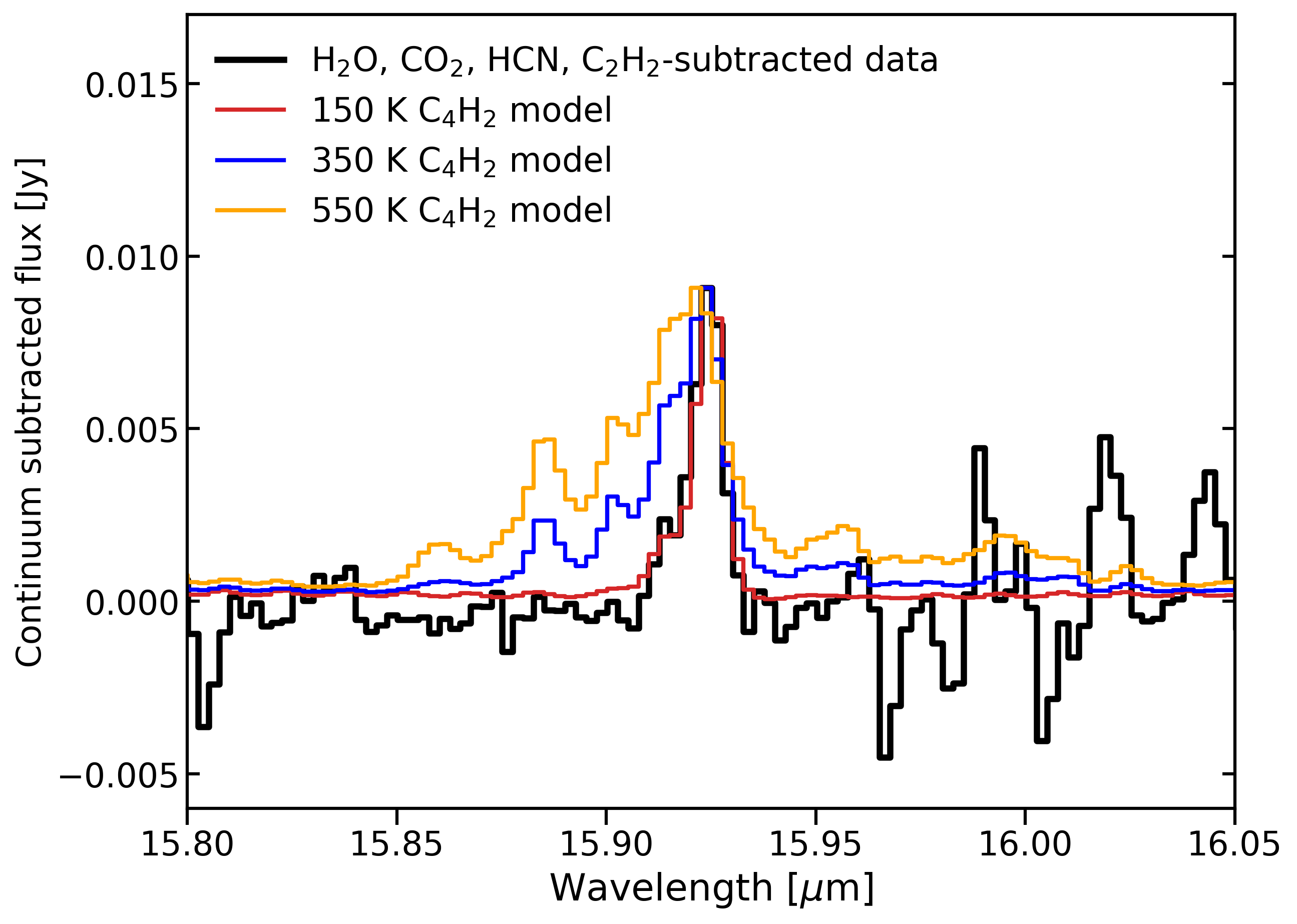}
    \caption{Comparison of slab models of \ce{C4H2} emission with varying temperatures. The column density is the same for the three models, corresponding to 10$^{14}$\,cm$^{-2}$. The emitting area is adjusted to match the peak of the emission feature. The best-fit model corresponds to the \reply{150}\,K line.}
    \label{fig:c4h2_alt}
\end{figure}

We show in \autoref{fig:c4h2_alt} the effects of fitting the $Q$-branch of \ce{C4H2} with higher temperatures than the best-fitting one. Temperatures higher than 200\,K broaden the Q-branch emission and do not reproduce the observed feature.  This implies that there is a more extended, warm reservoir of carbon-rich material.

\section{SED modeling}
\label{app:SED}

\begin{figure}
    \centering
    \includegraphics[width=8.5cm]{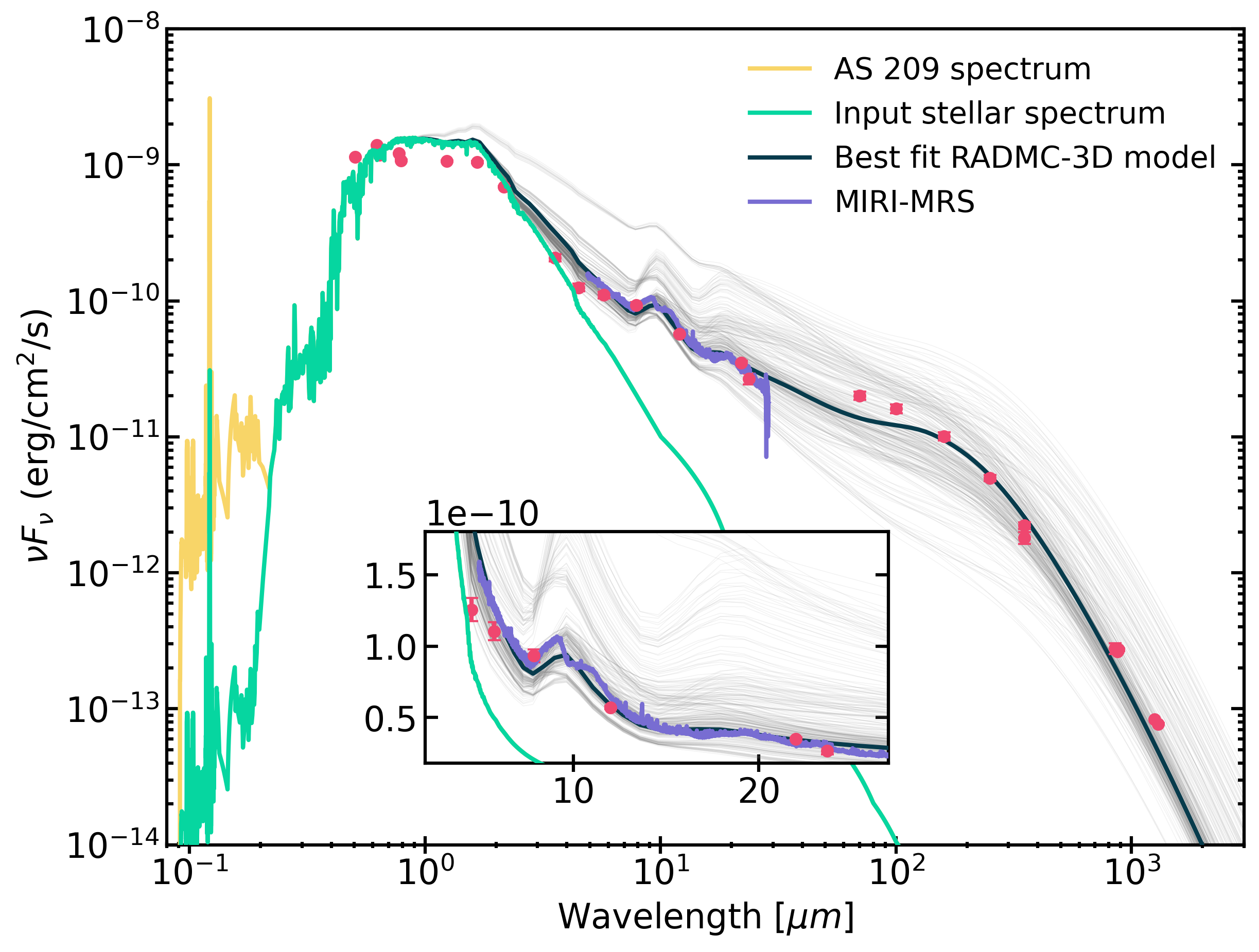}
    \caption{SED fits of the radiative transfer models using \texttt{RADMC-3D}. The input stellar UV flux for DoAr~33 is reduced relative to AS~209 for wavelengths $<$0.22$\mu$m. The best-fit model is shown as the solid black line, and the other models are shown in light grey. \reply{The inset shows the fit to the MIRI-MRS spectrum.}}
    \label{fig:SED}
\end{figure}

In order to accurately model the dust distribution of DoAr~33, we used the radiative transfer code \texttt{RADMC-3D} \citep{radmc3d}. This code allows us to input a dust distribution, stellar spectrum, and opacities, to calculate an SED. We used the DSHARP opacities \citep{Birnstiel_dsharp}, along with an AS209-like stellar spectrum \citep{Dionatos19, Zhang21_mapsco}, with reduced UV emission to account for the difference in accretion rates between DoAr~33 and AS209. We show the difference between the two spectra in \autoref{fig:SED}. For the dust, we assumed two populations of small and large grains, following the equations described in \S~\ref{sec:methods}. We varied all of the parameters listed in \autoref{tab:model_params}, assuming the following values: 

\begin{itemize}
    \item $R_c$: 18, 20, 22, 40, 100 au
    \item $\Sigma_c$: 13, 15, 17 g\,cm$^{-2}$
    \item $\gamma$: 0.7, 0.9 
    \item $h_c$: 0.04, 0.045, 0.05, 0.06, 0.1, 0.3
    \item $\psi$: 0.05, 0.08, 0.09, 0.1, 0.2, 0.3 
    \item $f$: 0.99, 0.999
    \item $\chi$: 0.2, 0.4, 0.7, 0.9 
\end{itemize}

We ran a model for each combination of parameters, and calculated their SEDs. To compare to observations, we used the photometry for DoAr~33 reported by \citet{liu2022}, which covers the wavelength space from 0.4-1300 $\mu$m. We used these points, along with the MIRI-MRS spectrum to perform a $\chi^2$ minimization. In the ranges where the spectrum and the photometry overlapped, we favored the fit to the spectrum. We found we were not able to match the 100$\mu$m region of the SED with a smooth dust profile, therefore, we increased the flaring of large dust grains to mimic the 9\,au gap feature found with mm-observations \citep{Huang18}. We tested the change of large dust settling at three locations in the disk: 9, 10, and 12 au. Inside of this radius, we assumed a settling of $\chi=0.1$, and outside of this we tested two settling factors $\chi=0.8,0.9$. We found the minimum $\chi^2$ was obtained in the model with $\chi=0.8$ outside 10\,au. The fit for this model, along with the photometry, MIRI MRS spectrum, and input stellar spectrum, are shown in \autoref{fig:SED}.

\begin{figure}
    \centering
    \includegraphics[width=8.5cm]{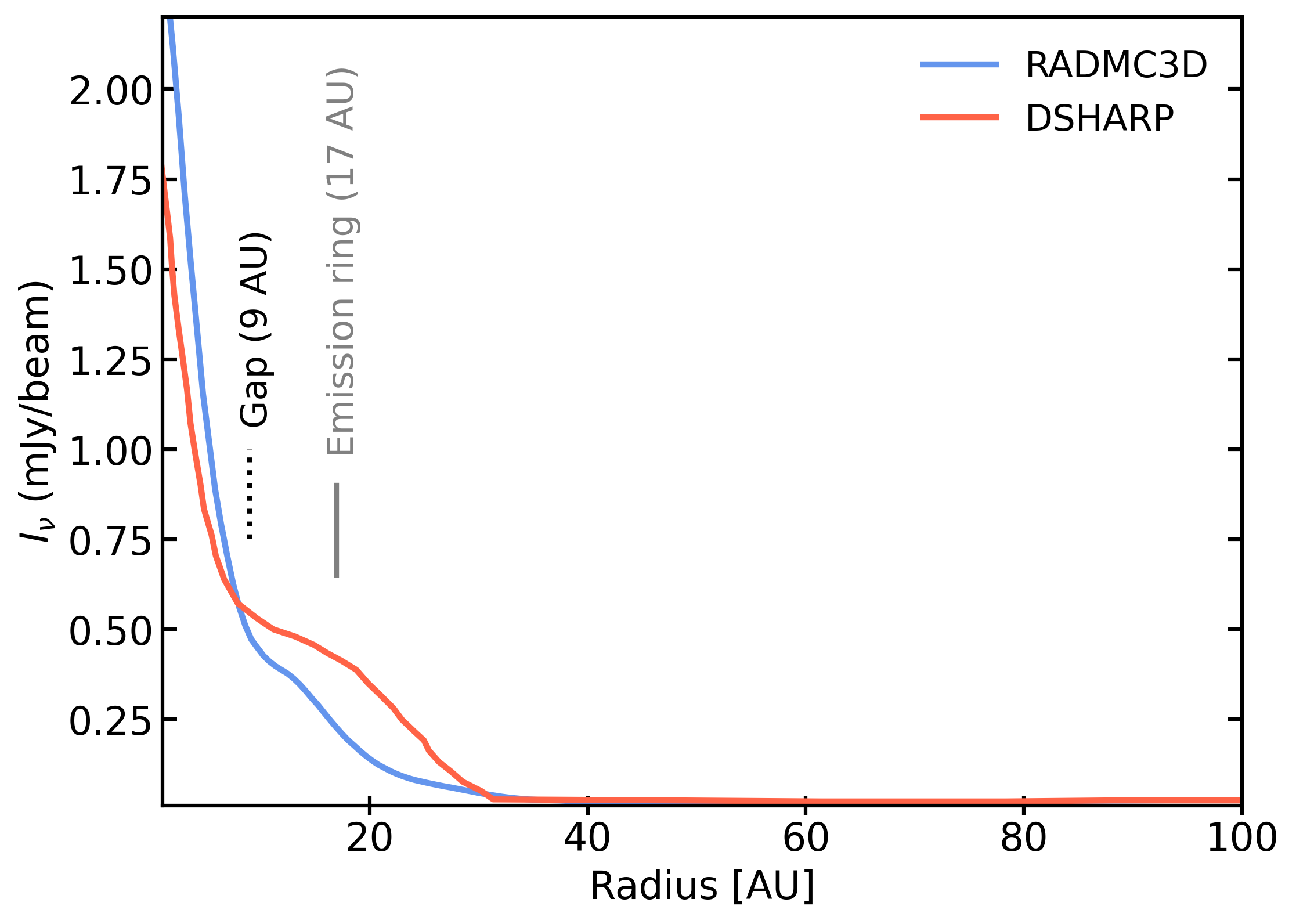}
    \caption{Deprojected and azimuthally averaged radial intensity profiles of the \texttt{RADMC-3D} model and the DSHARP observation \citep{Huang18}. }
    \label{fig:dsharp}
\end{figure}

We calculate a continuum image at $\lambda=1.25\,$mm which we convolve to the corresponding beam to compare with the observation from DSHARP \citep{Huang18}. The deprojected and azimuthally averaged radial intensity profile of the model is shown, along with the observation, in \autoref{fig:dsharp}.

\section{Slab model fits}
\label{app:slab_models}

\begin{figure*}
    \centering
    \includegraphics[width=14.5cm]{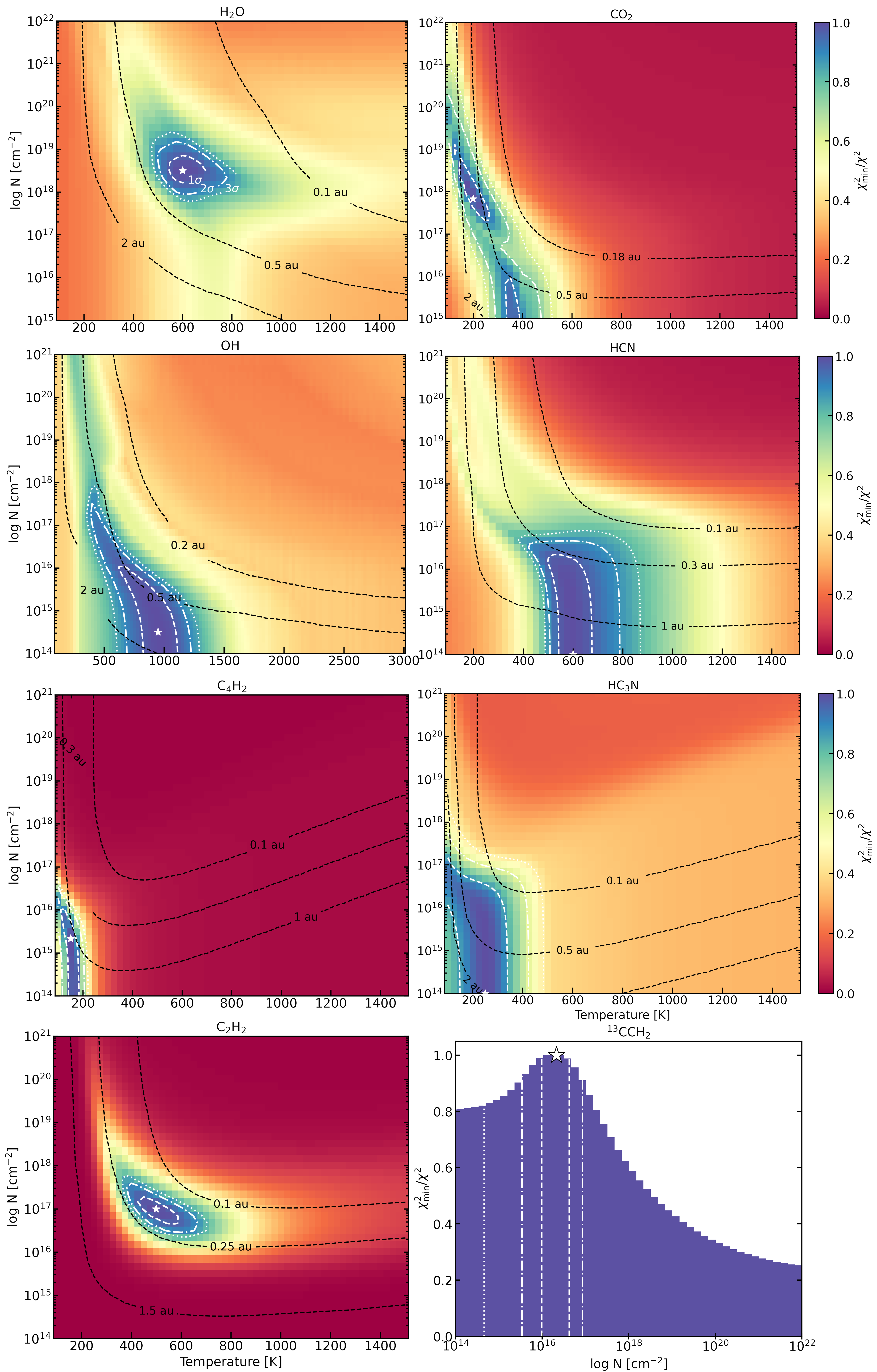}
    \caption{Grids of $\chi^2$ for the molecules fitted with LTE slab models. The color scale shows the reduced $\chi^2$, the white contours show the $1\sigma$, $2\sigma$ and $3\sigma$ uncertainty levels, and the dashed black contours correspond to the best-fit emitting radius for each pair. \reply{The bottom panel right corresponds to the \ce{^{13}C^{12}CH2} fit, where the temperature and emitting area are fixed to that of \ce{C2H2}, therefore, we fit only the column density.}}
    \label{fig:chi2}
\end{figure*}

\begin{figure*}
    \centering
    \includegraphics[width=13.5cm]{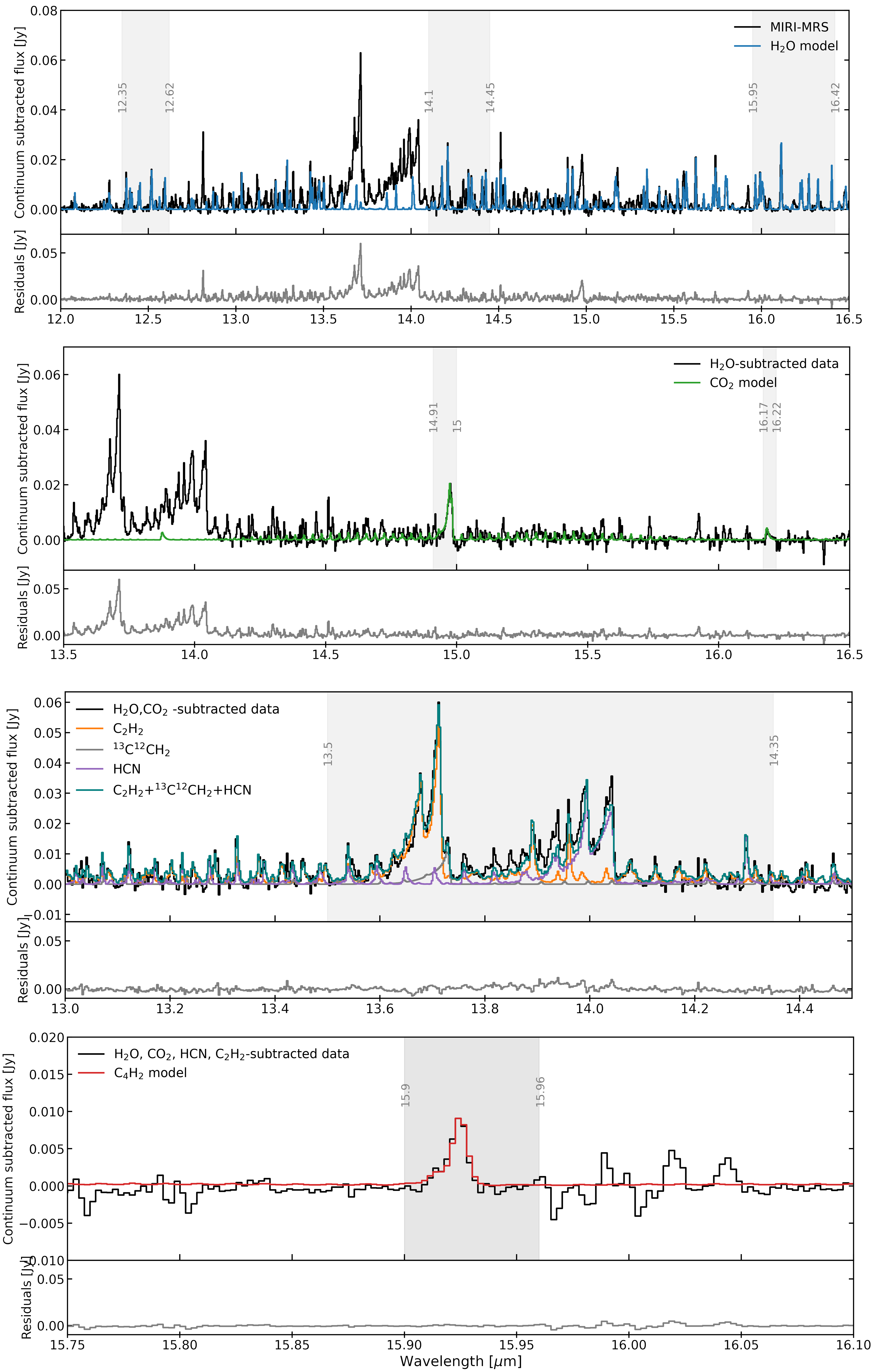}
    \caption{Best-fit slab model fits to the data for the molecules detected. The \ce{H2O} model is subtracted before fitting \ce{CO2}, and the \ce{CO2} model is subtracted before simultaneously fitting \ce{C2H2} and HCN. The shaded region in each plot shows the regions used to calculate the $\chi^2$. The grey subplots at the bottom show the residuals after subtracting the slab models.}
    \label{fig:slab_specs}
\end{figure*}

\begin{figure*}
    \centering
    \includegraphics[width=16cm]{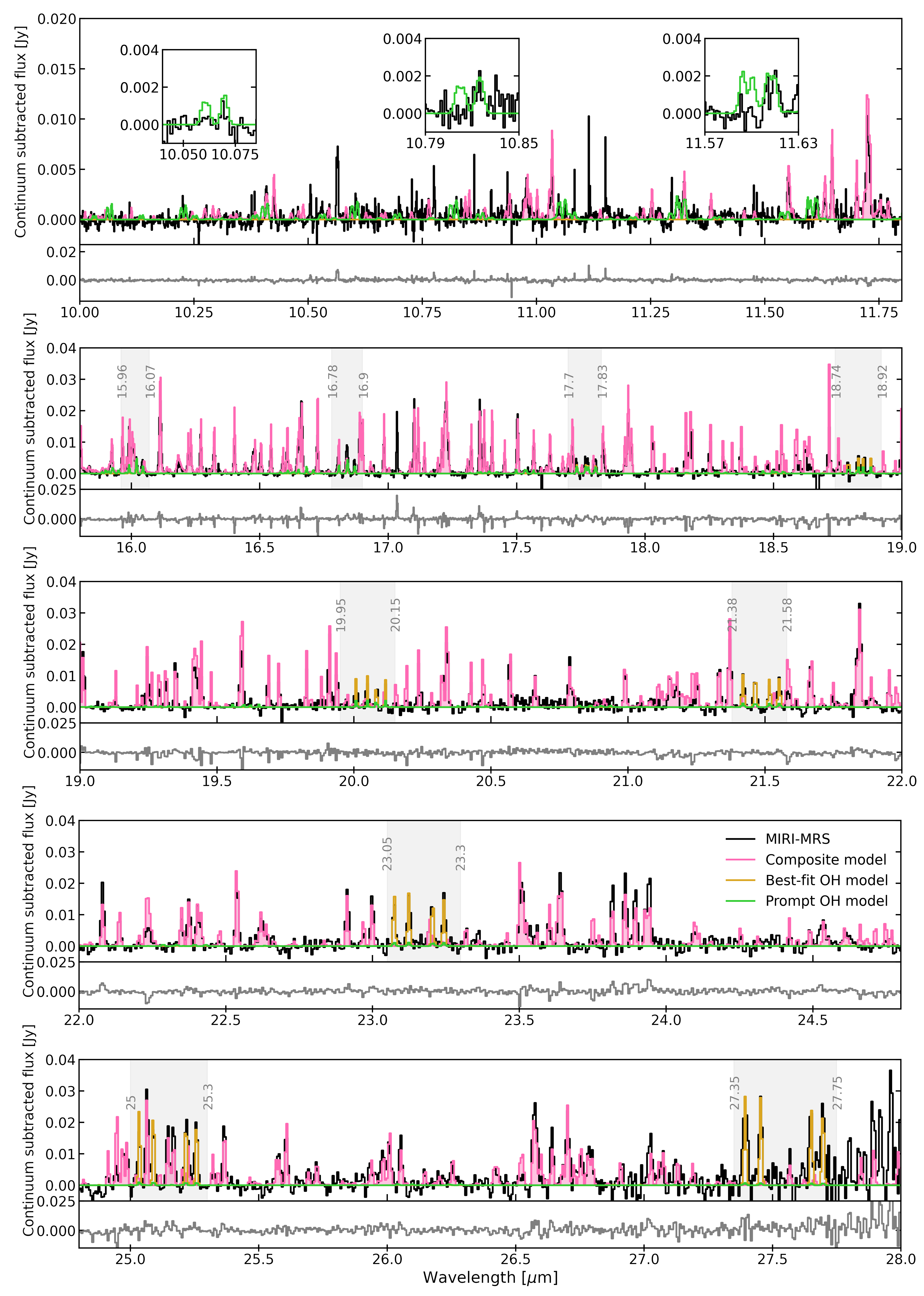}
    \caption{\reply{Best-fit slab model for OH, show in yellow. For reference, we show a 6000\,K OH model in green, to point out the asymmetry observed in the prompt emission lines in the top panel. The grey lines in the subplots show the residuals for each region. The shaded zones show the ranges used to calculate the $\chi^2$. Most of the emission from the composite model correspond to water lines, assuming the same parameters as the one retrieved in the $12-16\,\mu$m range.}}
    \label{fig:slab_specs_oh}
\end{figure*}

We calculate the reduced $\chi^2$ for each model following
\begin{equation}
    \chi^2 = \frac{1}{N}\sum_{i=0}^N\frac{(F_{\rm{obs},i}-F_{\rm{mod},i})^2}{\sigma^2}\,,
\end{equation}
\noindent where N represents the degrees of freedom, in this case corresponding to the quantity of resolution elements within the spectral windows used for conducting the fit minus the number of model parameters. We choose $\sigma$ to be the standard deviation in the region \reply{15.81-15.9}$\mu$m since it has no prominent emission features. We obtain a value of \reply{$\sigma=0.77$mJy}. The reduced $\chi^2$ values for each molecule are shown in \autoref{fig:chi2}.  The selected regions for calculating $\chi^2$ are shown in \autoref{fig:slab_specs} and \autoref{fig:slab_specs_oh}, along with the best fit models and their residuals. 

\section{Abundance maps from the thermochemical models}
\label{app:abundances}

\begin{figure*}
    \centering
    \includegraphics[width=17cm]{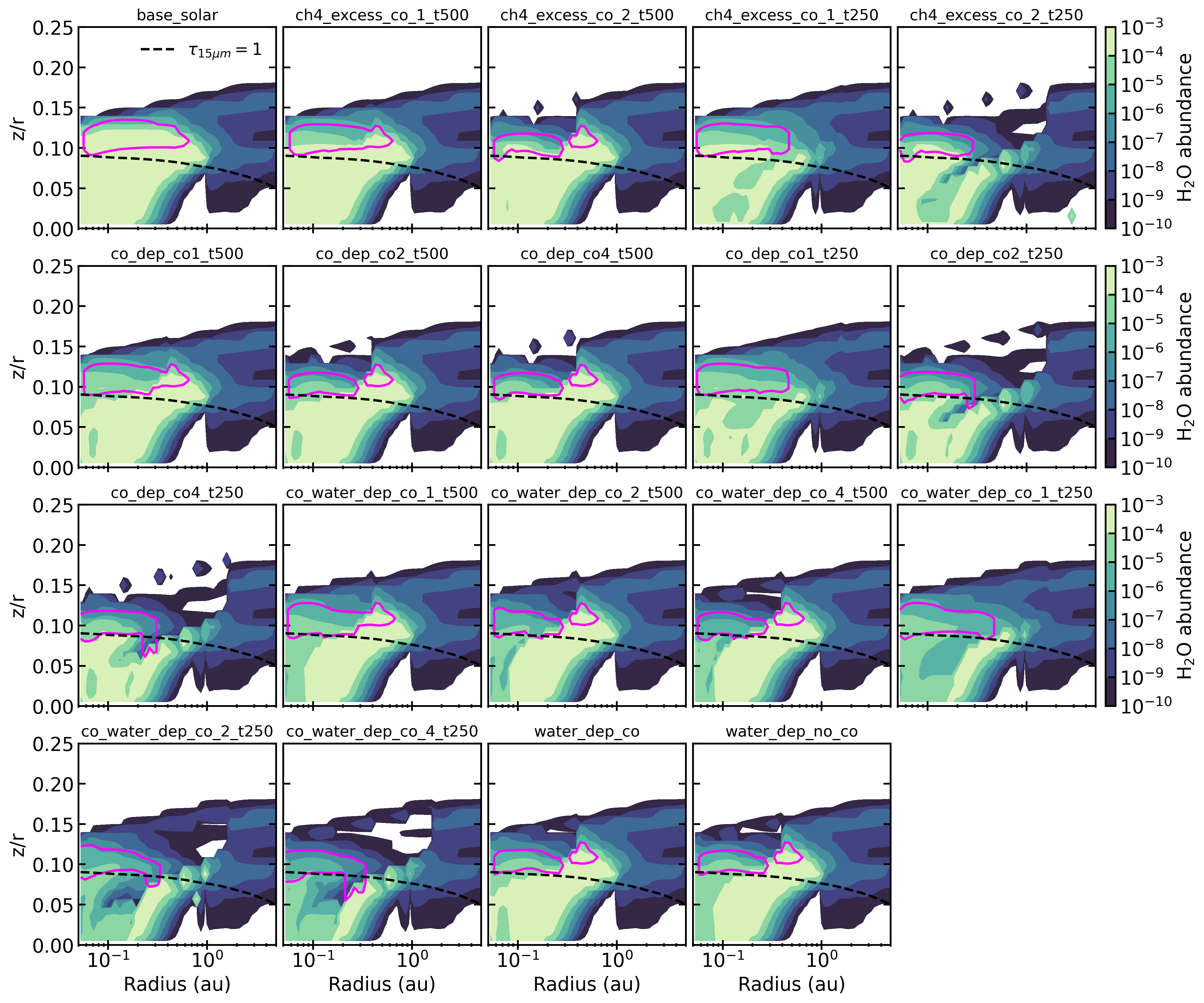}
    \caption{\ce{H2O} abundance maps for all of the thermochemical models. The magenta line corresponds to the contribution function for the \ce{H2O} transition in \autoref{tab:islat_transitions}. The titles correspond to those in \autoref{tab:dali_models}.}
    \label{fig:h2o_abundances}
\end{figure*}

\begin{figure*}
    \centering
    \includegraphics[width=17cm]{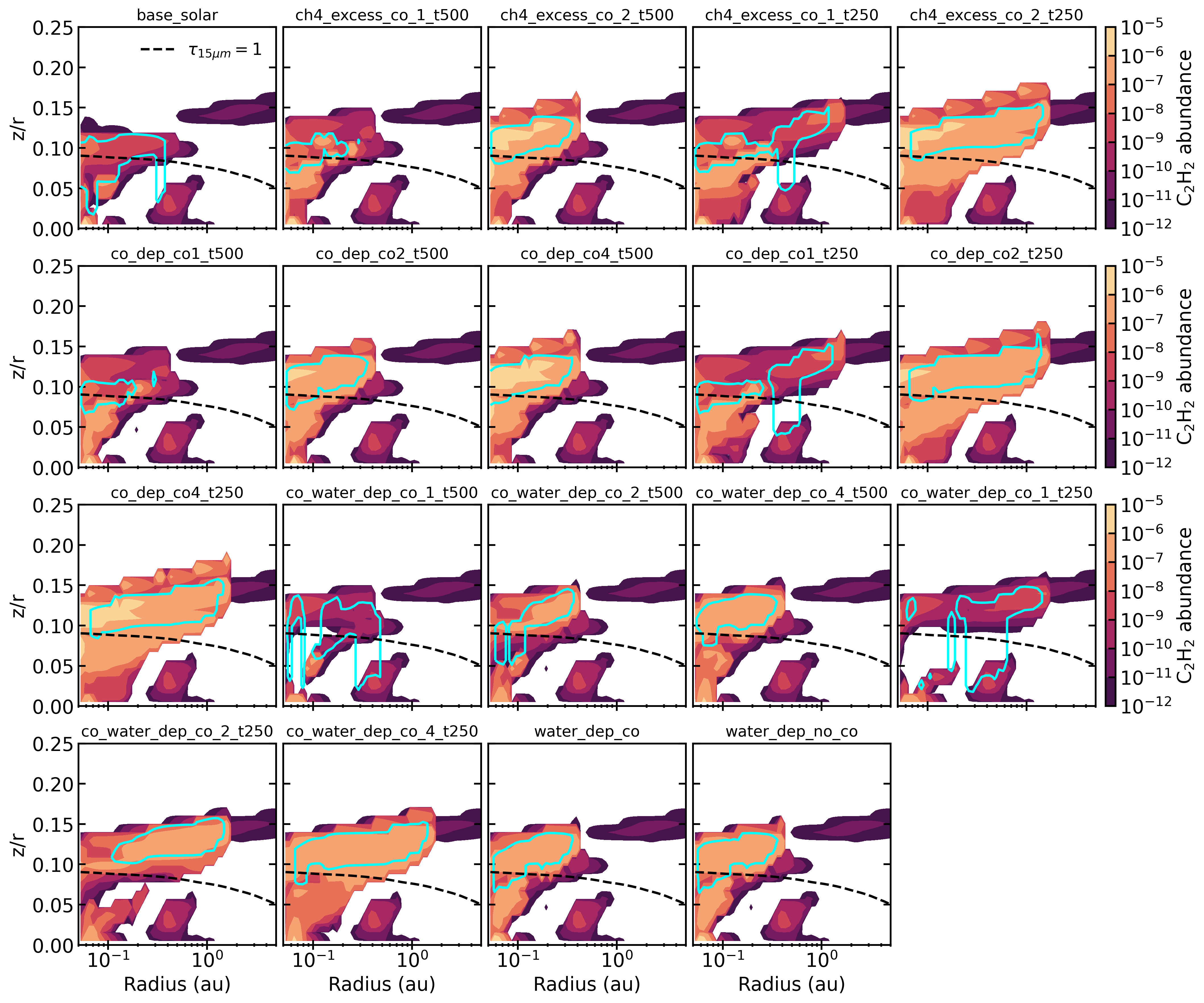}
    \caption{Same as \autoref{fig:h2o_abundances}, but for \ce{C2H2} abundances. The cyan line corresponds to the contribution function for the \ce{C2H2} transition in \autoref{tab:islat_transitions}.}
    \label{fig:c2h2_abundances}
\end{figure*}

\begin{figure*}
    \centering
    \includegraphics[width=17cm]{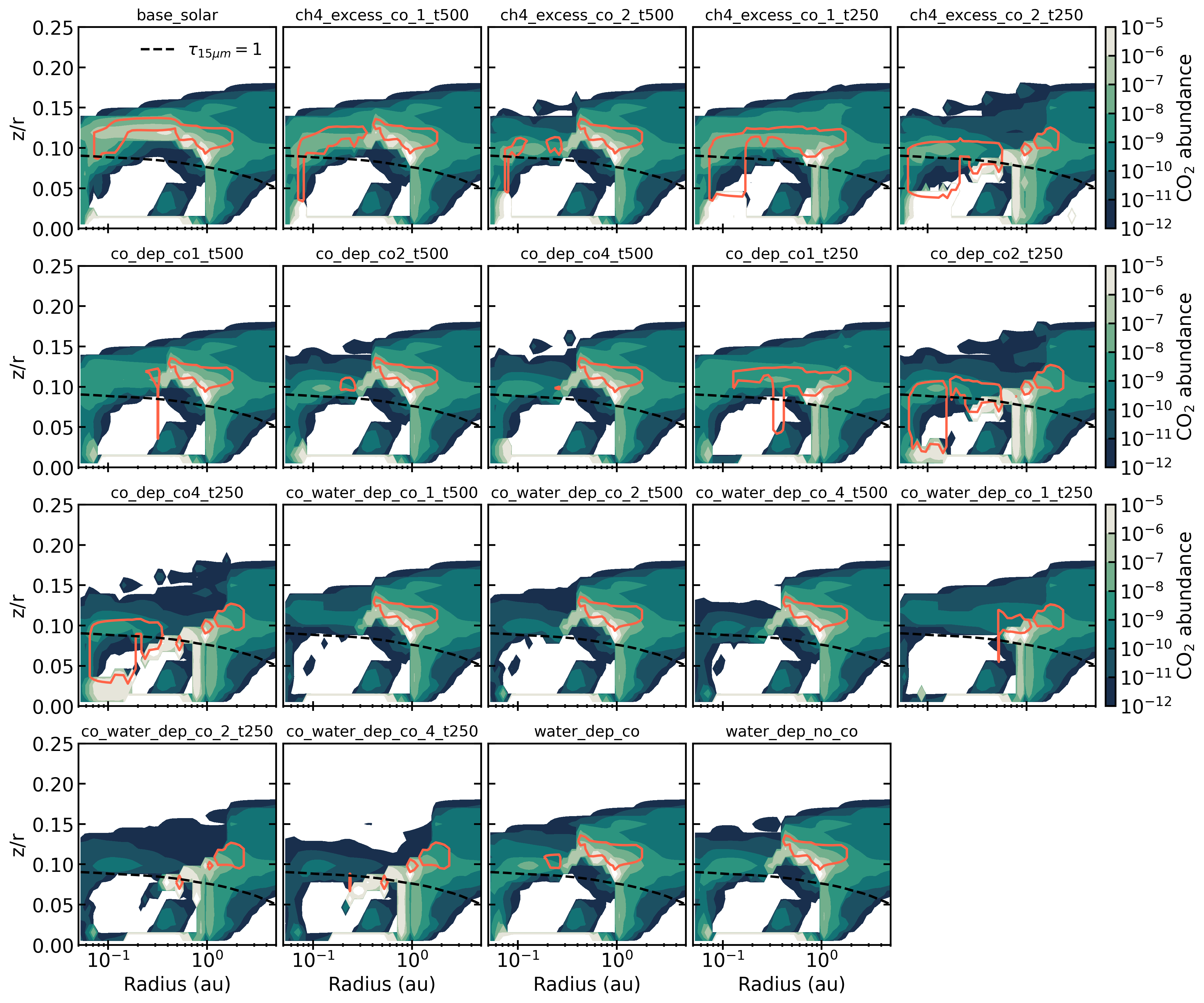}
    \caption{Same as \autoref{fig:h2o_abundances}, but for \ce{CO2} abundances. The orange line corresponds to the contribution function for the \ce{CO2} transition in \autoref{tab:islat_transitions}.}
    \label{fig:co2_abundances}
\end{figure*}

The abundance maps of \ce{H2O}, \ce{CO2} and \ce{C2H2} used to calculate the emitting columns of the thermochemical models are shown in \autoref{fig:h2o_abundances}, \autoref{fig:c2h2_abundances} and \autoref{fig:co2_abundances}, along with the $\tau=1$ surface at 15\,$\mu$m. We also include the contribution functions for each molecule, corresponding to the transitions in \autoref{tab:islat_transitions}.

\bibliography{a-main}{}

\begin{thebibliography}{}
\expandafter\ifx\csname natexlab\endcsname\relax\def\natexlab#1{#1}\fi
\providecommand{\url}[1]{\href{#1}{#1}}
\providecommand{\dodoi}[1]{doi:~\href{http://doi.org/#1}{\nolinkurl{#1}}}
\providecommand{\doeprint}[1]{\href{http://ascl.net/#1}{\nolinkurl{http://ascl.net/#1}}}
\providecommand{\doarXiv}[1]{\href{https://arxiv.org/abs/#1}{\nolinkurl{https://arxiv.org/abs/#1}}}

\bibitem[{{Aikawa} {et~al.}(1999){Aikawa}, {Umebayashi}, {Nakano}, \&
  {Miyama}}]{Aikawa1999}
{Aikawa}, Y., {Umebayashi}, T., {Nakano}, T., \& {Miyama}, S.~M. 1999, \apj,
  519, 705, \dodoi{10.1086/307400}

\bibitem[{{Alcal{\'a}} {et~al.}(2017){Alcal{\'a}}, {Manara}, {Natta}, {Frasca},
  {Testi}, {Nisini}, {Stelzer}, {Williams}, {Antoniucci}, {Biazzo}, {Covino},
  {Esposito}, {Getman}, \& {Rigliaco}}]{alcala2017}
{Alcal{\'a}}, J.~M., {Manara}, C.~F., {Natta}, A., {et~al.} 2017, \aap, 600,
  A20, \dodoi{10.1051/0004-6361/201629929}

\bibitem[{{Anderson} {et~al.}(2017){Anderson}, {Bergin}, {Blake}, {Ciesla},
  {Visser}, \& {Lee}}]{Anderson17}
{Anderson}, D.~E., {Bergin}, E.~A., {Blake}, G.~A., {et~al.} 2017, \apj, 845,
  13, \dodoi{10.3847/1538-4357/aa7da1}

\bibitem[{{Andrews} {et~al.}(2010){Andrews}, {Wilner}, {Hughes}, {Qi}, \&
  {Dullemond}}]{andrews2010}
{Andrews}, S.~M., {Wilner}, D.~J., {Hughes}, A.~M., {Qi}, C., \& {Dullemond},
  C.~P. 2010, \apj, 723, 1241, \dodoi{10.1088/0004-637X/723/2/1241}

\bibitem[{{Andrews} {et~al.}(2018){Andrews}, {Huang}, {P{\'e}rez}, {Isella},
  {Dullemond}, {Kurtovic}, {Guzm{\'a}n}, {Carpenter}, {Wilner}, {Zhang}, {Zhu},
  {Birnstiel}, {Bai}, {Benisty}, {Hughes}, {{\"O}berg}, \& {Ricci}}]{Andrews18}
{Andrews}, S.~M., {Huang}, J., {P{\'e}rez}, L.~M., {et~al.} 2018, \apj, 869,
  L41, \dodoi{10.3847/2041-8213/aaf741}

\bibitem[{{Arabhavi} {et~al.}(2024){Arabhavi}, {Kamp}, {Henning}, {van
  Dishoeck}, {Christiaens}, {Gasman}, {Perrin}, {G{\"u}del}, {Tabone},
  {Kanwar}, {Waters}, {Pascucci}, {Samland}, {Perotti}, {Bettoni}, {Grant},
  {Lagage}, {Ray}, {Vandenbussche}, {Absil}, {Argyriou}, {Barrado},
  {Boccaletti}, {Bouwman}, {Caratti o Garatti}, {Glauser}, {Lahuis}, {Mueller},
  {Olofsson}, {Pantin}, {Scheithauer}, {Morales-Calder{\'o}n}, {Franceschi},
  {Jang}, {Pawellek}, {Rodgers-Lee}, {Schreiber}, {Schwarz}, {Temmink},
  {Vlasblom}, {Wright}, {Colina}, \& {{\"O}stlin}}]{arabhavi2024}
{Arabhavi}, A.~M., {Kamp}, I., {Henning}, T., {et~al.} 2024, Science, 384,
  1086, \dodoi{10.1126/science.adi8147}

\bibitem[{{Argyriou} {et~al.}(2023){Argyriou}, {Glasse}, {Law}, {Labiano},
  {{\'A}lvarez-M{\'a}rquez}, {Patapis}, {Kavanagh}, {Gasman}, {Mueller},
  {Larson}, {Vandenbussche}, {Glauser}, {Royer}, {Dicken}, {Harkett},
  {Sargent}, {Engesser}, {Jones}, {Kendrew}, {Noriega-Crespo}, {Brandl},
  {Rieke}, {Wright}, {Lee}, \& {Wells}}]{argyriou2023}
{Argyriou}, I., {Glasse}, A., {Law}, D.~R., {et~al.} 2023, \aap, 675, A111,
  \dodoi{10.1051/0004-6361/202346489}

\bibitem[{{Astropy Collaboration} {et~al.}(2013){Astropy Collaboration},
  {Robitaille}, {Tollerud}, {Greenfield}, {Droettboom}, {Bray}, {Aldcroft},
  {Davis}, {Ginsburg}, {Price-Whelan}, {Kerzendorf}, {Conley}, {Crighton},
  {Barbary}, {Muna}, {Ferguson}, {Grollier}, {Parikh}, {Nair}, {Unther},
  {Deil}, {Woillez}, {Conseil}, {Kramer}, {Turner}, {Singer}, {Fox}, {Weaver},
  {Zabalza}, {Edwards}, {Azalee Bostroem}, {Burke}, {Casey}, {Crawford},
  {Dencheva}, {Ely}, {Jenness}, {Labrie}, {Lim}, {Pierfederici}, {Pontzen},
  {Ptak}, {Refsdal}, {Servillat}, \& {Streicher}}]{astropy:2013}
{Astropy Collaboration}, {Robitaille}, T.~P., {Tollerud}, E.~J., {et~al.} 2013,
  \aap, 558, A33, \dodoi{10.1051/0004-6361/201322068}

\bibitem[{{Astropy Collaboration} {et~al.}(2018){Astropy Collaboration},
  {Price-Whelan}, {Sip{\H{o}}cz}, {G{\"u}nther}, {Lim}, {Crawford}, {Conseil},
  {Shupe}, {Craig}, {Dencheva}, {Ginsburg}, {Vand erPlas}, {Bradley},
  {P{\'e}rez-Su{\'a}rez}, {de Val-Borro}, {Aldcroft}, {Cruz}, {Robitaille},
  {Tollerud}, {Ardelean}, {Babej}, {Bach}, {Bachetti}, {Bakanov}, {Bamford},
  {Barentsen}, {Barmby}, {Baumbach}, {Berry}, {Biscani}, {Boquien}, {Bostroem},
  {Bouma}, {Brammer}, {Bray}, {Breytenbach}, {Buddelmeijer}, {Burke},
  {Calderone}, {Cano Rodr{\'\i}guez}, {Cara}, {Cardoso}, {Cheedella}, {Copin},
  {Corrales}, {Crichton}, {D'Avella}, {Deil}, {Depagne}, {Dietrich}, {Donath},
  {Droettboom}, {Earl}, {Erben}, {Fabbro}, {Ferreira}, {Finethy}, {Fox},
  {Garrison}, {Gibbons}, {Goldstein}, {Gommers}, {Greco}, {Greenfield},
  {Groener}, {Grollier}, {Hagen}, {Hirst}, {Homeier}, {Horton}, {Hosseinzadeh},
  {Hu}, {Hunkeler}, {Ivezi{\'c}}, {Jain}, {Jenness}, {Kanarek}, {Kendrew},
  {Kern}, {Kerzendorf}, {Khvalko}, {King}, {Kirkby}, {Kulkarni}, {Kumar},
  {Lee}, {Lenz}, {Littlefair}, {Ma}, {Macleod}, {Mastropietro}, {McCully},
  {Montagnac}, {Morris}, {Mueller}, {Mumford}, {Muna}, {Murphy}, {Nelson},
  {Nguyen}, {Ninan}, {N{\"o}the}, {Ogaz}, {Oh}, {Parejko}, {Parley}, {Pascual},
  {Patil}, {Patil}, {Plunkett}, {Prochaska}, {Rastogi}, {Reddy Janga},
  {Sabater}, {Sakurikar}, {Seifert}, {Sherbert}, {Sherwood-Taylor}, {Shih},
  {Sick}, {Silbiger}, {Singanamalla}, {Singer}, {Sladen}, {Sooley},
  {Sornarajah}, {Streicher}, {Teuben}, {Thomas}, {Tremblay}, {Turner},
  {Terr{\'o}n}, {van Kerkwijk}, {de la Vega}, {Watkins}, {Weaver}, {Whitmore},
  {Woillez}, {Zabalza}, \& {Astropy Contributors}}]{astropy:2018}
{Astropy Collaboration}, {Price-Whelan}, A.~M., {Sip{\H{o}}cz}, B.~M., {et~al.}
  2018, \aj, 156, 123, \dodoi{10.3847/1538-3881/aabc4f}

\bibitem[{{Astropy Collaboration} {et~al.}(2022){Astropy Collaboration},
  {Price-Whelan}, {Lim}, {Earl}, {Starkman}, {Bradley}, {Shupe}, {Patil},
  {Corrales}, {Brasseur}, {N{"o}the}, {Donath}, {Tollerud}, {Morris},
  {Ginsburg}, {Vaher}, {Weaver}, {Tocknell}, {Jamieson}, {van Kerkwijk},
  {Robitaille}, {Merry}, {Bachetti}, {G{"u}nther}, {Aldcroft},
  {Alvarado-Montes}, {Archibald}, {B{'o}di}, {Bapat}, {Barentsen}, {Baz{'a}n},
  {Biswas}, {Boquien}, {Burke}, {Cara}, {Cara}, {Conroy}, {Conseil}, {Craig},
  {Cross}, {Cruz}, {D'Eugenio}, {Dencheva}, {Devillepoix}, {Dietrich},
  {Eigenbrot}, {Erben}, {Ferreira}, {Foreman-Mackey}, {Fox}, {Freij}, {Garg},
  {Geda}, {Glattly}, {Gondhalekar}, {Gordon}, {Grant}, {Greenfield}, {Groener},
  {Guest}, {Gurovich}, {Handberg}, {Hart}, {Hatfield-Dodds}, {Homeier},
  {Hosseinzadeh}, {Jenness}, {Jones}, {Joseph}, {Kalmbach}, {Karamehmetoglu},
  {Ka{l}uszy{'n}ski}, {Kelley}, {Kern}, {Kerzendorf}, {Koch}, {Kulumani},
  {Lee}, {Ly}, {Ma}, {MacBride}, {Maljaars}, {Muna}, {Murphy}, {Norman},
  {O'Steen}, {Oman}, {Pacifici}, {Pascual}, {Pascual-Granado}, {Patil},
  {Perren}, {Pickering}, {Rastogi}, {Roulston}, {Ryan}, {Rykoff}, {Sabater},
  {Sakurikar}, {Salgado}, {Sanghi}, {Saunders}, {Savchenko}, {Schwardt},
  {Seifert-Eckert}, {Shih}, {Jain}, {Shukla}, {Sick}, {Simpson},
  {Singanamalla}, {Singer}, {Singhal}, {Sinha}, {Sip{H{o}}cz}, {Spitler},
  {Stansby}, {Streicher}, {{{S}}umak}, {Swinbank}, {Taranu}, {Tewary},
  {Tremblay}, {Val-Borro}, {Van Kooten}, {Vasovi{'c}}, {Verma}, {de Miranda
  Cardoso}, {Williams}, {Wilson}, {Winkel}, {Wood-Vasey}, {Xue}, {Yoachim},
  {Zhang}, {Zonca}, \& {Astropy Project Contributors}}]{astropy:2022}
{Astropy Collaboration}, {Price-Whelan}, A.~M., {Lim}, P.~L., {et~al.} 2022,
  \apj, 935, 167, \dodoi{10.3847/1538-4357/ac7c74}

\bibitem[{{Banzatti} {et~al.}(2020){Banzatti}, {Pascucci}, {Bosman}, {Pinilla},
  {Salyk}, {Herczeg}, {Pontoppidan}, {Vazquez}, {Watkins}, {Krijt}, {Hendler},
  \& {Long}}]{Banzatti20}
{Banzatti}, A., {Pascucci}, I., {Bosman}, A.~D., {et~al.} 2020, \apj, 903, 124,
  \dodoi{10.3847/1538-4357/abbc1a}

\bibitem[{{Banzatti} {et~al.}(2023){Banzatti}, {Pontoppidan}, {Carr},
  {Jellison}, {Pascucci}, {Najita}, {Mu{\~n}oz-Romero}, {{\"O}berg}, {Kalyaan},
  {Pinilla}, {Krijt}, {Long}, {Lambrechts}, {Rosotti}, {Herczeg}, {Salyk},
  {Zhang}, {Bergin}, {Ballering}, {Meyer}, {Bruderer}, \& {Jdiscs
  Collaboration}}]{banzatti2023}
{Banzatti}, A., {Pontoppidan}, K.~M., {Carr}, J.~S., {et~al.} 2023, \apjl, 957,
  L22, \dodoi{10.3847/2041-8213/acf5ec}

\bibitem[{{Banzatti} {et~al.}(2024){Banzatti}, {Salyk}, {Pontoppidan}, {Carr},
  {Zhang}, {Arulanantham}, {Cleeves}, {Najita}, {Oberg}, {Pascucci}, {Blake},
  {Krijt}, {Munoz-Romero}, {Bergin}, {Cieza}, {Pinilla}, {Long}, {Mallaney},
  {Xie}, \& {the JDISCS Collaboration}}]{banzatti2024}
{Banzatti}, A., {Salyk}, C., {Pontoppidan}, K.~M., {et~al.} 2024, arXiv
  e-prints, arXiv:2409.16255.
\newblock \doarXiv{2409.16255}

\bibitem[{{Bergin} {et~al.}(2015){Bergin}, {Blake}, {Ciesla}, {Hirschmann}, \&
  {Li}}]{Bergin15}
{Bergin}, E.~A., {Blake}, G.~A., {Ciesla}, F., {Hirschmann}, M.~M., \& {Li}, J.
  2015, Proceedings of the National Academy of Science, 112, 8965,
  \dodoi{10.1073/pnas.1500954112}

\bibitem[{{Bergin} {et~al.}(2023){Bergin}, {Kempton}, {Hirschmann},
  {Bastelberger}, {Teal}, {Blake}, {Ciesla}, \& {Li}}]{Bergin23}
{Bergin}, E.~A., {Kempton}, E. M.~R., {Hirschmann}, M., {et~al.} 2023, \apjl,
  949, L17, \dodoi{10.3847/2041-8213/acd377}

\bibitem[{{Bergin} {et~al.}(2024){Bergin}, {Bosman}, {Teague}, {Calahan},
  {Willacy}, {Cleeves}, {Schwarz}, {Zhang}, \& {Bruderer}}]{Bergin2024}
{Bergin}, E.~A., {Bosman}, A., {Teague}, R., {et~al.} 2024, \apj, 965, 147,
  \dodoi{10.3847/1538-4357/ad3443}

\bibitem[{{Birnstiel} {et~al.}(2018){Birnstiel}, {Dullemond}, {Zhu}, {Andrews},
  {Bai}, {Wilner}, {Carpenter}, {Huang}, {Isella}, {Benisty}, {P{\'e}rez}, \&
  {Zhang}}]{Birnstiel_dsharp}
{Birnstiel}, T., {Dullemond}, C.~P., {Zhu}, Z., {et~al.} 2018, \apjl, 869, L45,
  \dodoi{10.3847/2041-8213/aaf743}

\bibitem[{{Booth} \& {Ilee}(2019)}]{Booth19}
{Booth}, R.~A., \& {Ilee}, J.~D. 2019, \mnras, 487, 3998,
  \dodoi{10.1093/mnras/stz1488}

\bibitem[{{Bosman} {et~al.}(2021{\natexlab{a}}){Bosman}, {Alarc{\'o}n},
  {Zhang}, \& {Bergin}}]{Bosman21}
{Bosman}, A.~D., {Alarc{\'o}n}, F., {Zhang}, K., \& {Bergin}, E.~A.
  2021{\natexlab{a}}, \apj, 910, 3, \dodoi{10.3847/1538-4357/abe127}

\bibitem[{{Bosman} {et~al.}(2022{\natexlab{a}}){Bosman}, {Bergin}, {Calahan},
  \& {Duval}}]{bosman2022_water}
{Bosman}, A.~D., {Bergin}, E.~A., {Calahan}, J., \& {Duval}, S.~E.
  2022{\natexlab{a}}, \apjl, 930, L26, \dodoi{10.3847/2041-8213/ac66ce}

\bibitem[{{Bosman} {et~al.}(2022{\natexlab{b}}){Bosman}, {Bergin}, {Calahan},
  \& {Duval}}]{bosman2022}
{Bosman}, A.~D., {Bergin}, E.~A., {Calahan}, J.~K., \& {Duval}, S.~E.
  2022{\natexlab{b}}, \apjl, 933, L40, \dodoi{10.3847/2041-8213/ac7d9f}

\bibitem[{{Bosman} {et~al.}(2022{\natexlab{c}}){Bosman}, {Bergin}, {Calahan},
  \& {Duval}}]{Bosman22_co2}
---. 2022{\natexlab{c}}, \apjl, 933, L40, \dodoi{10.3847/2041-8213/ac7d9f}

\bibitem[{{Bosman} {et~al.}(2021{\natexlab{b}}){Bosman}, {Bergin}, {Loomis},
  {Andrews}, {van't Hoff}, {Teague}, {{\"O}berg}, {Guzm{\'a}n}, {Walsh},
  {Aikawa}, {Alarc{\'o}n}, {Bae}, {Bergner}, {Booth}, {Cataldi}, {Cleeves},
  {Czekala}, {Huang}, {Ilee}, {Law}, {Le Gal}, {Liu}, {Long}, {M{\'e}nard},
  {Nomura}, {P{\'e}rez}, {Qi}, {Schwarz}, {Sierra}, {Tsukagoshi}, {Yamato},
  {Wilner}, \& {Zhang}}]{Bosman21_mapsco}
{Bosman}, A.~D., {Bergin}, E.~A., {Loomis}, R.~A., {et~al.} 2021{\natexlab{b}},
  \apjs, 257, 15, \dodoi{10.3847/1538-4365/ac1433}

\bibitem[{{Bouvier} \& {Appenzeller}(1992)}]{bouvier1992}
{Bouvier}, J., \& {Appenzeller}, I. 1992, \aaps, 92, 481

\bibitem[{{Bruderer}(2013)}]{bruderer2013}
{Bruderer}, S. 2013, \aap, 559, A46, \dodoi{10.1051/0004-6361/201321171}

\bibitem[{{Bruderer} {et~al.}(2012){Bruderer}, {van Dishoeck}, {Doty}, \&
  {Herczeg}}]{Bruderer12}
{Bruderer}, S., {van Dishoeck}, E.~F., {Doty}, S.~D., \& {Herczeg}, G.~J. 2012,
  \aap, 541, A91, \dodoi{10.1051/0004-6361/201118218}

\bibitem[{Bushouse {et~al.}(2024)Bushouse, Eisenhamer, Dencheva, Davies,
  Greenfield, Morrison, Hodge, Simon, Grumm, Droettboom, Slavich, Sosey, Pauly,
  Miller, Jedrzejewski, Hack, Davis, Crawford, Law, Gordon, Regan, Cara,
  MacDonald, Bradley, Shanahan, Jamieson, Teodoro, Williams, \&
  Pena-Guerrero}]{bushouse_2024_1_15_0}
Bushouse, H., Eisenhamer, J., Dencheva, N., {et~al.} 2024, JWST Calibration
  Pipeline, 1.15.0,  Zenodo, \dodoi{10.5281/zenodo.12556702}

\bibitem[{{Carr} \& {Najita}(2011)}]{Carr11}
{Carr}, J.~S., \& {Najita}, J.~R. 2011, \apj, 733, 102,
  \dodoi{10.1088/0004-637X/733/2/102}

\bibitem[{{Carr} \& {Najita}(2014)}]{carr2014}
---. 2014, \apj, 788, 66, \dodoi{10.1088/0004-637X/788/1/66}

\bibitem[{{Chubb} {et~al.}(2020){Chubb}, {Tennyson}, \&
  {Yurchenko}}]{chubb2020}
{Chubb}, K.~L., {Tennyson}, J., \& {Yurchenko}, S.~N. 2020, \mnras, 493, 1531,
  \dodoi{10.1093/mnras/staa229}

\bibitem[{{Cieza} {et~al.}(2010){Cieza}, {Schreiber}, {Romero}, {Mora},
  {Merin}, {Swift}, {Orellana}, {Williams}, {Harvey}, \& {Evans}}]{cieza2010}
{Cieza}, L.~A., {Schreiber}, M.~R., {Romero}, G.~A., {et~al.} 2010, \apj, 712,
  925, \dodoi{10.1088/0004-637X/712/2/925}

\bibitem[{{Dionatos} {et~al.}(2019){Dionatos}, {Woitke}, {G{\"u}del},
  {Degroote}, {Liebhart}, {Anthonioz}, {Antonellini}, {Baldovin-Saavedra},
  {Carmona}, {Dominik}, {Greaves}, {Ilee}, {Kamp}, {M{\'e}nard}, {Min},
  {Pinte}, {Rab}, {Rigon}, {Thi}, \& {Waters}}]{Dionatos19}
{Dionatos}, O., {Woitke}, P., {G{\"u}del}, M., {et~al.} 2019, \aap, 625, A66,
  \dodoi{10.1051/0004-6361/201832860}

\bibitem[{{Dominik} {et~al.}(2021){Dominik}, {Min}, \& {Tazaki}}]{dominik2021}
{Dominik}, C., {Min}, M., \& {Tazaki}, R. 2021, {OpTool: Command-line driven
  tool for creating complex dust opacities}, Astrophysics Source Code Library,
  record ascl:2104.010

\bibitem[{{Draine}(2011)}]{Draine_ism}
{Draine}, B.~T. 2011, {Physics of the Interstellar and Intergalactic Medium}

\bibitem[{{Dullemond} {et~al.}(2012){Dullemond}, {Juhasz}, {Pohl}, {Sereshti},
  {Shetty}, {Peters}, {Commercon}, \& {Flock}}]{radmc3d}
{Dullemond}, C.~P., {Juhasz}, A., {Pohl}, A., {et~al.} 2012, {RADMC-3D: A
  multi-purpose radiative transfer tool}, Astrophysics Source Code Library,
  record ascl:1202.015.
\newblock \doeprint{1202.015}

\bibitem[{{Duval} {et~al.}(2022){Duval}, {Bosman}, \& {Bergin}}]{duval2022}
{Duval}, S.~E., {Bosman}, A.~D., \& {Bergin}, E.~A. 2022, \apjl, 934, L25,
  \dodoi{10.3847/2041-8213/ac822b}

\bibitem[{{Furlan} {et~al.}(2011){Furlan}, {Luhman}, {Espaillat}, {D'Alessio},
  {Adame}, {Manoj}, {Kim}, {Watson}, {Forrest}, {McClure}, {Calvet}, {Sargent},
  {Green}, \& {Fischer}}]{furlan2011}
{Furlan}, E., {Luhman}, K.~L., {Espaillat}, C., {et~al.} 2011, \apjs, 195, 3,
  \dodoi{10.1088/0067-0049/195/1/3}

\bibitem[{{Gaia Collaboration} {et~al.}(2018){Gaia Collaboration}, {Brown},
  {Vallenari}, {Prusti}, {de Bruijne}, {Babusiaux}, {Bailer-Jones}, {Biermann},
  {Evans}, {Eyer}, {Jansen}, {Jordi}, {Klioner}, {Lammers}, {Lindegren},
  {Luri}, {Mignard}, {Panem}, {Pourbaix}, {Randich}, {Sartoretti}, {Siddiqui},
  {Soubiran}, {van Leeuwen}, {Walton}, {Arenou}, {Bastian}, {Cropper},
  {Drimmel}, {Katz}, {Lattanzi}, {Bakker}, {Cacciari}, {Casta{\~n}eda},
  {Chaoul}, {Cheek}, {De Angeli}, {Fabricius}, {Guerra}, {Holl}, {Masana},
  {Messineo}, {Mowlavi}, {Nienartowicz}, {Panuzzo}, {Portell}, {Riello},
  {Seabroke}, {Tanga}, {Th{\'e}venin}, {Gracia-Abril}, {Comoretto},
  {Garcia-Reinaldos}, {Teyssier}, {Altmann}, {Andrae}, {Audard},
  {Bellas-Velidis}, {Benson}, {Berthier}, {Blomme}, {Burgess}, {Busso},
  {Carry}, {Cellino}, {Clementini}, {Clotet}, {Creevey}, {Davidson}, {De
  Ridder}, {Delchambre}, {Dell'Oro}, {Ducourant},
  {Fern{\'a}ndez-Hern{\'a}ndez}, {Fouesneau}, {Fr{\'e}mat}, {Galluccio},
  {Garc{\'\i}a-Torres}, {Gonz{\'a}lez-N{\'u}{\~n}ez}, {Gonz{\'a}lez-Vidal},
  {Gosset}, {Guy}, {Halbwachs}, {Hambly}, {Harrison}, {Hern{\'a}ndez},
  {Hestroffer}, {Hodgkin}, {Hutton}, {Jasniewicz}, {Jean-Antoine-Piccolo},
  {Jordan}, {Korn}, {Krone-Martins}, {Lanzafame}, {Lebzelter}, {L{\"o}ffler},
  {Manteiga}, {Marrese}, {Mart{\'\i}n-Fleitas}, {Moitinho}, {Mora}, {Muinonen},
  {Osinde}, {Pancino}, {Pauwels}, {Petit}, {Recio-Blanco}, {Richards},
  {Rimoldini}, {Robin}, {Sarro}, {Siopis}, {Smith}, {Sozzetti}, {S{\"u}veges},
  {Torra}, {van Reeven}, {Abbas}, {Abreu Aramburu}, {Accart}, {Aerts},
  {Altavilla}, {{\'A}lvarez}, {Alvarez}, {Alves}, {Anderson}, {Andrei},
  {Anglada Varela}, {Antiche}, {Antoja}, {Arcay}, {Astraatmadja}, {Bach},
  {Baker}, {Balaguer-N{\'u}{\~n}ez}, {Balm}, {Barache}, {Barata}, {Barbato},
  {Barblan}, {Barklem}, {Barrado}, {Barros}, {Barstow}, {Bartholom{\'e}
  Mu{\~n}oz}, {Bassilana}, {Becciani}, {Bellazzini}, {Berihuete}, {Bertone},
  {Bianchi}, {Bienaym{\'e}}, {Blanco-Cuaresma}, {Boch}, {Boeche}, {Bombrun},
  {Borrachero}, {Bossini}, {Bouquillon}, {Bourda}, {Bragaglia}, {Bramante},
  {Breddels}, {Bressan}, {Brouillet}, {Br{\"u}semeister}, {Brugaletta},
  {Bucciarelli}, {Burlacu}, {Busonero}, {Butkevich}, {Buzzi}, {Caffau},
  {Cancelliere}, {Cannizzaro}, {Cantat-Gaudin}, {Carballo}, {Carlucci},
  {Carrasco}, {Casamiquela}, {Castellani}, {Castro-Ginard}, {Charlot},
  {Chemin}, {Chiavassa}, {Cocozza}, {Costigan}, {Cowell}, {Crifo}, {Crosta},
  {Crowley}, {Cuypers}, {Dafonte}, {Damerdji}, {Dapergolas}, {David}, {David},
  {de Laverny}, {De Luise}, {De March}, {de Martino}, {de Souza}, {de Torres},
  {Debosscher}, {del Pozo}, {Delbo}, {Delgado}, {Delgado}, {Di Matteo},
  {Diakite}, {Diener}, {Distefano}, {Dolding}, {Drazinos}, {Dur{\'a}n},
  {Edvardsson}, {Enke}, {Eriksson}, {Esquej}, {Eynard Bontemps}, {Fabre},
  {Fabrizio}, {Faigler}, {Falc{\~a}o}, {Farr{\`a}s Casas}, {Federici},
  {Fedorets}, {Fernique}, {Figueras}, {Filippi}, {Findeisen}, {Fonti},
  {Fraile}, {Fraser}, {Fr{\'e}zouls}, {Gai}, {Galleti}, {Garabato},
  {Garc{\'\i}a-Sedano}, {Garofalo}, {Garralda}, {Gavel}, {Gavras}, {Gerssen},
  {Geyer}, {Giacobbe}, {Gilmore}, {Girona}, {Giuffrida}, {Glass}, {Gomes},
  {Granvik}, {Gueguen}, {Guerrier}, {Guiraud}, {Guti{\'e}rrez-S{\'a}nchez},
  {Haigron}, {Hatzidimitriou}, {Hauser}, {Haywood}, {Heiter}, {Helmi}, {Heu},
  {Hilger}, {Hobbs}, {Hofmann}, {Holland}, {Huckle}, {Hypki}, {Icardi},
  {Jan{\ss}en}, {Jevardat de Fombelle}, {Jonker}, {Juh{\'a}sz}, {Julbe},
  {Karampelas}, {Kewley}, {Klar}, {Kochoska}, {Kohley}, {Kolenberg},
  {Kontizas}, {Kontizas}, {Koposov}, {Kordopatis}, {Kostrzewa-Rutkowska},
  {Koubsky}, {Lambert}, {Lanza}, {Lasne}, {Lavigne}, {Le Fustec}, {Le
  Poncin-Lafitte}, {Lebreton}, {Leccia}, {Leclerc}, {Lecoeur-Taibi},
  {Lenhardt}, {Leroux}, {Liao}, {Licata}, {Lindstr{\o}m}, {Lister}, {Livanou},
  {Lobel}, {L{\'o}pez}, {Managau}, {Mann}, {Mantelet}, {Marchal}, {Marchant},
  {Marconi}, {Marinoni}, {Marschalk{\'o}}, {Marshall}, {Martino}, {Marton},
  {Mary}, {Massari}, {Matijevi{\v{c}}}, {Mazeh}, {McMillan}, {Messina},
  {Michalik}, {Millar}, {Molina}, {Molinaro}, {Moln{\'a}r}, {Montegriffo},
  {Mor}, {Morbidelli}, {Morel}, {Morris}, {Mulone}, {Muraveva}, {Musella},
  {Nelemans}, {Nicastro}, {Noval}, {O'Mullane}, {Ord{\'e}novic},
  {Ord{\'o}{\~n}ez-Blanco}, {Osborne}, {Pagani}, {Pagano}, {Pailler},
  {Palacin}, {Palaversa}, {Panahi}, {Pawlak}, {Piersimoni}, {Pineau}, {Plachy},
  {Plum}, {Poggio}, {Poujoulet}, {Pr{\v{s}}a}, {Pulone}, {Racero}, {Ragaini},
  {Rambaux}, {Ramos-Lerate}, {Regibo}, {Reyl{\'e}}, {Riclet}, {Ripepi}, {Riva},
  {Rivard}, {Rixon}, {Roegiers}, {Roelens}, {Romero-G{\'o}mez}, {Rowell},
  {Royer}, {Ruiz-Dern}, {Sadowski}, {Sagrist{\`a} Sell{\'e}s}, {Sahlmann},
  {Salgado}, {Salguero}, {Sanna}, {Santana-Ros}, {Sarasso}, {Savietto},
  {Schultheis}, {Sciacca}, {Segol}, {Segovia}, {S{\'e}gransan}, {Shih},
  {Siltala}, {Silva}, {Smart}, {Smith}, {Solano}, {Solitro}, {Sordo}, {Soria
  Nieto}, {Souchay}, {Spagna}, {Spoto}, {Stampa}, {Steele},
  {Steidelm{\"u}ller}, {Stephenson}, {Stoev}, {Suess}, {Surdej}, {Szabados},
  {Szegedi-Elek}, {Tapiador}, {Taris}, {Tauran}, {Taylor}, {Teixeira},
  {Terrett}, {Teyssandier}, {Thuillot}, {Titarenko}, {Torra Clotet}, {Turon},
  {Ulla}, {Utrilla}, {Uzzi}, {Vaillant}, {Valentini}, {Valette}, {van Elteren},
  {Van Hemelryck}, {van Leeuwen}, {Vaschetto}, {Vecchiato}, {Veljanoski},
  {Viala}, {Vicente}, {Vogt}, {von Essen}, {Voss}, {Votruba}, {Voutsinas},
  {Walmsley}, {Weiler}, {Wertz}, {Wevers}, {Wyrzykowski}, {Yoldas},
  {{\v{Z}}erjal}, {Ziaeepour}, {Zorec}, {Zschocke}, {Zucker}, {Zurbach}, \&
  {Zwitter}}]{gaia2018}
{Gaia Collaboration}, {Brown}, A.~G.~A., {Vallenari}, A., {et~al.} 2018, \aap,
  616, A1, \dodoi{10.1051/0004-6361/201833051}

\bibitem[{{Gasman} {et~al.}(2023){Gasman}, {van Dishoeck}, {Grant}, {Temmink},
  {Tabone}, {Henning}, {Kamp}, {G{\"u}del}, {Lagage}, {Perotti}, {Christiaens},
  {Samland}, {Arabhavi}, {Argyriou}, {Abergel}, {Absil}, {Barrado},
  {Boccaletti}, {Bouwman}, {Caratti o Garatti}, {Geers}, {Glauser},
  {Guadarrama}, {Jang}, {Kanwar}, {Lahuis}, {Morales-Calder{\'o}n}, {Mueller},
  {Nehm{\'e}}, {Olofsson}, {Pantin}, {Pawellek}, {Ray}, {Rodgers-Lee},
  {Scheithauer}, {Schreiber}, {Schwarz}, {Vandenbussche}, {Vlasblom}, {Waters},
  {Wright}, {Colina}, {Greve}, \& {{\"O}stlin}}]{Gasman2023}
{Gasman}, D., {van Dishoeck}, E.~F., {Grant}, S.~L., {et~al.} 2023, \aap, 679,
  A117, \dodoi{10.1051/0004-6361/202347005}

\bibitem[{{Glassgold} \& {Najita}(2015)}]{Glassgold15}
{Glassgold}, A.~E., \& {Najita}, J.~R. 2015, \apj, 810, 125,
  \dodoi{10.1088/0004-637X/810/2/125}

\bibitem[{{Gordon} {et~al.}(2022){Gordon}, {Rothman}, {Hargreaves}, {Hashemi},
  {Karlovets}, {Skinner}, {Conway}, {Hill}, {Kochanov}, {Tan}, {Wcis{\l}o},
  {Finenko}, {Nelson}, {Bernath}, {Birk}, {Boudon}, {Campargue}, {Chance},
  {Coustenis}, {Drouin}, {Flaud}, {Gamache}, {Hodges}, {Jacquemart}, {Mlawer},
  {Nikitin}, {Perevalov}, {Rotger}, {Tennyson}, {Toon}, {Tran}, {Tyuterev},
  {Adkins}, {Baker}, {Barbe}, {Can{\`e}}, {Cs{\'a}sz{\'a}r}, {Dudaryonok},
  {Egorov}, {Fleisher}, {Fleurbaey}, {Foltynowicz}, {Furtenbacher}, {Harrison},
  {Hartmann}, {Horneman}, {Huang}, {Karman}, {Karns}, {Kassi}, {Kleiner},
  {Kofman}, {Kwabia-Tchana}, {Lavrentieva}, {Lee}, {Long}, {Lukashevskaya},
  {Lyulin}, {Makhnev}, {Matt}, {Massie}, {Melosso}, {Mikhailenko}, {Mondelain},
  {M{\"u}ller}, {Naumenko}, {Perrin}, {Polyansky}, {Raddaoui}, {Raston},
  {Reed}, {Rey}, {Richard}, {T{\'o}bi{\'a}s}, {Sadiek}, {Schwenke},
  {Starikova}, {Sung}, {Tamassia}, {Tashkun}, {Vander Auwera}, {Vasilenko},
  {Vigasin}, {Villanueva}, {Vispoel}, {Wagner}, {Yachmenev}, \&
  {Yurchenko}}]{gordon2022}
{Gordon}, I.~E., {Rothman}, L.~S., {Hargreaves}, R.~J., {et~al.} 2022, \jqsrt,
  277, 107949, \dodoi{10.1016/j.jqsrt.2021.107949}

\bibitem[{{Grant} {et~al.}(2023){Grant}, {van Dishoeck}, {Tabone}, {Gasman},
  {Henning}, {Kamp}, {G{\"u}del}, {Lagage}, {Bettoni}, {Perotti},
  {Christiaens}, {Samland}, {Arabhavi}, {Argyriou}, {Abergel}, {Absil},
  {Barrado}, {Boccaletti}, {Bouwman}, {Caratti o Garatti}, {Geers}, {Glauser},
  {Guadarrama}, {Jang}, {Kanwar}, {Lahuis}, {Morales-Calder{\'o}n}, {Mueller},
  {Nehm{\'e}}, {Olofsson}, {Pantin}, {Pawellek}, {Ray}, {Rodgers-Lee},
  {Scheithauer}, {Schreiber}, {Schwarz}, {Temmink}, {Vandenbussche},
  {Vlasblom}, {Waters}, {Wright}, {Colina}, {Greve}, {Justannont}, \&
  {{\"O}stlin}}]{grant2023}
{Grant}, S.~L., {van Dishoeck}, E.~F., {Tabone}, B., {et~al.} 2023, \apjl, 947,
  L6, \dodoi{10.3847/2041-8213/acc44b}

\bibitem[{{Grasser} {et~al.}(2021){Grasser}, {Ratzenb{\"o}ck}, {Alves},
  {Gro{\ss}schedl}, {Meingast}, {Zucker}, {Hacar}, {Lada}, {Goodman},
  {Lombardi}, {Forbes}, {Bomze}, \& {M{\"o}ller}}]{grasser2021}
{Grasser}, N., {Ratzenb{\"o}ck}, S., {Alves}, J., {et~al.} 2021, \aap, 652, A2,
  \dodoi{10.1051/0004-6361/202140438}

\bibitem[{{Gullbring} {et~al.}(2000){Gullbring}, {Calvet}, {Muzerolle}, \&
  {Hartmann}}]{Gullbring2000}
{Gullbring}, E., {Calvet}, N., {Muzerolle}, J., \& {Hartmann}, L. 2000, \apj,
  544, 927, \dodoi{10.1086/317253}

\bibitem[{{Hartmann} {et~al.}(2016){Hartmann}, {Herczeg}, \&
  {Calvet}}]{Hartmann16}
{Hartmann}, L., {Herczeg}, G., \& {Calvet}, N. 2016, \araa, 54, 135,
  \dodoi{10.1146/annurev-astro-081915-023347}

\bibitem[{{Heinzeller} {et~al.}(2011){Heinzeller}, {Nomura}, {Walsh}, \&
  {Millar}}]{Heinzeller11}
{Heinzeller}, D., {Nomura}, H., {Walsh}, C., \& {Millar}, T.~J. 2011, \apj,
  731, 115, \dodoi{10.1088/0004-637X/731/2/115}

\bibitem[{{Hily-Blant} {et~al.}(2019){Hily-Blant}, {Magalhaes de Souza},
  {Kastner}, \& {Forveille}}]{Hily-Blant2019}
{Hily-Blant}, P., {Magalhaes de Souza}, V., {Kastner}, J., \& {Forveille}, T.
  2019, \aap, 632, L12, \dodoi{10.1051/0004-6361/201936750}

\bibitem[{{Huang} {et~al.}(2018){Huang}, {Andrews}, {Dullemond}, {Isella},
  {P{\'e}rez}, {Guzm{\'a}n}, {{\"O}berg}, {Zhu}, {Zhang}, {Bai}, {Benisty},
  {Birnstiel}, {Carpenter}, {Hughes}, {Ricci}, {Weaver}, \& {Wilner}}]{Huang18}
{Huang}, J., {Andrews}, S.~M., {Dullemond}, C.~P., {et~al.} 2018, \apjl, 869,
  L42, \dodoi{10.3847/2041-8213/aaf740}

\bibitem[{{Jang} {et~al.}(2024){Jang}, {Waters}, {Kamp}, \&
  {Dullemond}}]{jang2024}
{Jang}, H., {Waters}, R., {Kamp}, I., \& {Dullemond}, C.~P. 2024, \aap, 687,
  A275, \dodoi{10.1051/0004-6361/202348630}

\bibitem[{{Jellison} {et~al.}(2024){Jellison}, {Johnson}, {Banzatti}, \&
  {Bruderer}}]{jellison2024}
{Jellison}, E., {Johnson}, M., {Banzatti}, A., \& {Bruderer}, S. 2024, arXiv
  e-prints, arXiv:2402.04060, \dodoi{10.48550/arXiv.2402.04060}

\bibitem[{Johnson {et~al.}(2024)Johnson, Banzatti, Fuller, \&
  Jellison}]{iSLAT_code}
Johnson, M., Banzatti, A., Fuller, J., \& Jellison, E. 2024, spexod/iSLAT:
  Second release, v4.03,  Zenodo, \dodoi{10.5281/zenodo.12167853}

\bibitem[{{Jones} {et~al.}(2023){Jones}, {{\'A}lvarez-M{\'a}rquez}, {Sloan},
  {Kavanagh}, {Argyriou}, {Law}, {Labiano}, {Patapis}, {Mueller}, {Larson},
  {Bright}, {Klaassen}, {Fox}, {Gasman}, {Geers}, {Glauser}, {Guillard},
  {Nayak}, {Noriega-Crespo}, {Ressler}, {Sargent}, {Temim}, {Vandenbussche}, \&
  {Garc{\'\i}a Mar{\'\i}n}}]{jones2023}
{Jones}, O.~C., {{\'A}lvarez-M{\'a}rquez}, J., {Sloan}, G.~C., {et~al.} 2023,
  \mnras, 523, 2519, \dodoi{10.1093/mnras/stad1609}

\bibitem[{{Kalyaan} {et~al.}(2021){Kalyaan}, {Pinilla}, {Krijt}, {Mulders}, \&
  {Banzatti}}]{kalyaan2021}
{Kalyaan}, A., {Pinilla}, P., {Krijt}, S., {Mulders}, G.~D., \& {Banzatti}, A.
  2021, \apj, 921, 84, \dodoi{10.3847/1538-4357/ac1e96}

\bibitem[{{Kalyaan} {et~al.}(2023){Kalyaan}, {Pinilla}, {Krijt}, {Banzatti},
  {Rosotti}, {Mulders}, {Lambrechts}, {Long}, \& {Herczeg}}]{Kalyaan23}
{Kalyaan}, A., {Pinilla}, P., {Krijt}, S., {et~al.} 2023, \apj, 954, 66,
  \dodoi{10.3847/1538-4357/ace535}

\bibitem[{{Kamp} {et~al.}(2024){Kamp}, {Galli}, \& {Rab}}]{Kamp2024}
{Kamp}, I., {Galli}, D., \& {Rab}, C. 2024, in Astrochemical Modeling:
  Practical Aspects of Microphysics in Numerical Simulations. Edited by Stefano
  Bovino and Tommaso Grassi, 283--306,
  \dodoi{10.1016/B978-0-32-391746-9.00019-5}

\bibitem[{{Kanwar} {et~al.}(2024{\natexlab{a}}){Kanwar}, {Kamp}, {Woitke},
  {Rab}, {Thi}, \& {Min}}]{kanwar2024a}
{Kanwar}, J., {Kamp}, I., {Woitke}, P., {et~al.} 2024{\natexlab{a}}, \aap, 681,
  A22, \dodoi{10.1051/0004-6361/202346262}

\bibitem[{{Kanwar} {et~al.}(2024{\natexlab{b}}){Kanwar}, {Kamp}, {Jang},
  {Waters}, {van Dishoeck}, {Christiaens}, {Arabhavi}, {Henning}, {G{\"u}del},
  {Woitke}, {Absil}, {Barrado}, {Caratti o Garatti}, {Glauser}, {Lahuis},
  {Scheithauer}, {Vandenbussche}, {Gasman}, {Grant}, {Kurtovic}, {Perotti},
  {Tabone}, \& {Temmink}}]{kanwar2024}
{Kanwar}, J., {Kamp}, I., {Jang}, H., {et~al.} 2024{\natexlab{b}}, \aap, 689,
  A231, \dodoi{10.1051/0004-6361/202450078}

\bibitem[{{Kemper} {et~al.}(2004){Kemper}, {Vriend}, \& {Tielens}}]{kemper2004}
{Kemper}, F., {Vriend}, W.~J., \& {Tielens}, A.~G.~G.~M. 2004, \apj, 609, 826,
  \dodoi{10.1086/421339}

\bibitem[{{Klarmann} {et~al.}(2018){Klarmann}, {Ormel}, \&
  {Dominik}}]{Klarmann18}
{Klarmann}, L., {Ormel}, C.~W., \& {Dominik}, C. 2018, \aap, 618, L1,
  \dodoi{10.1051/0004-6361/201833719}

\bibitem[{{Kress} {et~al.}(2010){Kress}, {Tielens}, \& {Frenklach}}]{Kress10}
{Kress}, M.~E., {Tielens}, A.~G.~G.~M., \& {Frenklach}, M. 2010, Advances in
  Space Research, 46, 44, \dodoi{10.1016/j.asr.2010.02.004}

\bibitem[{{Labiano} {et~al.}(2021){Labiano}, {Argyriou},
  {{\'A}lvarez-M{\'a}rquez}, {Glasse}, {Glauser}, {Patapis}, {Law}, {Brandl},
  {Justtanont}, {Lahuis}, {Mart{\'\i}nez-Galarza}, {Mueller}, {Noriega-Crespo},
  {Royer}, {Shaughnessy}, \& {Vandenbussche}}]{labiano2021}
{Labiano}, A., {Argyriou}, I., {{\'A}lvarez-M{\'a}rquez}, J., {et~al.} 2021,
  \aap, 656, A57, \dodoi{10.1051/0004-6361/202140614}

\bibitem[{{Li} {et~al.}(2021){Li}, {Bergin}, {Blake}, {Ciesla}, \&
  {Hirschmann}}]{Li21}
{Li}, J., {Bergin}, E.~A., {Blake}, G.~A., {Ciesla}, F.~J., \& {Hirschmann},
  M.~M. 2021, Science Advances, 7, 3632, \dodoi{10.1126/sciadv.abd3632}

\bibitem[{{Ligterink} {et~al.}(2024){Ligterink}, {Kipfer}, \&
  {Gavino}}]{Ligterink2024}
{Ligterink}, N.~F.~W., {Kipfer}, K.~A., \& {Gavino}, S. 2024, \aap, 687, A224,
  \dodoi{10.1051/0004-6361/202450405}

\bibitem[{{Liu} {et~al.}(2012){Liu}, {Madlener}, {Wolf}, {Wang}, \&
  {Ruge}}]{Liu12}
{Liu}, Y., {Madlener}, D., {Wolf}, S., {Wang}, H., \& {Ruge}, J.~P. 2012, \aap,
  546, A7, \dodoi{10.1051/0004-6361/201219336}

\bibitem[{{Liu} {et~al.}(2022){Liu}, {Linz}, {Fang}, {Henning}, {Wolf},
  {Flock}, {Rosotti}, {Wang}, \& {Li}}]{liu2022}
{Liu}, Y., {Linz}, H., {Fang}, M., {et~al.} 2022, \aap, 668, A175,
  \dodoi{10.1051/0004-6361/202244505}

\bibitem[{{Lynden-Bell} \& {Pringle}(1974)}]{lynden_bell1974}
{Lynden-Bell}, D., \& {Pringle}, J.~E. 1974, \mnras, 168, 603,
  \dodoi{10.1093/mnras/168.3.603}

\bibitem[{{Madhusudhan}(2012)}]{madhusudhan2012}
{Madhusudhan}, N. 2012, \apj, 758, 36, \dodoi{10.1088/0004-637X/758/1/36}

\bibitem[{{Mah} {et~al.}(2023){Mah}, {Bitsch}, {Pascucci}, \&
  {Henning}}]{Mah23}
{Mah}, J., {Bitsch}, B., {Pascucci}, I., \& {Henning}, T. 2023, \aap, 677, L7,
  \dodoi{10.1051/0004-6361/202347169}

\bibitem[{{Manara} {et~al.}(2023){Manara}, {Ansdell}, {Rosotti}, {Hughes},
  {Armitage}, {Lodato}, \& {Williams}}]{manara2023}
{Manara}, C.~F., {Ansdell}, M., {Rosotti}, G.~P., {et~al.} 2023, in
  Astronomical Society of the Pacific Conference Series, Vol. 534, Protostars
  and Planets VII, ed. S.~{Inutsuka}, Y.~{Aikawa}, T.~{Muto}, K.~{Tomida}, \&
  M.~{Tamura}, 539, \dodoi{10.48550/arXiv.2203.09930}

\bibitem[{{Manara} {et~al.}(2016){Manara}, {Rosotti}, {Testi}, {Natta},
  {Alcal{\'a}}, {Williams}, {Ansdell}, {Miotello}, {van der Marel}, {Tazzari},
  {Carpenter}, {Guidi}, {Mathews}, {Oliveira}, {Prusti}, \& {van
  Dishoeck}}]{Manara16}
{Manara}, C.~F., {Rosotti}, G., {Testi}, L., {et~al.} 2016, \aap, 591, L3,
  \dodoi{10.1051/0004-6361/201628549}

\bibitem[{{Meijerink} {et~al.}(2009){Meijerink}, {Pontoppidan}, {Blake},
  {Poelman}, \& {Dullemond}}]{meijerink2009}
{Meijerink}, R., {Pontoppidan}, K.~M., {Blake}, G.~A., {Poelman}, D.~R., \&
  {Dullemond}, C.~P. 2009, \apj, 704, 1471,
  \dodoi{10.1088/0004-637X/704/2/1471}

\bibitem[{{Miotello} {et~al.}(2022){Miotello}, {Kamp}, {Birnstiel}, {Cleeves},
  \& {Kataoka}}]{Miotello_ppvii}
{Miotello}, A., {Kamp}, I., {Birnstiel}, T., {Cleeves}, L.~I., \& {Kataoka}, A.
  2022, arXiv e-prints, arXiv:2203.09818.
\newblock \doarXiv{2203.09818}

\bibitem[{{Miotello} {et~al.}(2019){Miotello}, {Facchini}, {van Dishoeck},
  {Cazzoletti}, {Testi}, {Williams}, {Ansdell}, {van Terwisga}, \& {van der
  Marel}}]{Miotello19}
{Miotello}, A., {Facchini}, S., {van Dishoeck}, E.~F., {et~al.} 2019, \aap,
  631, A69, \dodoi{10.1051/0004-6361/201935441}

\bibitem[{{Mishra} \& {Li}(2015)}]{Mishra15}
{Mishra}, A., \& {Li}, A. 2015, \apj, 809, 120,
  \dodoi{10.1088/0004-637X/809/2/120}

\bibitem[{{Mu{\~n}oz-Romero} {et~al.}(2024){Mu{\~n}oz-Romero}, {{\"O}berg},
  {Banzatti}, {Pontoppidan}, {Andrews}, {Wilner}, {Bergin}, {Czekala}, {Law},
  {Salyk}, {Teague}, {Qi}, {Bergner}, {Huang}, {Walsh}, {Guzm{\'a}n},
  {Cleeves}, {Aikawa}, {Bae}, {Booth}, {Cataldi}, {Ilee}, {Le Gal}, {Long},
  {Loomis}, {Menard}, \& {Liu}}]{munoz-romero2024}
{Mu{\~n}oz-Romero}, C.~E., {{\"O}berg}, K.~I., {Banzatti}, A., {et~al.} 2024,
  \apj, 964, 36, \dodoi{10.3847/1538-4357/ad20e9}

\bibitem[{{Najita} {et~al.}(2013){Najita}, {Carr}, {Pontoppidan}, {Salyk}, {van
  Dishoeck}, \& {Blake}}]{Najita13}
{Najita}, J.~R., {Carr}, J.~S., {Pontoppidan}, K.~M., {et~al.} 2013, \apj, 766,
  134, \dodoi{10.1088/0004-637X/766/2/134}

\bibitem[{{Najita} {et~al.}(2010){Najita}, {Carr}, {Strom}, {Watson},
  {Pascucci}, {Hollenbach}, {Gorti}, \& {Keller}}]{Najita10}
{Najita}, J.~R., {Carr}, J.~S., {Strom}, S.~E., {et~al.} 2010, \apj, 712, 274,
  \dodoi{10.1088/0004-637X/712/1/274}

\bibitem[{{Nelson} {et~al.}(2013){Nelson}, {Gressel}, \&
  {Umurhan}}]{nelson2013}
{Nelson}, R.~P., {Gressel}, O., \& {Umurhan}, O.~M. 2013, \mnras, 435, 2610,
  \dodoi{10.1093/mnras/stt1475}

\bibitem[{Nieva \& Przybilla(2012)}]{Nieva12}
Nieva, M.~F., \& Przybilla, N. 2012, Astronomy and Astrophysics, 539, A143

\bibitem[{{{\"O}berg} {et~al.}(2011{\natexlab{a}}){{\"O}berg}, {Boogert},
  {Pontoppidan}, {van den Broek}, {van Dishoeck}, {Bottinelli}, {Blake}, \&
  {Evans}}]{oberg11_c2d}
{{\"O}berg}, K.~I., {Boogert}, A.~C.~A., {Pontoppidan}, K.~M., {et~al.}
  2011{\natexlab{a}}, \apj, 740, 109, \dodoi{10.1088/0004-637X/740/2/109}

\bibitem[{{{\"O}berg} {et~al.}(2011{\natexlab{b}}){{\"O}berg}, {Murray-Clay},
  \& {Bergin}}]{Oberg11_C_O}
{{\"O}berg}, K.~I., {Murray-Clay}, R., \& {Bergin}, E.~A. 2011{\natexlab{b}},
  ApJL, 743, L16, \dodoi{10.1088/2041-8205/743/1/L16}

\bibitem[{{Pascucci} {et~al.}(2009){Pascucci}, {Apai}, {Luhman}, {Henning},
  {Bouwman}, {Meyer}, {Lahuis}, \& {Natta}}]{Pascucci2009}
{Pascucci}, I., {Apai}, D., {Luhman}, K., {et~al.} 2009, \apj, 696, 143,
  \dodoi{10.1088/0004-637X/696/1/143}

\bibitem[{{Pascucci} {et~al.}(2023){Pascucci}, {Cabrit}, {Edwards}, {Gorti},
  {Gressel}, \& {Suzuki}}]{pascucci2023}
{Pascucci}, I., {Cabrit}, S., {Edwards}, S., {et~al.} 2023, in Astronomical
  Society of the Pacific Conference Series, Vol. 534, Protostars and Planets
  VII, ed. S.~{Inutsuka}, Y.~{Aikawa}, T.~{Muto}, K.~{Tomida}, \& M.~{Tamura},
  567, \dodoi{10.48550/arXiv.2203.10068}

\bibitem[{{Pascucci} {et~al.}(2013){Pascucci}, {Herczeg}, {Carr}, \&
  {Bruderer}}]{Pascucci13}
{Pascucci}, I., {Herczeg}, G., {Carr}, J.~S., \& {Bruderer}, S. 2013, \apj,
  779, 178, \dodoi{10.1088/0004-637X/779/2/178}

\bibitem[{{Pinilla} {et~al.}(2013){Pinilla}, {Birnstiel}, {Benisty}, {Ricci},
  {Natta}, {Dullemond}, {Dominik}, \& {Testi}}]{pinilla2013}
{Pinilla}, P., {Birnstiel}, T., {Benisty}, M., {et~al.} 2013, \aap, 554, A95,
  \dodoi{10.1051/0004-6361/201220875}

\bibitem[{{Pontoppidan} {et~al.}(2004){Pontoppidan}, {van Dishoeck}, \&
  {Dartois}}]{Pontoppidan2004}
{Pontoppidan}, K.~M., {van Dishoeck}, E.~F., \& {Dartois}, E. 2004, \aap, 426,
  925, \dodoi{10.1051/0004-6361:20041276}

\bibitem[{{Pontoppidan} {et~al.}(2024){Pontoppidan}, {Salyk}, {Banzatti},
  {Zhang}, {Pascucci}, {{\"O}berg}, {Long}, {Mu{\~n}oz-Romero}, {Carr},
  {Najita}, {Blake}, {Arulanantham}, {Andrews}, {Ballering}, {Bergin},
  {Calahan}, {Cobb}, {Colmenares}, {Dickson-Vandervelde}, {Dignan}, {Green},
  {Heretz}, {Herczeg}, {Kalyaan}, {Krijt}, {Pauly}, {Pinilla}, {Trapman}, \&
  {Xie}}]{pontoppidan2024}
{Pontoppidan}, K.~M., {Salyk}, C., {Banzatti}, A., {et~al.} 2024, \apj, 963,
  158, \dodoi{10.3847/1538-4357/ad20f0}

\bibitem[{{Rigby} {et~al.}(2023){Rigby}, {Perrin}, {McElwain}, {Kimble},
  {Friedman}, {Lallo}, {Doyon}, {Feinberg}, {Ferruit}, {Glasse}, {Rieke},
  {Rieke}, {Wright}, {Willott}, {Colon}, {Milam}, {Neff}, {Stark}, {Valenti},
  {Abell}, {Abney}, {Abul-Huda}, {Acton}, {Adams}, {Adler}, {Aguilar}, {Ahmed},
  {Albert}, {Alberts}, {Aldridge}, {Allen}, {Altenburg},
  {{\'A}lvarez-M{\'a}rquez}, {Alves de Oliveira}, {Andersen}, {Anderson},
  {Anderson}, {Argyriou}, {Armstrong}, {Arribas}, {Artigau}, {Arvai},
  {Atkinson}, {Bacon}, {Bair}, {Banks}, {Barrientes}, {Barringer}, {Bartosik},
  {Bast}, {Baudoz}, {Beatty}, {Bechtold}, {Beck}, {Bergeron}, {Bergkoetter},
  {Bhatawdekar}, {Birkmann}, {Blazek}, {Blome}, {Boccaletti}, {B{\"o}ker},
  {Boia}, {Bonaventura}, {Bond}, {Bosley}, {Boucarut}, {Bourque}, {Bouwman},
  {Bower}, {Bowers}, {Boyer}, {Bradley}, {Brady}, {Braun}, {Breda},
  {Bresnahan}, {Bright}, {Britt}, {Bromenschenkel}, {Brooks}, {Brooks},
  {Brown}, {Brown}, {Brown}, {Bunker}, {Burger}, {Bushouse}, {Cale}, {Cameron},
  {Cameron}, {Canipe}, {Caplinger}, {Caputo}, {Cara}, {Carey}, {Carniani},
  {Carrasquilla}, {Carruthers}, {Case}, {Catherine}, {Chance}, {Chapman},
  {Charlot}, {Charlow}, {Chayer}, {Chen}, {Cherinka}, {Chichester}, {Chilton},
  {Chonis}, {Clampin}, {Clark}, {Clark}, {Coe}, {Coleman}, {Comber}, {Comeau},
  {Connolly}, {Cooper}, {Cooper}, {Coppock}, {Correnti}, {Cossou}, {Coulais},
  {Coyle}, {Cracraft}, {Curti}, {Cuturic}, {Davis}, {Davis}, {Dean}, {DeLisa},
  {deMeester}, {Dencheva}, {Dencheva}, {DePasquale}, {Deschenes}, {Hunor
  Detre}, {Diaz}, {Dicken}, {DiFelice}, {Dillman}, {Dixon}, {Doggett},
  {Donaldson}, {Douglas}, {DuPrie}, {Dupuis}, {Durning}, {Easmin}, {Eck},
  {Edeani}, {Egami}, {Ehrenwinkler}, {Eisenhamer}, {Eisenhower}, {Elie},
  {Elliott}, {Elliott}, {Ellis}, {Engesser}, {Espinoza}, {Etienne}, {Etxaluze},
  {Falini}, {Feeney}, {Ferry}, {Filippazzo}, {Fincham}, {Fix}, {Flagey},
  {Florian}, {Flynn}, {Fontanella}, {Ford}, {Forshay}, {Fox}, {Franz}, {Fu},
  {Fullerton}, {Galkin}, {Galyer}, {Garc{\'\i}a Mar{\'\i}n}, {Gardner},
  {Gardner}, {Garland}, {Garrett}, {Gasman}, {Gaspar}, {Gaudreau}, {Gauthier},
  {Geers}, {Geithner}, {Gennaro}, {Giardino}, {Girard}, {Giuliano},
  {Glassmire}, {Glauser}, {Glazer}, {Godfrey}, {Golimowski}, {Gollnitz},
  {Gong}, {Gonzaga}, {Gordon}, {Gordon}, {Goudfrooij}, {Greene}, {Greenhouse},
  {Grimaldi}, {Groebner}, {Grundy}, {Guillard}, {Gutman}, {Ha}, {Haderlein},
  {Hagedorn}, {Hainline}, {Haley}, {Hami}, {Hamilton}, {Hammel}, {Hansen},
  {Harkins}, {Harr}, {Hart}, {Hart}, {Hartig}, {Hashimoto}, {Haskins},
  {Hathaway}, {Havey}, {Hayden}, {Hecht}, {Heller-Boyer}, {Henriques}, {Henry},
  {Hermann}, {Hernandez}, {Hesman}, {Hicks}, {Hilbert}, {Hines}, {Hoffman},
  {Holfeltz}, {Holler}, {Hoppa}, {Hott}, {Howard}, {Howard}, {Hunter},
  {Hunter}, {Hurst}, {Husemann}, {Hustak}, {Ilinca Ignat}, {Illingworth},
  {Irish}, {Jackson}, {Jahromi}, {Jakobsen}, {James}, {James}, {Januszewski},
  {Jenkins}, {Jirdeh}, {Johnson}, {Johnson}, {Jones}, {Jones}, {Jones},
  {Jones}, {Jordan}, {Jordan}, {Jurczyk}, {Jurling}, {Kaleida}, {Kalmanson},
  {Kammerer}, {Kang}, {Kao}, {Karakla}, {Kavanagh}, {Kelly}, {Kendrew},
  {Kennedy}, {Kenny}, {Keski-kuha}, {Keyes}, {Kidwell}, {Kinzel}, {Kirk},
  {Kirkpatrick}, {Kirshenblat}, {Klaassen}, {Knapp}, {Knight}, {Knollenberg},
  {Koehler}, {Koekemoer}, {Kovacs}, {Kulp}, {Kumari}, {Kyprianou}, {La Massa},
  {Labador}, {Labiano}, {Lagage}, {Lajoie}, {Lallo}, {Lam}, {Lamb}, {Lambros},
  {Lampenfield}, {Langston}, {Larson}, {Law}, {Lawrence}, {Lee}, {Leisenring},
  {Lepo}, {Leveille}, {Levenson}, {Levine}, {Levy}, {Lewis}, {Lewis},
  {Libralato}, {Lightsey}, {Link}, {Liu}, {Lo}, {Lockwood}, {Logue}, {Long},
  {Long}, {Loomis}, {Lopez-Caniego}, {Lorenzo Alvarez}, {Love-Pruitt}, {Lucy},
  {Luetzgendorf}, {Maghami}, {Maiolino}, {Major}, {Malla}, {Malumuth},
  {Manjavacas}, {Mannfolk}, {Marrione}, {Marston}, {Martel}, {Maschmann},
  {Masci}, {Masciarelli}, {Maszkiewicz}, {Mather}, {McKenzie}, {McLean},
  {McMaster}, {Melbourne}, {Mel{\'e}ndez}, {Menzel}, {Merz}, {Meyett}, {Meza},
  {Miskey}, {Misselt}, {Moller}, {Morrison}, {Morse}, {Moseley}, {Mosier},
  {Mountain}, {Mueckay}, {Mueller}, {Mullally}, {Murphy}, {Murray}, {Murray},
  {Mustelier}, {Muzerolle}, {Mycroft}, {Myers}, {Myrick}, {Nanavati}, {Nance},
  {Nayak}, {Naylor}, {Nelan}, {Nickson}, {Nielson}, {Nieto-Santisteban},
  {Nikolov}, {Noriega-Crespo}, {O'Shaughnessy}, {O'Sullivan}, {Ochs}, {Ogle},
  {Oleszczuk}, {Olmsted}, {Osborne}, {Ottens}, {Owens}, {Pacifici}, {Pagan},
  {Page}, {Park}, {Parrish}, {Patapis}, {Paul}, {Pauly}, {Pavlovsky}, {Pedder},
  {Peek}, {Pena-Guerrero}, {Penanen}, {Perez}, {Perna}, {Perriello},
  {Phillips}, {Pietraszkiewicz}, {Pinaud}, {Pirzkal}, {Pitman}, {Piwowar},
  {Platais}, {Player}, {Plesha}, {Pollizi}, {Polster}, {Pontoppidan},
  {Porterfield}, {Proffitt}, {Pueyo}, {Pulliam}, {Quirt}, {Quispe Neira},
  {Ramos Alarcon}, {Ramsay}, {Rapp}, {Rapp}, {Rauscher}, {Ravindranath},
  {Rawle}, {Regan}, {Reichard}, {Reis}, {Ressler}, {Rest}, {Reynolds}, {Rhue},
  {Richon}, {Rickman}, {Ridgaway}, {Ritchie}, {Rix}, {Robberto}, {Robinson},
  {Robinson}, {Robinson}, {Rock}, {Rodriguez}, {Rodriguez Del Pino}, {Roellig},
  {Rohrbach}, {Roman}, {Romelfanger}, {Rose}, {Roteliuk}, {Roth}, {Rothwell},
  {Rowlands}, {Roy}, {Royer}, {Royle}, {Rui}, {Rumler}, {Runnels}, {Russ},
  {Rustamkulov}, {Ryden}, {Ryer}, {Sabata}, {Sabatke}, {Sabbi}, {Samuelson},
  {Sapp}, {Sappington}, {Sargent}, {Sauer}, {Scheithauer}, {Schlawin},
  {Schlitz}, {Schmitz}, {Schneider}, {Schreiber}, {Schulze}, {Schwab}, {Scott},
  {Sembach}, {Shanahan}, {Shaughnessy}, {Shaw}, {Shawger}, {Shay}, {Sheehan},
  {Shen}, {Sherman}, {Shiao}, {Shih}, {Shivaei}, {Sienkiewicz}, {Sing},
  {Sirianni}, {Sivaramakrishnan}, {Skipper}, {Sloan}, {Slocum}, {Slowinski},
  {Smith}, {Smith}, {Smith}, {Smith}, {Snyder}, {Soh}, {Sohn}, {Soto},
  {Spencer}, {Stallcup}, {Stansberry}, {Starr}, {Starr}, {Stewart},
  {Stiavelli}, {Straughn}, {Strickland}, {Stys}, {Summers}, {Sun}, {Sunnquist},
  {Swade}, {Swam}, {Swaters}, {Swoish}, {Taylor}, {Taylor}, {Te Plate}, {Tea},
  {Teague}, {Telfer}, {Temim}, {Thatte}, {Thompson}, {Thompson}, {Thomson},
  {Tikkanen}, {Tippet}, {Todd}, {Toolan}, {Tran}, {Trejo}, {Truong},
  {Tsukamoto}, {Tustain}, {Tyra}, {Ubeda}, {Underwood}, {Uzzo}, {Van Campen},
  {Vandal}, {Vandenbussche}, {Vila}, {Volk}, {Wahlgren}, {Waldman}, {Walker},
  {Wander}, {Warfield}, {Warner}, {Wasiak}, {Watkins}, {Weaver}, {Weilert},
  {Weiser}, {Weiss}, {Weissman}, {Welty}, {West}, {Wheate}, {Wheatley},
  {Wheeler}, {White}, {Whiteaker}, {Whitehouse}, {Whiteleather}, {Whitman},
  {Williams}, {Willmer}, {Willoughby}, {Wilson}, {Wirth}, {Wislowski}, {Wolf},
  {Wolfe}, {Wolff}, {Workman}, {Wright}, {Wu}, {Wu}, {Wymer}, {Yates},
  {Yeager}, {Yeates}, {Yerger}, {Yoon}, {Young}, {Yu}, {Zak}, {Zeidler},
  {Zhou}, {Zielinski}, {Zincke}, \& {Zonak}}]{rigby2023}
{Rigby}, J., {Perrin}, M., {McElwain}, M., {et~al.} 2023, \pasp, 135, 048001,
  \dodoi{10.1088/1538-3873/acb293}

\bibitem[{{Salyk}(2022)}]{spectools2022}
{Salyk}, C. 2022, {csalyk/spectools\_ir: First release}, v1.0.0,  Zenodo,
  \dodoi{10.5281/zenodo.5818682}

\bibitem[{{Salyk} {et~al.}(2011){Salyk}, {Pontoppidan}, {Blake}, {Najita}, \&
  {Carr}}]{salyk2011}
{Salyk}, C., {Pontoppidan}, K.~M., {Blake}, G.~A., {Najita}, J.~R., \& {Carr},
  J.~S. 2011, \apj, 731, 130, \dodoi{10.1088/0004-637X/731/2/130}

\bibitem[{{Sano} {et~al.}(2000){Sano}, {Miyama}, {Umebayashi}, \&
  {Nakano}}]{sano2000}
{Sano}, T., {Miyama}, S.~M., {Umebayashi}, T., \& {Nakano}, T. 2000, \apj, 543,
  486, \dodoi{10.1086/317075}

\bibitem[{{Sch{\"o}ier} {et~al.}(2005){Sch{\"o}ier}, {van der Tak}, {van
  Dishoeck}, \& {Black}}]{schoier2005}
{Sch{\"o}ier}, F.~L., {van der Tak}, F.~F.~S., {van Dishoeck}, E.~F., \&
  {Black}, J.~H. 2005, \aap, 432, 369, \dodoi{10.1051/0004-6361:20041729}

\bibitem[{{Schwarz} {et~al.}(2024){Schwarz}, {Henning}, {Christiaens},
  {Gasman}, {Samland}, {Perotti}, {Jang}, {Grant}, {Tabone},
  {Morales-Calder{\'o}n}, {Kamp}, {van Dishoeck}, {G{\"u}del}, {Lagage},
  {Barrado}, {Caratti o Garatti}, {Glauser}, {Ray}, {Vandenbussche}, {Waters},
  {Arabhavi}, {Kanwar}, {Olofsson}, {Rodgers-Lee}, {Schreiber}, \&
  {Temmink}}]{schwarz2024}
{Schwarz}, K.~R., {Henning}, T., {Christiaens}, V., {et~al.} 2024, \apj, 962,
  8, \dodoi{10.3847/1538-4357/ad1393}

\bibitem[{{Stevenson}(1990)}]{Stevenson1990}
{Stevenson}, D.~J. 1990, \apj, 348, 730, \dodoi{10.1086/168282}

\bibitem[{{Tabone} {et~al.}(2024){Tabone}, {van Dishoeck}, \&
  {Black}}]{tabone2024}
{Tabone}, B., {van Dishoeck}, E.~F., \& {Black}, J.~H. 2024, arXiv e-prints,
  arXiv:2406.14560, \dodoi{10.48550/arXiv.2406.14560}

\bibitem[{{Tabone} {et~al.}(2021){Tabone}, {van Hemert}, {van Dishoeck}, \&
  {Black}}]{tabone2021}
{Tabone}, B., {van Hemert}, M.~C., {van Dishoeck}, E.~F., \& {Black}, J.~H.
  2021, \aap, 650, A192, \dodoi{10.1051/0004-6361/202039549}

\bibitem[{{Tabone} {et~al.}(2023){Tabone}, {Bettoni}, {van Dishoeck},
  {Arabhavi}, {Grant}, {Gasman}, {Henning}, {Kamp}, {G{\"u}del}, {Lagage},
  {Ray}, {Vandenbussche}, {Abergel}, {Absil}, {Argyriou}, {Barrado},
  {Boccaletti}, {Bouwman}, {Caratti o Garatti}, {Geers}, {Glauser},
  {Justannont}, {Lahuis}, {Mueller}, {Nehm{\'e}}, {Olofsson}, {Pantin},
  {Scheithauer}, {Waelkens}, {Waters}, {Black}, {Christiaens}, {Guadarrama},
  {Morales-Calder{\'o}n}, {Jang}, {Kanwar}, {Pawellek}, {Perotti}, {Perrin},
  {Rodgers-Lee}, {Samland}, {Schreiber}, {Schwarz}, {Colina}, {{\"O}stlin}, \&
  {Wright}}]{tabone2023}
{Tabone}, B., {Bettoni}, G., {van Dishoeck}, E.~F., {et~al.} 2023, Nature
  Astronomy, 7, 805, \dodoi{10.1038/s41550-023-01965-3}

\bibitem[{{Temmink} {et~al.}(2024{\natexlab{a}}){Temmink}, {van Dishoeck},
  {Grant}, {Tabone}, {Gasman}, {Christiaens}, {Samland}, {Argyriou}, {Perotti},
  {G{\"u}del}, {Henning}, {Lagage}, {Abergel}, {Absil}, {Barrado}, {Caratti o
  Garatti}, {Glauser}, {Kamp}, {Lahuis}, {Olofsson}, {Ray}, {Scheithauer},
  {Vandenbussche}, {Waters}, {Arabhavi}, {Jang}, {Kanwar},
  {Morales-Calder{\'o}n}, {Rodgers-Lee}, {Schreiber}, {Schwarz}, \&
  {Colina}}]{temmink2024}
{Temmink}, M., {van Dishoeck}, E.~F., {Grant}, S.~L., {et~al.}
  2024{\natexlab{a}}, \aap, 686, A117, \dodoi{10.1051/0004-6361/202348911}

\bibitem[{{Temmink} {et~al.}(2024{\natexlab{b}}){Temmink}, {van Dishoeck},
  {Gasman}, {Grant}, {Tabone}, {G{\"u}del}, {Henning}, {Barrado}, {Caratti o
  Garatti}, {Glauser}, {Kamp}, {Arabhavi}, {Jang}, {Kurtovic}, {Perotti},
  {Schwarz}, \& {Vlasblom}}]{temmink2024b}
{Temmink}, M., {van Dishoeck}, E.~F., {Gasman}, D., {et~al.}
  2024{\natexlab{b}}, \aap, 689, A330, \dodoi{10.1051/0004-6361/202450355}

\bibitem[{{Turner} {et~al.}(2014){Turner}, {Fromang}, {Gammie}, {Klahr},
  {Lesur}, {Wardle}, \& {Bai}}]{turner2014}
{Turner}, N.~J., {Fromang}, S., {Gammie}, C., {et~al.} 2014, in Protostars and
  Planets VI, ed. H.~{Beuther}, R.~S. {Klessen}, C.~P. {Dullemond}, \&
  T.~{Henning}, 411--432, \dodoi{10.2458/azu_uapress_9780816531240-ch018}

\bibitem[{{Villenave} {et~al.}(2020){Villenave}, {M{\'e}nard}, {Dent},
  {Duch{\^e}ne}, {Stapelfeldt}, {Benisty}, {Boehler}, {van der Plas}, {Pinte},
  {Telkamp}, {Wolff}, {Flores}, {Lesur}, {Louvet}, {Riols}, {Dougados},
  {Williams}, \& {Padgett}}]{Villenave20}
{Villenave}, M., {M{\'e}nard}, F., {Dent}, W.~R.~F., {et~al.} 2020, \aap, 642,
  A164, \dodoi{10.1051/0004-6361/202038087}

\bibitem[{{Walsh} {et~al.}(2015){Walsh}, {Nomura}, \& {van
  Dishoeck}}]{walsh2015}
{Walsh}, C., {Nomura}, H., \& {van Dishoeck}, E. 2015, \aap, 582, A88,
  \dodoi{10.1051/0004-6361/201526751}

\bibitem[{{Watson} {et~al.}(2009){Watson}, {Leisenring}, {Furlan}, {Bohac},
  {Sargent}, {Forrest}, {Calvet}, {Hartmann}, {Nordhaus}, {Green}, {Kim},
  {Sloan}, {Chen}, {Keller}, {d'Alessio}, {Najita}, {Uchida}, \&
  {Houck}}]{watson2009}
{Watson}, D.~M., {Leisenring}, J.~M., {Furlan}, E., {et~al.} 2009, \apjs, 180,
  84, \dodoi{10.1088/0067-0049/180/1/84}

\bibitem[{{Wells} {et~al.}(2015){Wells}, {Pel}, {Glasse}, {Wright},
  {Aitink-Kroes}, {Azzollini}, {Beard}, {Brandl}, {Gallie}, {Geers}, {Glauser},
  {Hastings}, {Henning}, {Jager}, {Justtanont}, {Kruizinga}, {Lahuis}, {Lee},
  {Martinez-Delgado}, {Mart{\'\i}nez-Galarza}, {Meijers}, {Morrison},
  {M{\"u}ller}, {Nakos}, {O'Sullivan}, {Oudenhuysen}, {Parr-Burman}, {Pauwels},
  {Rohloff}, {Schmalzl}, {Sykes}, {Thelen}, {van Dishoeck}, {Vandenbussche},
  {Venema}, {Visser}, {Waters}, \& {Wright}}]{wells2015}
{Wells}, M., {Pel}, J.~W., {Glasse}, A., {et~al.} 2015, \pasp, 127, 646,
  \dodoi{10.1086/682281}

\bibitem[{{Werner} {et~al.}(2004){Werner}, {Roellig}, {Low}, {Rieke}, {Rieke},
  {Hoffmann}, {Young}, {Houck}, {Brandl}, {Fazio}, {Hora}, {Gehrz}, {Helou},
  {Soifer}, {Stauffer}, {Keene}, {Eisenhardt}, {Gallagher}, {Gautier}, {Irace},
  {Lawrence}, {Simmons}, {Van Cleve}, {Jura}, {Wright}, \&
  {Cruikshank}}]{werner2004}
{Werner}, M.~W., {Roellig}, T.~L., {Low}, F.~J., {et~al.} 2004, \apjs, 154, 1,
  \dodoi{10.1086/422992}

\bibitem[{{Woitke} {et~al.}(2009){Woitke}, {Kamp}, \& {Thi}}]{woitke09}
{Woitke}, P., {Kamp}, I., \& {Thi}, W.-F. 2009, \aap, 501, 383,
  \dodoi{10.1051/0004-6361/200911821}

\bibitem[{{Woods} \& {Willacy}(2009)}]{Woods2009}
{Woods}, P.~M., \& {Willacy}, K. 2009, \apj, 693, 1360,
  \dodoi{10.1088/0004-637X/693/2/1360}

\bibitem[{{Wright} {et~al.}(2023){Wright}, {Rieke}, {Glasse}, {Ressler},
  {Garc{\'\i}a Mar{\'\i}n}, {Aguilar}, {Alberts}, {{\'A}lvarez-M{\'a}rquez},
  {Argyriou}, {Banks}, {Baudoz}, {Boccaletti}, {Bouchet}, {Bouwman}, {Brandl},
  {Breda}, {Bright}, {Cale}, {Colina}, {Cossou}, {Coulais}, {Cracraft}, {De
  Meester}, {Dicken}, {Engesser}, {Etxaluze}, {Fox}, {Friedman}, {Fu},
  {Gasman}, {G{\'a}sp{\'a}r}, {Gastaud}, {Geers}, {Glauser}, {Gordon},
  {Greene}, {Greve}, {Grundy}, {G{\"u}del}, {Guillard}, {Haderlein},
  {Hashimoto}, {Henning}, {Hines}, {Holler}, {Detre}, {Jahromi}, {James},
  {Jones}, {Justtanont}, {Kavanagh}, {Kendrew}, {Klaassen}, {Krause},
  {Labiano}, {Lagage}, {Lambros}, {Larson}, {Law}, {Lee}, {Libralato}, {Lorenzo
  Alverez}, {Meixner}, {Morrison}, {Mueller}, {Murray}, {Mycroft}, {Myers},
  {Nayak}, {Naylor}, {Nickson}, {Noriega-Crespo}, {{\"O}stlin}, {O'Sullivan},
  {Ottens}, {Patapis}, {Penanen}, {Pietraszkiewicz}, {Ray}, {Regan},
  {Roteliuk}, {Royer}, {Samara-Ratna}, {Samuelson}, {Sargent}, {Scheithauer},
  {Schneider}, {Schreiber}, {Shaughnessy}, {Sheehan}, {Shivaei}, {Sloan},
  {Tamas}, {Teague}, {Temim}, {Tikkanen}, {Tustain}, {van Dishoeck},
  {Vandenbussche}, {Weilert}, {Whitehouse}, \& {Wolff}}]{wright2023}
{Wright}, G.~S., {Rieke}, G.~H., {Glasse}, A., {et~al.} 2023, \pasp, 135,
  048003, \dodoi{10.1088/1538-3873/acbe66}

\bibitem[{{Xie} {et~al.}(2023){Xie}, {Pascucci}, {Long}, {Pontoppidan},
  {Banzatti}, {Kalyaan}, {Salyk}, {Liu}, {Najita}, {Pinilla}, {Arulanantham},
  {Herczeg}, {Carr}, {Bergin}, {Ballering}, {Krijt}, {Blake}, {Zhang},
  {{\"O}berg}, {Green}, \& {Jdiscs Collaboration}}]{Xie2023}
{Xie}, C., {Pascucci}, I., {Long}, F., {et~al.} 2023, \apjl, 959, L25,
  \dodoi{10.3847/2041-8213/ad0ed9}

\bibitem[{{Yoshida} {et~al.}(2024){Yoshida}, {Nomura}, {Furuya}, {Teague},
  {Law}, {Tsukagoshi}, {Lee}, {Rab}, {{\"O}berg}, \& {Loomis}}]{Yoshida2024}
{Yoshida}, T.~C., {Nomura}, H., {Furuya}, K., {et~al.} 2024, \apj, 966, 63,
  \dodoi{10.3847/1538-4357/ad2fb4}

\bibitem[{{Youdin} \& {Lithwick}(2007)}]{youdin2007}
{Youdin}, A.~N., \& {Lithwick}, Y. 2007, \icarus, 192, 588,
  \dodoi{10.1016/j.icarus.2007.07.012}

\bibitem[{{Zhang} {et~al.}(2021){Zhang}, {Booth}, {Law}, {Bosman}, {Schwarz},
  {Bergin}, {{\"O}berg}, {Andrews}, {Guzm{\'a}n}, {Walsh}, {Qi}, {van't Hoff},
  {Long}, {Wilner}, {Huang}, {Czekala}, {Ilee}, {Cataldi}, {Bergner}, {Aikawa},
  {Teague}, {Bae}, {Loomis}, {Calahan}, {Alarc{\'o}n}, {M{\'e}nard}, {Le Gal},
  {Sierra}, {Yamato}, {Nomura}, {Tsukagoshi}, {P{\'e}rez}, {Trapman}, {Liu}, \&
  {Furuya}}]{Zhang21_mapsco}
{Zhang}, K., {Booth}, A.~S., {Law}, C.~J., {et~al.} 2021, \apjs, 257, 5,
  \dodoi{10.3847/1538-4365/ac1580}

\end{thebibliography}
\bibliographystyle{aasjournal}

\end{CJK*}
\end{document}